\documentclass[aps,prd,twocolumn,nofootinbib,superscriptaddress]{revtex4-1}
\usepackage{graphics}
\usepackage{epsfig}
\usepackage[english]{babel}
\usepackage{amsmath} 
\usepackage{verbatim} 
\usepackage{mathrsfs}
\usepackage{amsfonts}
\usepackage{amssymb}
\usepackage[latin1]{inputenc}
\usepackage{hyperref}

\newcommand{\tc}{\eta_k^c}
\newcommand{\beq}{\begin{equation}}
\newcommand{\eeq}{\end{equation}}
\newcommand{\nk}{\textbf{k}}
\newcommand{\dphi}{\delta \phi}
\newcommand{\x}{\textbf{x}}

\newcommand{\bra}{\langle}
\newcommand{\ket}{\rangle}

\newcommand{\mH}{\mathcal{H}}

\newcommand{\zk}{|z_k|}

\newcommand{\barr}{\begin{eqnarray}}
\newcommand{\earr}{\end{eqnarray}}
\newcommand{\bea}{\begin{eqnarray*}}
\newcommand{\eea}{\end{eqnarray*}}

\newcommand{\nn}{\nonumber \\}

\newcommand{\half}{\frac{1}{2}}

\newcommand{\cre}{\hat{a}^{\dagger}}    

\newcommand{\ann}{\hat{a}}

\newcommand{\mR}{\mathcal{R}}

\newcommand{\psiind}{\Psi_{\nk}^{\text{ind} }}
\newcommand{\psinew}{\Psi_{\nk}^{\text{newt} }}
\newcommand{\psiwig}{\Psi_{\nk}^{\text{wig} }}

\begin{document}

\title{Inflation including collapse of the wave function: The quasi-de 
Sitter case}

\author{Gabriel Le\'{o}n}
\affiliation{Departamento de F\'{\i}sica, Facultad de Ciencias Exactas y 
Naturales, Universidad de Buenos Aires, Ciudad Universitaria - PabI, Buenos 
Aires 1428, Argentina}

\author{Susana J. Landau}
\affiliation{Departamento de F\'{\i}sica, Facultad de Ciencias Exactas y
Naturales, Universidad de Buenos Aires and IFIBA, CONICET, Ciudad Universitaria 
- PabI, Buenos Aires 1428, Argentina}

\author{Mar\'{\i}a P\'{\i}a Piccirilli}
\affiliation{Grupo de Astrof\'{\i}sica, Relatividad y Cosmolog\'{\i}a, Facultad 
de Ciencias Astron\'{o}micas y Geof\'{\i}sicas, Universidad Nacional de La 
Plata, Paseo del Bosque S/N 1900 La Plata, Pcia de Buenos Aires, Argentina}

\begin{abstract}
The precise physical mechanism describing the emergence of the seeds of cosmic 
structure from a perfect isotropic and homogeneous universe has not been fully 
explained by the standard version of inflationary  models. To handle this 
shortcoming, D. Sudarsky and collaborators have developed a proposal: \emph{the 
self-induced collapse hypothesis.} In this scheme, the objective collapse of 
the inflaton wave function is responsible for the emergence of inhomogeneity 
and anisotropy at all scales. In previous papers, the proposal was developed 
with an almost exact de Sitter space-time approximation for the background that 
led to a perfect scale-invariant power spectrum. In the present article, we 
consider a full  quasi-de Sitter expansion and calculate the primordial power 
spectrum for three different choices of the self-induced collapse. The 
consideration of a quasi-de Sitter background allow us to distinguish 
departures from an exact scale-invariant power spectrum that are due to the 
inclusion of  the collapse hypothesis. These deviations are also different from 
the prediction of standard inflationary models with running spectral index. 
Comparison with the primordial power spectrum and the CMB temperature  
fluctuation spectrum preferred by the latest observational data is also 
discussed. From the analysis performed in this work, it follows that most of 
the collapse schemes analyzed in this paper are viable candidates to explain 
present observations of the CMB fluctuation spectrum.
\end{abstract}

\maketitle

\section{Introduction}
\label{intro}
Recent observations of the Cosmic Microwave Background (CMB) radiation are one  
of the most powerful tools to study the early universe and also to provide a 
method to settle the value of the cosmological parameters. In the last 20 
years, there has been a lot of improvements in the measurements precision  of 
the CMB 
radiation anisotropies. Furthermore, the agreement between theory and 
observations has strengthened the theoretical status of inflationary scenarios 
among cosmologists.

In the standard inflationary paradigm, the emergence of all structures in our 
universe like galaxies and galaxy clusters are described by  a featureless 
stage 
represented by a background Friedmann-Robertson-Walker (FRW) cosmology with a 
nearly exponential expansion driven by the potential of a single scalar field 
and from its quantum fluctuations characterized by a simple vacuum state. 
However, when this scenario is considered more carefully, a conceptual problem 
emerges regarding a change in the initial symmetries of the universe.  This is, 
from a highly homogeneous and isotropic initial state that characterizes the 
quantum perturbations of both the classical 
background inflaton and space-time, the universe ends in a state with ``real'' 
inhomogeneities and anisotropies. In other words, if one considers quantum 
mechanics as a fundamental theory, then it is appropriate to use it in order to 
describe the universe as a whole; therefore, every classical description of the 
universe shall be associated to an imprecise highly complex quantum mechanical 
state. Moreover, the observed universe, at certain scales, does exhibit 
inhomogeneous and anisotropic features; consequently, its quantum description 
in 
terms of a quantum state must encode these non-symmetrical aspects. On the 
other 
hand, the dynamics of the standard inflationary paradigm does not contain any 
aspect that can be responsible for breaking the initial symmetries of the early 
quantum state, which happened to be perfectly homogeneous and isotropic. In 
this 
sense we consider that the standard inflationary paradigm is incomplete. D. 
Sudarsky and collaborators 
\cite{PSS06,Shortcomings,US08,Leon10,DT11,Leon11,LSS12,LLS13,pia,CPS13} 
have developed one proposal to handle these shortcomings. The way  to deal with 
the problem is  introducing a new ingredient into the inflationary account of 
the origin of cosmic seeds: \emph{the self-induced collapse hypothesis}, i.e. a 
scheme in which an internally induced collapse of the wave function of the 
inflaton field is the mechanism by which inhomogeneities and anisotropies arise 
at each particular length scale.

The collapse proposal was inspired by previous ideas of R. Penrose and L.  
Di\'osi 
\cite{penrose1996,diosi1987,diosi1989,ghirardi1985}, which regarded the 
collapse 
of the wave function as an actual physical process induced by gravity (instead 
of just an artifact of the description of Physics). At this point of the 
discussion, we do not know exactly what kind of physical mechanism would lie 
behind what, at the semiclassical level we are working, looks like a 
spontaneous 
collapse of the wave function. We assume that the effect is caused by unknown 
quantum aspects of gravitation. Essentially, the collapse hypothesis simply 
sustains that something intrinsic to the system, i.e. independent of external 
agents (e.g. observers), induces the collapse or reduction of the quantum 
mechanical state of the system. Various proposals of that sort have been 
developed
\cite{ghirardi1985,pearle1989,bassi2003,bassi2013}, and might well be 
compatible with the self-induced collapse of the inflaton's wave function that 
we are considering. The proposal is, at this stage of the analysis, a purely 
phenomenological scheme. It does not attempt to explain the process in terms of 
some specific new physical theory, but
merely gives a rather general parametrization of the quantum transition 
involved. We will refer to this phenomenological model as the collapse scheme. 

Here, it is worthwhile to mention that the previous conceptual problem is 
sometimes known in the literature as the quantum-to-classical transition of the 
primordial perturbations. As a matter of fact, a partially understanding of 
such issue has been gained by relying on the decoherence framework and the fact 
that the initial vacuum state of the inflaton evolves into a highly squeezed 
state \cite{kiefer,grishchuk}; in particular, it is usually argued that the 
predictions from the quantum theory, characterizing the inflaton fluctuations, 
are indistinguishable from those of a theory in which the random fluctuations 
are the result of a classical stochastic process \cite{jmartin2007}. However, 
this argument alone cannot explain the fact that a single (classical and 
random) outcome emerges from the quantum theory \cite{Mukhanov2005}. In other 
words, decoherence (and the squeezing of a quantum state) cannot solve the 
quantum measurement problem \cite{adler,schlosshauer}, which in the 
cosmological setting is amplified, i.e. there is no clear way to define 
entities such as observers, measurement devices, environmental degrees of 
freedom, etc. Other cosmologists seem to adopt the Everett  ``many-worlds'' 
interpretation of quantum mechanics when confronted with the 
quantum-to-classical transition in the inflationary universe (although a 
detailed and precise formulation of such posture is still not well represented  
in the literature). In the Everettian formulation, reality is 
made of a connected weave of ever-splitting worlds, each one realizing one of 
the alternatives that is opened by what we would call a quantum-mechanical 
measurement. Regarding this point, we want to mention that 
Everett's interpretation has not completely solved the measurement problem yet. 
In fact, there is a mapping between what in that approach would be called the 
splittings of the  worlds, and what would be called the ``measurements'' in 
the Copenhagen interpretation. Therefore, every question that can be made in 
the orthodox interpretation has a corresponding one in the Everettian one. That 
is, the specific  issues regarding the measurement problem would be: When and 
why does a world splitting occur?, under what circumstances does it occur? 
What constitutes a trigger? Furthermore, even if one could bypass those 
conundrums, a precise derivation of the Born rule (a crucial aspect to connect 
the theory's predictions with the experiments) and a clear justification for 
the choice of a particular basis in which the splitting takes place is unknown 
within  Everett's interpretation \cite{kent,stapp}.

Given the previous discussion, an objective reduction of the quantum 
state, characterizing the inflaton field, seems to be a plausible option for 
addressing the problem at hand. The detailed analysis of the original 
proposal at the conceptual and technical level can be consulted in Refs. 
\cite{PSS06,Shortcomings}.

Furthermore, even though there are well-known models characterizing the 
self-induced collapse of the wave function \cite{bassi2003,bassi2013} in a 
generic 
(non-cosmological) context, the relativistic framework for the objective 
collapse 
models  is still under development \cite{bedingham2010,pearle2014}. In this 
work, we 
will follow a more pragmatical approach and characterize the post-collapse 
quantum 
state by the expectation values of the field and its momentum conjugate, 
without 
relying on some particular collapse mechanism. On the other hand, there are 
still 
various possibilities regarding the description of the quantum expectation 
values in 
the post-collapse state, we will refer to these various prescriptions as 
\emph{collapse schemes}.  In a first attempt \cite{PSS06,US08}, two different 
schemes were considered: one in which, after the collapse, both expectation 
values of the field and momentum variable are randomly distributed within their 
respective ranges of uncertainties in the pre-collapsed state, and another one 
in which it is only the conjugate momentum that changes its expectation value 
from zero to a value in its corresponding range, as a result of the collapse.  
In later works \cite{US08,Leon11}, another scheme was considered motivated by 
the correlation between the field variable and its conjugated momentum  
existing 
in the pre-collapse state; in this scheme the collapse is characterized in 
terms 
of the Wigner function.

In previous papers \cite{US08,LSS12} the collapse proposal was 
developed using an ``almost'' exact de Sitter expansion factor during the 
inflationary period [i.e., it was assumed that the (cosmic) time derivative of 
the  Hubble factor $\dot H$ was exactly a constant]. In consequence, the 
primordial power spectrum, modified by the collapse hypothesis, resulted in a 
form $P(k) = A_s  C(k) $, this is, a power spectrum with scalar spectral index 
$n_s = 1$, and with  $C(k)$ a function representing the modification induced by 
the collapse hypothesis; in particular, the function $C(k)$ depends on the time 
of collapse of each mode of the inflaton field. Furthermore, the temperature 
and 
polarization power spectrum of the CMB were also modified, therefore, the 
proposal predictions  could be tested against current observational data 
\cite{LSS12}. In such way, it was possible to recover an exact scale-invariant 
power spectrum for some values of the collapse parameters that are related to 
the time of collapse of each mode of the inflaton field.

However, recent data reported by \emph{Planck} \cite{Planckcls13} and WMAP 
\cite{wmap9cls}, constrain the value of the spectral scalar index to $n_s = 
0.9603 \pm 0.0073$. Therefore, in this  paper, we go one step beyond and 
calculate the primordial power spectrum for different collapse schemes in a 
full 
quasi-de Sitter background, i.e. by considering small time variations of 
$\dot H$. In this manner, we will obtain an expression of the form $P(k) = A_s 
k^{n_s - 1 } Q(k)$; naturally, with $n_s \neq 1$, and also with $Q(k)$ a 
function introduced by the collapse hypothesis,  which can be reduced to the 
conventional phenomenological expression for some values of the collapse 
parameters.
We emphasize that this work is not only a matter of technical improvement, but 
also  helps to separate the features in the power spectrum that can be 
attributed to the collapse of the wave function and the aspects that are only 
due to considering a quasi-de Sitter background. That is, in previous works  
the 
prediction of the form  $P(k) = A_s  C(k) $ does not allow one to exactly 
identify the dependence on $k$ attributed to the collapse hypothesis and to the 
spectral index; meanwhile, in the prediction obtained in the present 
manuscript, $P(k) = A_s k^{n_s - 1 } Q(k)$, one can plainly recognize and 
separate the two kind of 
dependencies. 
We think that this last feature is important since, at the 
phenomenological level, one wishes to learn more about the unknown collapse 
mechanism, so it is of great significance to clearly identify its 
characteristics within the theoretical predictions that will be confronted with 
observational data. 

Additionally, our prediction for the power spectrum allows 
departures from the traditional inflationary approach that can be tested 
observationally. Moreover, as we will show later in the paper, since our model 
is conceptually different from the standard one, we cannot follow the 
traditional method of simply evaluate the power spectrum, obtained from pure de 
Sitter inflation, at the so-called ``horizon crossing'' and, in this way, 
achieve a power spectrum that is equivalent in shape to the corresponding one 
calculated in a quasi-de Sitter stage.  In order for our model to yield a 
conceptually consistent prediction, we must perform the full calculation in a 
quasi-de Sitter background.

On the other hand, in a recent work \cite{pia} we have analyzed the collapse 
hypothesis in the case that the collapse occurs
during the radiation epoch; and  we have shown that if one considers an almost 
exact de Sitter expansion for the inflationary era, then the model cannot 
account for the present observational CMB temperature fluctuation spectrum. 
This 
last statement also applies to the case where the collapse happens during 
inflation (which is the case of the present paper). In fact,  a simple 
calculation of  the $\chi^2$ value for the models presented in Ref. 
\cite{LSS12}, where an almost exact de Sitter background was considered using  
the CMB temperature data released by Planck \cite{Planckcls13}, the CMB 
polarization data reported by WMAP \cite{wmap9cls}, CMB temperature data for 
high values of $l$ reported by the Atacama Cosmology Project (ACT) \cite{ACTcls} 
and the South Pole Telescope (SPT) \cite{SPT12}, yields a value that is several 
orders of magnitude above the expected reasonable value. The difference between 
the analysis performed in the former paper and this simple calculation is that 
the data obtained by the Planck, ACT and SPT collaborations are much more 
accurate for small angular scales than the 7-years release of WMAP.

The paper is organized as follows: In Sec. \ref{einstein}, we review the 
semiclassical gravity approach, in which only the perturbations of the inflaton 
field are quantized and  obtain the corresponding linearized Einstein's 
equations. In Sec. \ref{quantum}, we perform the quantization of the inflaton 
field in a quasi-de Sitter background. Furthermore, we  relate the initial 
curvature perturbation  with the parameters characterizing each collapse 
scheme; 
we consider  three different choices for the quantum collapse. 
In Sec. \ref{oquantities}, we relate the CMB observational quantities 
with the primordial spectrum modified with the collapse hypothesis in a 
quasi-de 
Sitter background. In Sec. \ref{comparison}, we compare the primordial power 
spectrum obtained in Sec. \ref{oquantities} with the phenomenological 
expression 
in standard inflationary models. 
In Sec. \ref{plots}, we plot the primordial power spectrum obtained in this 
paper 
for some particular values of the collapse parameters and compare it with the 
primordial spectrum preferred by the data. In Sect. \ref{Cls} we present the 
prediction for the CMB temperature fluctuation spectrum and show that there are 
different predictions corresponding to the three collapse schemes proposed 
along with distinct values for the time of collapse. Finally, in Sec. 
\ref{conclusions}, we summarize the main results of the paper and present the 
conclusions.

Regarding notation and conventions, we will work with signature  $(-,+,+,+)$ 
for 
the metric; primes over functions will denote derivatives with respect to the 
conformal time $\eta$, and we will use units  where $c=\hbar=1$ but keep the 
gravitational constant $G$.

 \section{Semiclassical gravity and linearized Einstein's 
equations}\label{einstein}

The purpose of this section is to present our view regarding the relation 
between the space-time description in terms of the metric and the quantum 
degrees of freedom (DOF) of the matter fields, represented by the inflaton. 
First, we will introduce the physical point of view of such relation and then 
derive the corresponding equations.

We have proposed that the mechanism by which primordial anisotropies and 
inhomogeneities arise is a self-induced collapse of the 
inflaton wave function. As a consequence, the post-collapse quantum state 
must not be an homogeneous and isotropic state, this is, it is not an 
eigen-state of the linear and angular momentum operators. One could then assume 
that the post-collapse state was obtained from a particular collapse mechanism, 
and then compute the corresponding observables in that state. The question now 
would be: In the context of the inflationary scenario, what are the appropriate 
observables that result from the quantum theory? 

One possible approach would be to assume that both metric and 
matter perturbations are described  by a quantum field theory constructed on a 
classical unperturbed background; in the inflationary universe, this approach 
corresponds to the quantization of the so-called Mukhanov-Sasaki variable. 
Henceforth, if one assumes a particular collapse mechanism, which  somehow 
modifies the standard unitary evolution given by Schr\"{o}dinger equation, then 
the dynamics of the Mukhanov-Sasaki variable would induce non-standard 
predictions for the observational quantities (e.g. the spectrum of the 
temperature anisotropies). This scheme was developed in Refs. 
\cite{DLS11,jmartin,tpsingh,dasGW} for the inflationary universe.

Another potential prescription to relate the quantum DOF with the observational 
quantities, is to rely on the semiclassical gravity approximation represented 
by 
Einstein semiclassical equations $G_{ab} = 8 \pi G  \bra \hat{T}_{ab} \ket$; 
within this framework, the metric is described in a classical way, while the 
matter DOF are modeled by a quantum field theory in a curved classical 
background. Then, during inflation,  the semiclassical equations enable to 
relate the inflaton quantum  perturbations with the corresponding ones from the 
classical metric. However, assuming a particular collapse mechanism, which 
again 
can be envisioned as a modification of the Schr\"{o}dinger equation, would not 
alter 
the evolution of the metric perturbation; indeed, the dynamics of the modes 

characterizing the quantum field, representing the inflaton,  would be 
modified, 
 but the metric perturbation is always a classical object, and, thus, its 
evolution is not dictated by the modified  Schr\"{o}dinger equation. Assuming a 
particular collapse mechanism, would only modify the initial conditions of the 
motion equation for the metric perturbation, which again is always described at 
the classical level; in the context of inflation, this was analyzed in Ref. 
\cite{CPS13} (by assuming pure de Sitter inflation). Nevertheless, the initial 
condition for the motion equation of the metric perturbation, will contain the 
information regarding that a collapse of the wave function of the inflaton has 
occurred.

One main advantage of using the semiclassical picture is that the description 
and 
treatment of the metric (both the background and its perturbations) remains 
``classical'' at all times. As a consequence, there is no issue with the 
``quantum-to-classical transition" in the sense that one needs to justify going 
from ``metric operators'' (e.g. $\hat \Psi$) to classical metric variables 
(such 
as $\Psi$).  The fact that the space-time remains classical is  particularly 
important in the context of models involving dynamical reduction of the wave 
function, as such collapse or reduction is regarded as a physical process 
taking 
place in time and, therefore, it is clear that a setting allowing  
consideration 
of 
full space-time notions is preferred over, say, the timeless settings usually 
encountered in canonical approaches to quantum gravity (see Ref. \cite{isham} 
for a 
complete review on the problem of time in quantum gravity).

Another aspect of the semiclassical approximation, is that it is not valid 
during the collapse. The reason is that, as is well-known, introducing a 
dynamical collapse generically violates the conservation of energy, so the 
divergence of the energy-momentum tensor does not vanish, $\nabla_a \bra 
\hat{T}^{ab} \ket \neq 0$. If the divergence of the energy-momentum tensor does 
not vanish, and it is equated to the Einstein tensor, then of course the 
latter's divergence does not vanish either, $\nabla_a G^{ab} \neq 0$, which 
evidently is a problem since we know that the divergence of the Einstein tensor 
must be zero. Therefore, during the collapse, we cannot say how the modified 
dynamics of the quantum fields, provided by a collapse mechanism,  affects the 
classical metric perturbations that are directly related to the observables. 
However, the  semiclassical gravity approximation is  valid before and after 
the 
collapse, which correspond to the cases of interest for the present work.

More specifically, before the collapse,  the initial state of the universe, 
characterized by a few $e$-folds after inflation has begun, is described by 
both 
the homogeneous and isotropic  (H\&I) Bunch-Davies vacuum and the H\&I 
Friedmann-Robertson-Walker space-time. Afterwards, at some point during the 
inflationary 
epoch, the quantum state of the matter fields spontaneously changes 
to a new quantum state (i.e. the post-collapse state); this change is induced 
by 
some physical (but unknown) mechanism. However, the state resulting from the 
collapse needs not to share the same symmetries as the initial state. After the 
collapse, the gravitational DOF are assumed to be, once more, accurately 
described by Einstein's semiclassical equations. Nevertheless,  $\bra 
\hat{T}_{ab} \ket$ evaluated in the new state  does not  generically posses 
the 
symmetries of the pre-collapse state; hence, we are led to a new geometry that 
is no longer H\&I. We have presented just a very brief summary of the 
semiclassical picture and its relation with the collapse hypothesis during 
inflation. The full formalism has been  developed in Ref. \cite{DT11} and we 
invite the reader to consult such reference. We also should mention that we are 
not advocating that  semiclassical gravity must be regarded as a fundamental 
theory, we are using it as an appropriate approximation given the energy scales 
of the inflationary universe ($\sim 10^{16}$ GeV), i.e. we are treating it as a 
quantum field theory in a curved space-time.

It is also worthwhile to mention a few remarks regarding the tensor modes 
and the semiclassical gravity approach. Last year, the BICEP2 collaboration  
  reported a detection of the primordial $B$-mode polarization consistent 
with the prediction of standard inflationary models
\cite{BICEP2}. However, other authors pointed out that without an accurate dust 
map,  it is 
not possible to discern  between  dust polarization and polarization due to 
primordial gravity waves \cite{Mortonson14,Flauger14,Liu14}. More recently, the 
Planck collaboration analysis of the angular power spectrum of polarized dust 
emission at intermediate and high Galactic latitudes \cite{Planckdust} showed 
that the dust extrapolated power spectrum (obtained by extrapolating Planck 353 
GHz data to 150 GHz) is of  the same magnitude as the $B$-mode polarization 
power spectrum reported by the BICEP collaboration. 
Finally, a recent joint analysis of the BICEP2/Keck and Planck collaborations 
shows no evidence for primordial $B$-mode polarization at low $l$, meanwhile 
for high $l$, they have found evidence for $B$-modes that are originated 
by gravitational lensing \cite{BICEPKECKPLANCK}.

On the other hand, 
in our approach, the source of the curvature perturbations lies in the quantum 
inhomogeneities of the inflaton field (after the collapse). Once the collapse 
has taken place, the inhomogeneities of the inflaton feed into the 
gravitational 
DOF leading to perturbations in the metric components, in particular is a 
direct 
source of the scalar perturbations. However, the metric itself is not a source 
of the self-induced collapse. Therefore, as the scalar field does not act as a 
source for the metric tensor modes, at least not at first order considered 
here, 
the analysis concerning the amplitude of the primordial gravitational waves 
should be done at the second order in the perturbations; such analysis is 
beyond 
the scope of this paper and would be the subject of 
future research. Furthermore, in Ref. \cite{Markkanen2014},  the 
semiclassical gravity approximation plus a collapse of the inflaton's wave 
function results in an undetectable amplitude for the primordial gravitational 
waves; however, those authors consider that the state collapses on a spacelike 
hypersurface for all wavelengths modes, this contrasts with our view in which 
the time of collapse depends on the mode's wavelength.

Now that  we have established the conceptual relation between the matter and 
geometry fields, we will proceed to find the main equation which will 
illustrate 
this connection under the collapse hypothesis.

In the inflationary regime, the dominant type of matter is modeled by a scalar 
field $\phi$ called the inflaton with a potential $V$ responsible for the 
accelerating expansion of the universe. At the end of the inflationary epoch, 
the universe follows the standard Big Bang evolution, the transition mechanism 
is provided by the reheating period.

The inflationary universe  is  modeled  by the action of a scalar field 
minimally coupled to gravity:

\barr\label{actioncol}
S[\phi,g_{ab}] &=& \int d^4x \sqrt{-g} \bigg[ \frac{1}{16 \pi G} R[g] \nonumber 
\\
&-&\half \nabla_a \phi \nabla_b \phi g^{ab} - V[\phi] \bigg].
\earr
Varying Eq. \eqref{actioncol} with respect to the metric yields the field 
equations $G_{ab} = 8 \pi G T_{ab}$.


We proceed the analysis, in the standard fashion, separating the metric and the 
scalar field into a background (which is perfectly homogeneous and isotropic) 
plus  
a small perturbation, i.e. $g_{ab} = g_{ab}^{(0)} + \delta g_{ab}$ and, using 
conformal coordinates,  $\phi = \phi_0 (\eta) + \dphi (\x,\eta)$. 

Focusing first on the background, the space-time is characterized by a flat FRW 
space-time. 
%
Einstein equations for the background  $G_{00}^{(0)}=8\pi G T_{00}^{(0)}=8\pi 
G a^2 \rho$   yield  Friedmann equations: $3\mH^2 = 8 \pi G a^2 \rho,$
where $\mH \equiv a'(\eta)/a(\eta)$. The major contribution to the energy 
density $\rho$ comes from the inflaton potential $V$. 
%


In the slow-roll inflationary model, the conformal Hubble factor is 
characterized by
$\mH \simeq {-1}/{[\eta(1-\epsilon_H)]}$, with $ \epsilon_H \equiv 1 - 
{\mH'}/{\mH^2}$ the Hubble slow-roll parameter, which during inflation $1 
\gg \epsilon_H \simeq$ constant.  Note that in Ref. \cite{PSS06}, $\mH = 
-1/\eta$, this is, the background space-time is strictly de Sitter and leads to 
a final spectrum with $n_s=1$; this is different from the present paper where 
we 
will be considering a quasi-de Sitter background and that will lead us to a 
value for the scalar spectral index $n_s \neq 1$.

Furthermore, since we will work with the slow-roll approximation, then the 
motion equation for the background field  can be approximated by $3 \mH \phi_0' 
=-a^2 \partial_\phi V$. Additionally, it is convenient to introduce 
the potential slow-roll (PSR) parameters 

\beq\label{PSR}
\epsilon_V \equiv \frac{M_P^2}{2} \left( \frac{\partial_\phi V}{V}\right)^2, 
\qquad \delta_V \equiv M_P^2  \left( \frac{\partial_{\phi \phi}^2 V}{V}\right).
\eeq
Thus, by assuming $\epsilon_V, \delta_V \ll 1$, one identifies the region in 
the 
potential where the slow-roll approximation is valid. Furthermore, during 
slow-roll inflation $\epsilon_H \simeq \epsilon_V$.

The normalization of the scale factor will be set to $a=1$ at the present 
cosmological time. The inflationary era would come to an end at $\eta = \eta_r 
\approx -10^{-22}$ Mpc, that is, the conformal time during the inflationary era 
is in the range $-\infty < \eta < \eta_r$.

 Next  we focus on the perturbations. It is usual to decompose the metric 
fluctuations in terms of its scalar, vector and tensor components. In the case 
of interest for this article, we concern ourselves only with the scalar 
perturbations. The scalar metric perturbations in a FRW background space-time 
are generically described by the line element:
\barr
ds^2 &=& a^2 (\eta) \big\{  -(1-2\varphi) d\eta^2 + 2 (\partial_i B) dx^i d\eta
+\nonumber \\
&+& [ (1-2\psi) \delta_{ij} + 2 \partial_i \partial_j E] dx^i dx^j \big\}.
\earr
Since we will work in the semiclassical framework it is convenient to work with 
the gauge-invariant quantities known as the Bardeen potentials, defined as 
$\Phi \equiv \varphi + \frac{1}{a} [a (B-E')]'$ and $\Psi \equiv \psi +\mH 
(E'-B) $.  In a similar way, the perturbations of the inflaton can be modeled 
by the gauge-invariant fluctuation of the scalar field $\dphi^{(\textrm{GI})} 
(\eta,\x) = \dphi + \phi_0' (B-E')$.

In  \ref{appA} is shown that combining the perturbed Einstein 
equations (in the absence of anisotropic stress) and the slow-roll motion 
equation, one obtains:

\beq\label{25x}
\nabla^2 \Psi +\mu \Psi =  4 \pi G \phi_0' \dphi^{'(\textrm{GI})},
\eeq
where $\mu \equiv \mH^2-\mH'$. In Fourier space, Eq. \eqref{25x} reads 

\beq\label{25b2}
\Psi_{\nk} (\eta) = \sqrt{\frac{\epsilon_V}{2}} \frac{H}{M_P (k^2-\mu)} a 
\dphi'_{\nk} (\eta)^{(\textrm{GI})},
\eeq
with $H$ the Hubble factor and $M_P^2\equiv1/8 \pi G$ the reduced Planck's 
mass; 
also, we have used the definition of $\epsilon_V$, Friedmann 
equation and the slow-roll approximation for $\phi_0'$. Additionally, from the 
definition of $\mu$, one can check that $\mu = \epsilon_H \mH^2$. During most 
of 
the inflationary regime, the inequality $k^2 \gg \mu$ is satisfied (both when 
$|k\eta| \gg 1$ and $|k \eta| \ll 1$), it gets violated when $\epsilon_H$ 
starts 
departing from being a constant an approaching unity; in other words when 
inflation is coming to an end. However, since modes of observational interest 
are bigger than the Hubble radius ($|k\eta| \ll 1$) while the inflationary 
phase 
is still going on, the approximation $k^2 \gg \mu$ remains valid. Thus, Eq, 
\eqref{25b2} is approximated by

\beq\label{master0}
\Psi_{\nk} (\eta) \simeq  \sqrt{\frac{\epsilon_H}{2}} \frac{H}{M_P k^2} a 
\dphi'_{\nk} (\eta)^{(\textrm{GI})}.
\eeq
Finally, since we mentioned that we will rely on the semiclassical gravity 
framework, Eq. \eqref{master0} can be generalized to

\beq\label{master}
\Psi_{\nk} (\eta) \simeq  \sqrt{\frac{\epsilon_H}{2}} \frac{H}{M_P k^2} a \bra 
\hat{\dphi'}_{\nk} (\eta)^{(\textrm{GI})} \ket.
\eeq
Equation \eqref{master} is the main result of this section; the difference with 
respect to the perfect de Sitter case will be reflected in the motion equation 
for $\hat{\dphi'}_{\nk} (\eta)^{(\textrm{GI})}$, and also, in the fact that $H$ 
and $\epsilon_H$ are strictly not constant. 

Let us note that Eq. \eqref{master} is expressed in terms of gauge-invariant 
quantities $\Psi_{\nk}$ and $\hat{\dphi'}_{\nk} (\eta)^{(\textrm{GI})}$. 
Nevertheless, in the longitudinal gauge, $\Psi$ represents the curvature 
perturbation, and is related to $\dphi$ in the exact same way as in Eq. 
\eqref{master} \cite{brandenberger1993}. Thus, expression \eqref{master} serves 
to illustrate what we mentioned at the beginning of the section, namely, that 
the quantum treatment is all in the matter fields, which during inflation is 
dominated by the inflaton, while the curvature perturbation is always a 
classical quantity. 

 \section{Quantum theory of fluctuations,  collapse 
schemes and the primordial curvature perturbation}\label{quantum}
 
 In this section, we will present the quantum theory of the field variables and 
characterize the collapse proposal. As mentioned in the introduction, we will 
focus on the pragmatical approach first exposed in \cite{PSS06}; more on this 
pragmatical approach can be consulted in 
Refs. \cite{US08,Leon10,DLS11}.  Next, we will introduce the collapse schemes 
and finally we will find the expression for the curvature perturbation for the 
three schemes considered.

\subsection{Quantum theory of perturbations}\label{quantumperts}

Given that we are working within the semiclassical gravity framework, in which 
only the matter fields are quantized, and that the self-induced collapse  
generates the curvature perturbation, our fundamental quantum variable will be 
the fluctuation of the inflaton field $\dphi (\x,\eta)$; thus, we will consider 
the quantum theory of $\dphi (\x,\eta)$ in a curved background described by a 
quasi-de Sitter space-time. Furthermore,  it will be easier to work  with the 
rescaled field variable $y=a\dphi$; consequently, we can expand the action 
\eqref{actioncol} up to second order in the rescaled variable (i.e. up to 
second order in the scalar field fluctuations)

\barr\label{acciony}
\delta S^{(2)} &=&  \int d^4x \frac{1}{2} 
\bigg[ y'^2 - (\nabla y)^2 + \left(\frac{a'}{a} \right)^2 y^2 \nonumber  \\
&-& 2 \left(\frac{a'}{a} \right) y y' -y^2a^2  \partial_{\phi \phi}^2 V\bigg].
\earr

The canonical momentum conjugated to $y$ is $\pi \equiv \partial \delta 
\mathcal{L}^{(2)}/\partial y' = y'-(a'/a)y=a\dphi'$. The field and momentum 
variables are promoted to operators satisfying the equal time commutator 
relations $[\hat{y}(\x,\eta), \hat{\pi}(\x',\eta)] = i\delta (\x-\x')$ and 
$[\hat{y}(\x,\eta), \hat{y}(\x',\eta)] = [\hat{\pi}(\x,\eta), 
\hat{\pi}(\x',\eta)] = 0$. Expanding the field operator in 
Fourier modes yields

\beq
\hat{y}(\eta,\x) = \frac{1}{L^3} \sum_{\nk} \hat{y}_{\nk} (\eta) e^{i \nk \cdot 
\x},  
\eeq
and analogous expression for $\hat{\pi}(\eta,\x)$, note that  the sum is over 
the wave vectors $\vec k$ satisfying $k_i L=2\pi n_i$ 
for $i=1,2,3$ with $n_i$ integer and $\hat y_{\nk} (\eta) \equiv y_k(\eta) 
\ann_{\nk} + y_k^*(\eta) \cre_{-\nk}$ and  $\hat \pi_{\nk} (\eta) \equiv 
g_k(\eta) \ann_{\nk} + g_{k}^*(\eta) \cre_{-\nk}$, with $g_k(\eta) = y_k'(\eta) 
- \mH y_k (\eta)$.  From the previous expression 
it is clear that we are taking the quantization on a finite cubic box of length 
$L$, at the end of the calculations we will go to the continuum limit ($L \to 
\infty$, $\vec k \to $ cont.). The equation of motion for  $y_k(\eta)$ derived 
from action \eqref{acciony} is

\beq\label{ykmov}
y''_k(\eta) + \left(k^2 - \frac{a''}{a} + a^2 \partial^2_{\phi \phi} V \right) 
y_k(\eta)=0.
\eeq
Since $\mH = -1/[\eta(1-\epsilon_H)]$, one finds that $a''/a \simeq 
(2+3\epsilon_H)/\eta^2$; additionally, using the definition of $\delta_V$, 
Friedmann's equation and the explicit form of $\mH$, one has $a^2 
\partial^2_{\phi \phi} V \simeq 3 \delta_V / \eta^2$. Therefore, the motion 
equation Eq. \eqref{ykmov} is rewritten as

\beq\label{ykmov2}
y''_k(\eta) + \left(k^2 - \frac{2+3(\epsilon_H - \delta_V)}{\eta^2} \right) 
y_k(\eta)=0.
\eeq

The solution of Eq. \eqref{ykmov2}  is fixed by the canonical commutation 
relations between $\hat y$ and $\hat \pi$, which give 
$[\hat{a}_{\nk},\hat{a}^\dag_{\nk'}] = L^3 \delta_{\nk,\nk'}$, thus $y_k(\eta)$ 
must satisfy $y_k g_k^* - y_k^* g_k = i$ for all $k$ at some time $\eta$.
The choice of $y_k(\eta)$ corresponds to the choice of a vacuum state for the 
field, which in the present case would be the so-called Bunch-Davies vacuum  
characterized by

\beq\label{nucolapso}
y_k (\eta) = \left( \frac{-\pi \eta}{4} \right)^{1/2} e^{i[\nu + 1/2] (\pi/2)} 
H^{(1)}_\nu (-k\eta), 
\eeq
where $ \nu \equiv 3/2 + \epsilon_H -\delta_V$  and $H^{(1)}_\nu (-k\eta)$ is 
the Hankel function of the first kind of order 
$\nu$.\footnote{The Hankel functions of the first kind are defined as 
$H^{(1)}_\nu (x) \equiv J_\nu (x) + i Y_\nu (x)$ with $J_\nu$ and $Y_\nu$ the 
Bessel functions of the first and second kind respectively.} The solution 
involves a phase $e^{i[\nu + 1/2] (\pi/2)}$ that we will drop from 
calculations as it has no observational consequence. 

We note that in the case of an exact de Sitter universe, the motion equation  
would correspond to setting $\epsilon_H=\delta_V=0$ in Eq. \eqref{ykmov2}, and 
indeed that was the case studied in Refs. \cite{PSS06,US08}. The fact that the 
motion equation \eqref{ykmov2} now involves the slow-roll parameters will lead, 
as we will show in the rest of this article, to a prediction for the scalar 
spectral index that is generically $n_s \neq 1$.

The self-induced collapse hypothesis is based on considering that the collapse 
acts similar to a ``measurement'' (clearly, there is no external observer or 
detector involved). This lead us to consider Hermitian operators, which in 
ordinary quantum mechanics are the ones susceptible of direct measurement. 
Therefore, we separate $\hat y_{\nk} (\eta)$ and $\hat \pi_{\nk} (\eta)$ into 
their real and imaginary parts $\hat y_{\nk} (\eta)=\hat y_{\nk}{}^R (\eta) +i 
\hat y_{\nk}{}^I (\eta)$ and $\hat \pi_{\nk} (\eta) =\hat \pi_{\nk}{}^R (\eta) 
+i \hat \pi_{\nk}{}^I (\eta)$ in this way the operators $\hat y_{\nk}^{R, I} 
(\eta)$ and $\hat \pi_{\nk}^{R, I} (\eta)$ are  Hermitian operators. Thus, 

\begin{subequations}\label{operadoresRI}
\beq
\hat{y}_{\nk}^{R,I} (\eta) = \sqrt{2} \textrm{Re}[y_k(\eta) 
\hat{a}_{\nk}^{R,I}], 
\eeq
\beq
\hat{\pi}_{\nk}^{R,I} (\eta) = \sqrt{2} 
\textrm{Re}[g_k(\eta) \hat{a}_{\nk}^{R,I}],
\eeq
\end{subequations}
where $\hat{a}_{\nk}^R \equiv (\hat{a}_{\nk} + \hat{a}_{-\nk})/\sqrt{2}$, 
$\hat{a}_{\nk}^I \equiv -i (\hat{a}_{\nk} - \hat{a}_{-\nk})/\sqrt{2}$. 

The commutation relations for the $\hat{a}_{\nk}^{R,I}$ are non-standard

\beq\label{creanRI}
[\hat{a}_{\nk}^{R,I},\hat{a}_{\nk'}^{R,I \dag}] = L^3 (\delta_{\nk,\nk'} \pm 
\delta_{\nk,-\nk'}), 
\eeq
where the $+$ and the $-$ sign corresponds to the commutator with the $R$ and 
$I$ labels respectively; all other commutators vanish.

Up to this point,  we have  proceeded  in  a  similar way  to the traditional 
one,  except that we are  treating at the quantum level  only the scalar field 
and  not the  metric  fluctuation.  It is also   worthwhile  to   emphasize  
that the  vacuum state   defined  by $ \ann_{\nk}{}^{R,I} |0\ket =0$  is  100\% 
translational  and rotationally invariant (the formal proof was presented in 
Appendix A of Ref. \cite{LLS13}).

Our next task is to connect the quantum theory of the inflaton perturbations 
with the primordial curvature perturbation. We proceed by choosing to 
work in the longitudinal gauge, and express Eq. \eqref{master},  in terms of 
the expectation value of the conjugated momentum, this is,

\beq\label{masterpi}
\Psi_{\nk} (\eta) \simeq  \sqrt{\frac{\epsilon_H}{2}} \frac{H}{M_P k^2}  \bra 
\hat{\pi}_{\nk} (\eta) \ket .
\eeq
It is clear that, in the vacuum state, $ \bra \hat{\pi}_{\nk} (\eta) \ket =0$, 
which implies $\Psi_{\nk} =0$, i.e. there are no perturbations of the symmetric 
background space-time. It is only after the collapse has taken place 
($|\Theta\ket \neq |0\ket$)  that $\bra \hat{\pi}_{\nk} (\eta) \ket_\Theta \neq 
0$ generically and $\Psi_{\nk} \neq 0$; thus, the primordial inhomogeneities 
and 
anisotropies are born from the quantum collapse.

It is also important to note that the quantum collapse affects all modes 
$\nk$ of the inflaton, this is, the collapse takes the original vacuum 
state $| 0 \ket$ to a new quantum state: 

\beq
| \Theta \ket = \ldots | \Theta_{-\nk_2} \ket \otimes | 
\Theta_{-\nk_1} \ket \otimes | \Theta_{\nk_0} \ket \otimes | 
\Theta_{\nk_1} \ket \otimes | \Theta_{\nk_2} \ket \ldots
\eeq

Given Eq. \eqref{masterpi}, which was provided by the semiclassical framework, 
and also that all modes of the inflaton field are now in a post-collapse state 
$|\Theta \ket$, we can clearly see that the expectation value $\bra 
\hat{\pi}_{\nk} (\eta) \ket$ serves 
as a source for $\Psi_{\nk}$ for all $\nk $.  Once the collapse has 
created 
all modes $\Psi_{\nk}$, we can divide them in two types: 

\begin{enumerate}
\item Modes with an associated proper wavelength bigger than the Hubble radius 
at the time of collapse, we will call these the super-horizon modes; i.e. their 
corresponding Fourier (comoving) modes satisfy $k \ll \mH(\tc)$.

\item Modes with an associated proper wavelength smaller than the Hubble 
radius at the time of collapse, we will call these the sub-horizon modes; i.e. 
their corresponding Fourier (comoving) modes satisfy $k \gg \mH(\tc)$.
\end{enumerate}

In order to continue with the model, we will consider that the collapse is 
somehow analogous to an imprecise measurement\footnote{An imprecise measurement 
of an observable is one in which one does not end with an exact eigenstate  of 
that observable but  rather with a state that is  only peaked around the 
eigenvalue. Thus, we could consider measuring a  certain particle's position 
and 
momentum so as to end up with a state that is a wave packet with both position 
and momentum defined to a limited extent and, which certainly, does not  entail 
a conflict with Heisenberg's uncertainty bound.} of the operators $\hat 
y_{\nk}^{R, I} (\eta)$ and $\hat \pi_{\nk}^{R, I} (\eta)$. This is, we need to 
specify the dynamics of the expectation values $\bra  \hat{y}^{R, I}_{\nk} 
(\eta)   \ket$ and $\bra  \hat{\pi}^{R, I}_{\nk} (\eta)   \ket$, evaluated in 
the post-collapse state.  Furthermore, the analytical expression for  $\bra  
\hat{y}^{R, I}_{\nk} (\eta)   \ket$ and $\bra \hat{\pi}^{R, I}_{\nk} (\eta)   
\ket$ should  depend on  $\bra  \hat{y}^{R, I}_{\nk} (\tc)   \ket$ and $\bra  
\hat{\pi}^{R, I}_{\nk} (\tc)   \ket$, i.e. the expectation values evaluated at 
the time of collapse. In the next 
subsection we will show the precise manner to specify  $\bra  \hat{y}^{R, 
I}_{\nk} (\tc)   \ket$ and $\bra  \hat{\pi}^{R, I}_{\nk} (\tc) \ket$.

\subsection{Collapse schemes}\label{esquemas-colapso}

At this point in the analysis we need to characterize in a more precise 
manner the collapse of the wave function. 
Evidently, it would be desirable to provide a physical mechanism for the 
collapse. Nevertheless, there are some aspects that need to be addressed first. 
A full workable relativistic collapse mechanism is still unknown, however, some 
relativistic models have been recently proposed \cite{bedingham2010,pearle2014} 
and are still under development. 

On the other hand, some non-relativistic objective collapse models have been 
analyzed previously in the 
literature \cite{ghirardi1985,pearle1989,bassi2003,bassi2013}. In particular, 
the Continuous Spontaneous Localization (CSL) model \cite{pearle1989} is based 
on a non-linear stochastic modification of the standard Schr\"{o}dinger 
equation, in this way, spontaneous and random collapses of the wave function 
occur all the time, to all particles, regardless they are isolated or 
interacting. The idea behind the CSL model, sometimes referred to as the 
``amplification mechanism,'' is that the collapses must be rare for microscopic 
systems, in order to not alter their quantum behavior as described by the 
Schr\"{o}dinger equation, but at the same time, their effect must increase when 
several particles are hold together forming a macroscopic system. 

The CSL model has been applied to the inflationary universe in previous works 
\cite{jmartin,tpsingh,dasGW,CPS13,LB15}; nevertheless, the results obtained in 
those works are different among each other (e.g. in Ref. \cite{CPS13} the 
amplitude of primordial gravitational waves is zero at first-order in the 
perturbations, meanwhile in Ref. \cite{dasGW,LB15} this amplitude is similar 
to the predicted by the traditional inflationary model). The reason is the 
conceptual approach taken to address the subject, specifically, the treatment 
of the metric perturbations (in \cite {CPS13} the metric is always classical, 
while in \cite{dasGW,LB15} metric perturbations are quantized). Furthermore, 
there are still a few limitations on the CSL inflationary model that need 
to be investigated in detail, for instance: (i) CSL model is actually 
non-relativistic, but \cite{jmartin,tpsingh,dasGW,CPS13,LB15} assume a field 
theoretic CSL-like version, mode by mode, in momentum space, without a physical
justification; (ii) the CSL amplification mechanism is absent, or introduced as 
an \textit{ad hoc} assumption. It is clear that further research is needed in 
order to consider a complete collapse mechanism and its successful 
implementation to the inflationary universe. 

In the present manuscript, we are interested in analyzing the characteristics 
of 
the observational predictions when considering a generic self-induced collapse. 
Thus,  we will not consider a particular collapse mechanism and instead proceed 
in a pragmatical way. We will assume that whatever the collapse mechanism is 
behind, at the end of the collapse process, which can be associated  to the 
time 
of collapse, we can characterize the expectation values of the field and the 
momentum, 
evaluated at the post-collapse state. More precisely, we assume that the effect 
of the collapse on a state is  analogous  to some  sort of  approximate 
measurement; in other words,   after the  collapse, the expectation values of 
the field and momentum 
operators  in each mode  will  be related to the uncertainties  of the  initial 
state. In the vacuum state, $\hat{y}_{\nk}$ and $\hat{\pi}_{\nk}$ individually 
are distributed according to Gaussian wave functions centered at 0 with spread 
$(\Delta \hat{y}_{\nk})^2_0$ and $(\Delta\hat{\pi}_{\nk})^2_0$, respectively.  
We  could   consider  various  possibilities  for such relations; we will refer 
as ``collapse schemes'' to the different ways of characterizing the expectation 
values. In past works \cite{PSS06,US08} three different schemes were 
considered. 
These schemes were called \emph{independent, Newtonian} and \emph{Wigner} 
collapse schemes. In the following, we will describe them briefly.

\subsubsection{Independent scheme}

In this scheme one assumes that the expectation values of the field's mode 
$\hat{y}^{R,I}_{\nk}$, and their conjugate momentum 
$\hat{\pi}^{R,I}_{\nk}$, acquire independent values randomly. The 
expectation at the time of collapse is assumed to be of the form

\begin{subequations}\label{esquemaind}
\beq 
\bra \hat{y}^{R,I}_{\nk}(\eta^c_{\nk})\ket  = x_{\nk,1}^{R,I}
  \sqrt{\left(\Delta \hat{y}^{R,I}_{\nk} (\tc) \right)^2_0},
  \eeq
  \beq
  \bra \hat{\pi}^{R,I}_{\nk}(\tc) \ket = x_{\nk,2}^{R,I}
  \sqrt{\left(\Delta \hat{\pi}^{R,I}_{\nk} (\tc) \right)^2_0}.
  \eeq
\end{subequations}

In this scheme the expectation value jumps to a random value $x_{\nk}^{(R,I)}$ 
multiplied by the uncertainty of the vacuum state of the field. The random 
variables $x_{\nk,1}^{(R,I)}$, $x_{\nk,2}^{(R,I)}$ are selected from a Gaussian 
distribution centered at zero, with unity spread, and are statistically 
uncorrelated, that is the rationale of the name. This means that we are 
ignoring 
the natural correlation that exists in the conjugate fields in the pre-collapse 
state. In \ref{appB1}, is shown the explicit form of $\left(\Delta 
\hat{y}^{R,I}_{\nk} (\tc) \right)^2_0$ and $\left(\Delta 
\hat{\pi}^{R,I}_{\nk} (\tc) \right)^2_0$ within this collapse scheme.

\subsubsection{Newtonian collapse scheme}

This scheme is motivated by the fact that in the equation for the Newtonian 
potential, Eq. \eqref{masterpi}, only the expectation value of 
$\hat{\pi}_{\nk}$ 
appears. Thus, one is led to consider a scheme where the collapse affects only 
the conjugated momentum variable, this is

\beq\label{esquemanewt}
 \bra \hat{y}^{R,I}_{\nk}(\eta^c_{\nk})\ket  = 0, \qquad
  \bra \hat{\pi}^{R,I}_{\nk}(\tc) \ket = x_{\nk,2}^{R,I}
  \sqrt{\left(\Delta \hat{\pi}^{R,I}_{\nk} (\tc) \right)^2_0}.
\end{equation}
As in the previous case, $x_{\nk,2}^{(R,I)}$ represents a random Gaussian 
variable normalized and centered at zero. The quantity 
$\left(\Delta \hat{\pi}^{R,I}_{\nk} (\tc) \right)^2_0$ within this 
collapse scheme is the same as in the \emph{independent} scheme 
(see \ref{appB1}).

\subsubsection{Wigner collapse scheme}

The last collapse scheme, analyzed in detail in Refs. \cite{US08,Leon11}, is 
motivated by considering the correlation between $\hat{y}^{R,I}$ and
$\hat{\pi}^{R,I}$ existing in the pre-collapse state and characterize it in 
terms of the Wigner function (one knows from Heisenberg's uncertainty 
principle that the field and momentum variables should be correlated). 

The Wigner function of the vacuum state is a bi-dimensional Gaussian function. 
The assumption is that, at a certain (conformal) time $\tc$, the part of 
the state characterizing the mode $k$ will collapse, leading to a new state in 
which the expectation value of the fields will be characterized by
\begin{subequations}\label{esquemawig}
 \beq
\bra \hat{y}^{R,I}_{\nk}(\eta^c_{\nk})\ket  = x_{\nk}^{R,I} \Lambda_k (\tc) 
\cos \Theta_k (\tc), 
\eeq
\beq
  \bra \hat{\pi}^{R,I}_{\nk}(\tc) \ket = x_{\nk}^{R,I} k \Lambda_k (\tc) 
\sin \Theta_k (\tc),
\eeq
\end{subequations}
where $x_{\nk}^{R,I}$ is a random variable, characterized by a Gaussian 
probability distribution function  centered at zero with  spread one. The 
parameter $\Lambda_k (\tc)$ represents the major semi-axis of the ellipse 
characterizing the bi-dimensional Wigner function that can be considered a 
Gaussian in two dimensions; this is,  the ellipse corresponds to the boundary 
of 
the region in ``phase space'' where the Wigner function has a magnitude larger 
than 1/2 its maximum value. The other parameter, represented by $\Theta_k 
(\tc)$, is the angle between that axis and the $\hat{y}_{\nk}^{R,I}$ axis. The 
details involving the Wigner function and the collapse scheme can be consulted 
in Ref. \cite{US08}. The parameters $\Lambda_k$ and $\Theta_k$ depend on the 
time of collapse, one can follow the analysis presented in Ref \cite{US08} in 
order to find an expression for $\Lambda_k$ and $\Theta_k$ in terms of the time 
of collapse, but bearing in mind that, for the present manuscript, we 
are considering a quasi-de Sitter universe [see \ref{appB1} for an 
explicit expression of $\Lambda_k (\tc) $ and $\Theta_k (\tc)]$ . 

\subsection{The curvature perturbation for the three collapse 
schemes}\label{curvatura3esquemas}

So far, we have established the relation between the curvature perturbation 
and 
the quantum matter fields [see Eq. \eqref{masterpi}] and characterized the 
collapse by means of the expectation values of the field and its momentum, i.e. 
by introducing the collapse schemes. The next aim is to present an explicit 
expression for the curvature perturbation in terms of the parameters 
characterizing each collapse scheme. In order to attain that goal, we must 
first 
find an expression for the evolution of the expectation values of the fields. 
In 
fact, as can be seen from Eq. \eqref{masterpi}, we will only be concerned with 
the expectation value of the conjugated momentum $\bra \hat{\pi}_{\nk} (\eta) 
\ket$. In \ref{appB2}, we show that

\barr\label{expecpitexto}
\bra \hat{\pi}_{\nk}^{R,I} (\eta) \ket_\Theta &=& F(k\eta,z_k)  \bra 
\hat{y}_{\nk}^{R,I} (\tc) \ket_\Theta \nonumber \\
&+& G(k\eta,z_k) \bra \hat{\pi}_{\nk}^{R,I} (\tc) \ket_\Theta,
\earr
with the definitions of $F(k\eta,z_k)$  and $G(k\eta,z_k)$ also in 
 \ref{appB2}. 
The parameter $z_k$ is defined as $z_k \equiv k\tc$; thus, $z_k$ is directly 
associated to the time of collapse $\tc$. 

Finally, substituting Eq. \eqref{expecpitexto} in Eq. \eqref{masterpi}, we can 
find an expression for the curvature perturbation (in the longitudinal gauge).

\barr\label{masterpsi}
\Psi_{\nk} (\eta) &=&  \sqrt{\frac{\epsilon_V}{2}} \frac{H}{M_P k^2} \nonumber 
\\
&\times& \bigg[ F(k\eta,z_k) \left( \bra \hat{y}_{\nk}^{R} (\tc) \ket_\Theta  + 
i \bra \hat{y}_{\nk}^{I} (\tc) \ket_\Theta \right) \nonumber \\ 
&+& G(k\eta,z_k) \left( \bra \hat{\pi}_{\nk}^{R} (\tc) \ket_\Theta + i \bra 
\hat{\pi}_{\nk}^{I} (\tc) \ket_\Theta \right) \bigg].
\earr
We can see from Eq. \eqref{masterpsi} how the curvature perturbation depends on 
the three  collapse schemes through the quantities $\bra \hat{y}_{\nk}^{I} 
(\tc) \ket_\Theta $ and $\bra \hat{\pi}_{\nk}^{I} (\tc) \ket_\Theta$ . 
Henceforth, we have three different expressions for $\Psi_{\nk} (\eta)$ 
corresponding to the three previously introduced collapse schemes  
characterized 
in Eqs. \eqref{esquemaind}, \eqref{esquemanewt}, \eqref{esquemawig} (explicit 
expressions of $\Psi_{\nk}$ in the three schemes are given in 
\ref{appB2}).

One useful gauge-invariant quantity often encountered in the literature is the 
variable $\mathcal{R} (x)$.  The field $\mathcal{R}(x)$ is a field representing 
the curvature  perturbation in the comoving gauge.  Its Fourier's transform,  
represented by $\mR_{\nk}$, is constant for modes ``outside the horizon'' 
(irrespectively of the cosmological epoch), i.e. for modes with $k \ll \mH = 
aH$ (and assuming adiabatic perturbations). This is, the value of $\mR_{\nk}$ 
during inflation (in the limit $k \ll \mH$) would remain unchanged at all 
times, until the mode 
``re-enters the horizon,'' namely when $k\simeq \mH$. 

On the other hand, the curvature perturbation $\Psi$ in the longitudinal 
gauge, is also constant 
for modes outside the horizon during any given cosmological epoch but not 
during 
the transition between epochs. In fact, during the transition from the 
inflationary stage to the radiation dominated stage, $\Psi$ is amplified by a 
factor of $1/\epsilon_V$ \cite{brandenberger1993,Deruelle1995}. This behavior 
differs from the one of $\mathcal{R}$, which remains constant in spite of the 
epoch transition.

The curvature perturbation in the comoving gauge  $\mathcal{R}$ and the 
curvature perturbation in the longitudinal gauge $\Psi$ are related 
as $\mathcal{R} \equiv \Psi + (2/3)(\mH^{-1} {\Psi}' + \Psi)/(1+\omega)$, with 
$\omega \equiv p/\rho$. Therefore, for modes such that $k \ll \mH$,  during the 
inflationary epoch $\omega+1 \simeq 2 \epsilon_V/3$,  one has

\beq\label{R}
\lim_{ k \ll \mH} R_{\nk} \simeq \lim_{ k \ll \mH} \frac{\Psi_{\nk} (\eta)}{ 
\epsilon_V},
\eeq
with $\Psi_{\nk} (\eta)$, calculated during inflation, in the limit such that 
the modes are well outside the ``horizon'' (i.e. in the regime where 
$|k\eta| \ll 1$).

Therefore by expanding the expressions $\Psi_{\nk}$, within the three collapse 
schemes, to the lowest order in $|k\eta|$, and by 
making use of Eq. \eqref{R}, we can find $\mR_{\nk}$. Thus, after performing 
such expansion, the comoving curvature perturbation is

\begin{subequations}\label{R3esquemas}
\beq\label{psiindexp}
\mR^{\text{ind}}_{\nk}  \simeq  R_k \bigg[ M(|z_k|) X_{\nk,1} + N(|z_k|) 
X_{\nk,2} \bigg] |k\eta|^{3/2-\nu},
\eeq
\beq\label{psinewexp}
\mR^{\text{newt}}_{\nk}  \simeq R_k  N(|z_k|) X_{\nk,2} |k\eta|^{3/2-\nu},
\eeq
\beq\label{psiwigexp}
\mR^{\text{wig}}_{\nk}   \simeq R_k  W (|z_k|)  X_{\nk} |k\eta|^{3/2-\nu}.
\eeq
\end{subequations}

The functions $M(|z_k|)$, $N(|z_k|)$ and $W(|z_k|)$ are defined in 
\ref{appB3} and the amplitude 
\\
$R_k \equiv  \sqrt{ {L^3 \pi}/{ \epsilon_V }} {H 
2^{\nu-11/2}\Gamma (\nu-1)}/{ M_P k^{3/2}} $;
\\
also we have introduced the definitions $z_k \equiv k\tc$ and $X_{\nk} \equiv 
x_{\nk}^R+ i x_{\nk}^I$.

Equations \eqref{R3esquemas} are the 
main result of this section. They relate the initial curvature perturbation, 
which is associated with the temperature anisotropies in the CMB, with the 
parameters characterizing each collapse scheme, i.e. the time of collapse and 
the random variables. There is no analogous expression in the traditional 
inflationary paradigm, in which by relying on some ``quantum-to-classical'' 
arguments (see Refs. \cite{PSS06,Shortcomings} for a detailed discussion on the 
conceptual problems regarding such arguments), one is able to go from $\hat 
\mR_{\nk}$ to $\mR_{\nk}$ but without a clear identification of the physical 
(and probably random) process that originated the classical curvature 
perturbation.

We strongly remark that the random variables corresponding to each collapse 
scheme are fixed after the collapse of the wave function has occurred. In other 
words, if we somehow knew their exact value, we would be able to predict the 
exact value for $\mR_{\nk}$; notice that we have not even mentioned notions 
such  as average over an ensemble of universes or some related concepts. 
Nevertheless, we will do make use of the statistical properties of the random 
variables to be 
able to make theoretical predictions for the observational quantities, e.g. the 
power spectrum and the spectral index; this will be the focus of the next 
section.

 \section{An equivalent power-spectrum for the curvature 
perturbation}\label{oquantities}

 The focus of this section is to find an equivalent expression for what it is 
commonly referred to as the primordial power spectrum for the scalar 
perturbations. In the 
standard inflationary paradigm, such expression is given by\footnote{Actually, 
in the literature, one finds two kinds of power spectrum: the dimensional power 
spectrum $\mathcal{P} (k)$ and the dimensionless power spectrum $P(k)$; the 
latter is defined in terms of the former by  $ P(k) \equiv  (k^3/2\pi^2) 
\mathcal{P}(k) $} $P(k) = {A_s} k^{n_s-1}$, where ${A_s}$ is the amplitude  and 
$n_s$ is known as the spectral index of the scalar perturbations. Thus, in this 
section we will find a similar expression in which we will identify the 
amplitude and the scalar index within the collapse model.

We begin by showing how the observational quantities can be related with the 
parameters characterizing the collapse. The temperature anisotropies $\delta 
T/T_0$ of the CMB are clearly the most direct observational quantity available, 
with $T_0$ the mean temperature. Expanding $\delta T/T_0$ using spherical 
harmonics, the coefficients $a_{lm}$ are 

\beq\label{alm0}
a_{lm} = \int \Theta (\hat n) Y_{lm}^\star (\theta,\varphi) d\Omega,
 \eeq
with $\hat n = (\sin \theta \sin \varphi, \sin \theta \cos \varphi, \cos 
\theta)$ and $\theta,\varphi$ the coordinates on the celestial two-sphere; we 
have also defined $\Theta (\hat {n}) \equiv \delta T (\hat n)/ T_0$. Assuming 
instantaneous recombination, the relation between the primordial perturbations 
and the observed CMB temperature anisotropies is

\beq\label{mastertemp}
\Theta (\hat n) = [\Psi + \frac{1}{4} \delta_\gamma] (\eta_D) + \hat n \cdot 
\vec{v}_\gamma (\eta_D) + 2 \int_{\eta_D}^{\eta_0} \Psi'(\eta) d\eta, 
\eeq
where $\eta_D$ is the time of decoupling; $\delta_\gamma$ and 
$\vec{v}_\gamma$ are the density perturbations and velocity of the radiation 
fluid. 

It is common practice to decompose the temperature anisotropies in Fourier modes

\beq
\Theta (\hat n) = \sum_{\nk} \frac{\Theta (\nk)}{L^3} e^{i \nk \cdot R_D \hat 
n},
\eeq
with $R_D$ the radius of the last scattering surface. Afterwards, one solves 
the 
fluid motion equations with the initial condition provided by the curvature 
perturbation during inflation. Furthermore, using that $e^{i \nk \cdot R_D \hat 
n} = 4 \pi  \sum_{lm} i^l j_l (kR_D) Y_{lm} (\theta,\varphi) Y_{lm}^\star (\hat 
k )$,  expression \eqref{alm0} can be rewritten as

\beq\label{alm1}
a_{lm} = \frac{4 \pi i^l}{L^3} \sum_{\nk} j_l (kR_D) Y_{lm}^\star(\hat k) 
\Theta 
(\nk),
\eeq
with $j_l (kR_D)$ the spherical Bessel function of order $l$.

The linear evolution that relates the initial curvature perturbation 
$\mR_{\nk}$ and the temperature anisotropies $\Theta (\nk)$ is summarized in 
the 
transfer function $T(k)$, in other words, $T(k)$ is the result of solving the 
fluid motion equations (for each mode)  with the initial condition provided by 
the curvature perturbation $\mR_{\nk}$ and then make use of Eq. 
\eqref{mastertemp} to relate it  with the temperature anisotropies. Thus 
$\Theta 
(\nk) = T(k) \mR_{\nk}$.

Consequently, the coefficients $a_{lm}$, in terms of the modes $\mR_{\nk}$, are 
given by 
 
\beq\label{alm2}
a_{lm} = \frac{4 \pi i^l}{ L^3} \sum_{\nk} j_l (kR_D) Y_{lm}^\star(\hat k) T 
(k) 
\mR_{\nk},
\eeq
with $\mR_{\nk}$ during inflation, and in the limit $k \ll \mH$ or equivalently 
$ |k\eta| \ll 1$.

By substituting  Eqs. \eqref{R3esquemas}, corresponding to the explicit form of 
$\mR_{\nk}$ for each collapse scheme, in Eq. \eqref{alm2},   one can see how 
the coefficients $a_{lm}$ are directly related 
to the random variables $X_{\nk}$. Consequently, the coefficients $a_{lm}$ are 
in effect a sum  of  random complex  numbers (i.e. a sum over $\nk$ where each 
term is characterized by the random variables  $ X_{\nk}$, which is itself a 
complex random number), leading to what can be considered effectively as a 
two-dimensional (i.e. a  complex plane)  random walk. As is well known, one 
cannot give a perfect estimate for the direction of the final displacement  
resulting from the random walk. Nevertheless, one might give an estimate for 
the 
length of the displacement for the random walk. Such length is naturally 
associated with the magnitude $|a_{lm}|^2$; hence, we can make an estimate for 
the most likely value of $|a_{lm}|^2$ and interpret it as the theoretical 
prediction for the observed $|a_{lm}|^2$. Moreover, since the collapse is being 
modeled by a random process, we can consider a set of possible realizations of 
such process characterizing the universe in an unique manner, i.e.,  
characterized by the random variables $X_{\nk}$. If the probability 
distribution 
function of $X_{\nk}$ is Gaussian, then we can identify the most likely value 
$|a_{lm}|^2_{\text{ML}}$ with the mean value $\overline{|a_{lm}|^2}$ of all 
possible realizations; this is, $|a_{lm}|^2_{\text{ML}}= 
\overline{|a_{lm}|^2}$. 
The most likely value $|a_{lm}|^2_{\text{ML}}$ in each collapse scheme is 
explicitly given in  \ref{appC}.

Moreover, we need to make some further 
assumptions regarding the ensemble average of the product of the random 
variables for each collapse scheme. We will assume that $x_{\nk}^{R,I}$ 
variables are uncorrelated, this is, they satisfy

\beq\label{distribucionx}
\overline{x^R_{\nk,s} x^R_{\nk',s}} = \delta_{\nk,\nk'} + \delta_{\nk,-\nk'}, 
\quad \overline{x^I_{\nk,s} x^I_{\nk',s}} = \delta_{\nk,\nk'} - 
\delta_{\nk,-\nk'},
\eeq
with the label $s$ indicating the particular scheme associated to the random 
variables.

%
%
%
%
%


Note that we have taken into account that the real and imaginary parts of the 
random variables  are independent for every scheme. Additionally, we have 
considered the correlation between the modes $\nk$ and $-\nk$ in accordance 
with 
the commutation relation given by $[\hat{a}^{R}_{\nk},\hat{a}^{R \dag}_{\nk'}]$ 
and $[\hat{a}^{I}_{\nk},\hat{a}^{I \dag}_{\nk'}]$ [see Eq. \eqref{creanRI}]. 
%

The observational data is presented in terms of a quantity $C_l$ called the 
angular power spectrum. The definition of $C_l$ is given in terms of the 
coefficients $a_{lm}$ as $C_l = (2l+1)^{-1} \sum_m |a_{lm}|^2$. Therefore, we 
can use the prediction for $|a_{lm}|^2_{\text{ML}}$ for each collapse scheme 
and 
give a 
theoretical prediction for $C_l$ for the three collapse schemes considered. 
Thus, Eq. \eqref{distribucionx} and using  $|a_{lm}|^2_{\text{ML}}$ for each 
collapse scheme, we obtain

\beq\label{clcolapso}
C_l = { 4 \pi} \int_0^\infty \frac{dk}{k} j_l^2 (kR_D) T(k)^2 
\frac{\mathcal{C}}{\pi^2} Q(|z_k|) k^{3-2\nu},
\eeq
where

\beq\label{C}
\mathcal{C}\equiv \frac{\pi}{ M_P^2 \epsilon_H} \left( 2^{\nu-11/2} 
\Gamma(\nu-1) H |\eta|^{3/2-\nu}    \right)^2,
\eeq
and we have taken the limit $L\to \infty$ and $\nk \to$ continuum in order to 
go from sums over discrete $\nk$ to integrals over $\nk$. The function 
$Q(|z_k|)$ varies for each collapse scheme. For the \textit{independent} scheme

\beq\label{C1}
Q(|z_k|)^{\text{ind}} = M^2(|z_k|) + N^2(|z_k|);
\eeq
for the \textit{Newtonian} scheme

\beq\label{C2}
Q(|z_k|)^{\text{newt}} =   N^2(|z_k|),
\eeq
and for the \textit{Wigner} scheme
\beq\label{C3}
Q(|z_k|)^{\text{wig}} =   W^2(|z_k|);
\eeq
the definitions of $M,N,W$ are in  \ref{appB3}.

In the standard inflationary paradigm, a well-known result is that the power 
spectrum $P (k)$ for the perturbation $\mR_{\nk}$ and the $C_l$ are related by

\beq\label{cl}
C_l = {4 \pi} \int_0^\infty \frac{dk}{k}  j_l^2 (kR_D) T(k)^2 P(k).
\eeq

Thus, by comparing Eq. \eqref{clcolapso} with \eqref{cl} we can extract an 
``equivalent power spectrum'' for each collapse scheme (more details can be 
found in \ref{appD}). The form of the power 
spectrum, within the three collapse schemes, has a generic form

\beq\label{pscolapso}
P(k) = \frac{\mathcal{C}}{\pi^2} Q (|z_k|) k^{3-2\nu},
\eeq

Equation \eqref{pscolapso} is the main result of this section. In the next 
section, we will compare our prediction with the one given by the traditional 
approach and make a couple of remarks regarding the physical implications of 
our 
prediction. 

We would like to end this section by noting that our prediction for the power 
spectrum was extracted from what in principle are observable quantities, i.e 
the 
$C_l$'s. In fact, our model gives a direct theoretical prediction for the 
observed $C_l$, Eq. \eqref{clcolapso}, and then from such expression we 
``read'' what can be identified as the ``power spectrum'' in the traditional 
approach of 
inflation. This is conceptually different from the orthodox approach 
\cite{Mukhanov2005} in which the power spectrum is obtained directly from the 
two-point correlation function $\bra 0 | \hat \Psi (\nk) \hat \Psi(\nk') | 0 
\ket$. In contrast, our power spectrum is obtained from $\overline{\bra 
\hat{\pi}_{\nk} \ket \bra \hat{\pi}_{\nk '}\ket}$, where the expectation values 
are evaluated at the post-collapse state, this is, we have never relied on the 
calculation of the two-point quantum correlation function. In 
\ref{appD}
we show in detail the calculation of the CMB temperature angular power 
spectrum and its relation with the scalar power spectrum, this serves to 
present the reader why our proposal does not rely on the quantum two-point 
correlation function.

 \section{Comparisons between the traditional and the collapse power spectrum 
}\label{comparison}

 Let us recall that the standard prediction for the scalar power spectrum for 
the curvature perturbations, within single field slow-roll inflation, is $P(k) 
=A_s k^{n_s-1}$, where

\beq\label{Ayns}
A_s=\frac{2^{2\tilde \nu-4}|\Gamma(\tilde \nu)|^2  H^2  |\eta|^{3-2\tilde \nu}  
}{\pi^3 M_P^2 \epsilon_H} , \qquad n_s-1 =  -6 \epsilon_H + 2 \delta_V,
\eeq
with $\tilde \nu \equiv 3/2 + 3 \epsilon_H - \delta_V$. On the other hand, 
within our model, $\nu =3/2 + \epsilon_H - \delta_V$ [see Eq. 
\eqref{nucolapso}]; therefore, the equivalent scalar spectral index is $n_s - 1 
= -2 \epsilon_H + 2 \delta_V$.

As has been analyzed in Ref. \cite{Kinney2005}, the quantity $A_s$, namely the 
amplitude, is  approximately a time independent function [i.e., $d/d\eta \{H^2 
|\eta|^{3-2\nu}/\epsilon_H  \}= \mathcal{O}(\epsilon_H^2,\delta_V^2)$] for all 
$\eta$, this is, even if $H$, $\epsilon_H$ and $|\eta|^{3-2\tilde \nu}$ are 
time 
dependent, their combination, as it appears in $A_s$, is essentially constant. 
Since $A_s$ is nearly time independent, it is customary to express $P(k)$ in 
terms 
of the value of the conserved quantities when the mode crossed the horizon, 
$k=aH$. In other words, one chooses to express the value of the power spectrum, 
which is written in terms of a conserved quantity at first order in the 
slow-roll parameters,  as a time independent function of wavenumber $k$,

\beq\label{pstradicional}
P(k) = \frac{ 2^{2\tilde \nu-4}|\Gamma(\tilde \nu)|^2 }{\pi^3 M_P^2}  
\frac{H^2_\star (k)}{\epsilon_H^\star (k)}  
\eeq
where

\beq\label{pshc}
\frac{H^2_\star (k)}{\epsilon_H^\star (k)} \equiv \left( \frac{H^2 
|\eta|^{3-2\tilde \nu}}{\epsilon_H} \right) \bigg|_{k=aH},
\eeq
then, one computes the spectral index as $n_s - 1 = d \ln P(k) / d \ln k$ using 
Eqs. \eqref{pstradicional} and \eqref{pshc}. 

On the other hand, in the collapse model, the equivalent power spectrum is 
given 
by $P(k) =\mathcal{C}/\pi^2 Q(|z_k|) \\
k^{3-2\nu}$, Eq \eqref{pscolapso}. A 
few 
remarks are in order: 

First, the quantity $\mathcal{C}$ is of the same 
structure 
as $A_s$ in the traditional approach [see Eq. \eqref{C} and Eq. \eqref{Ayns}]. 
In 
other words, is nearly a time independent function, that is to say  its 
derivative with 
respect to the conformal time is of second order in the slow-roll 
parameters. One could follow the 
standard procedure and re-express $\mathcal{C}$ when the mode crossed the 
horizon; however, within our approach, the curvature perturbation and, thus, 
the 
power spectrum is non-vanishing only after the time of collapse.

Specifically, once the collapse has occurred (or more precisely, the 
collapse mechanism has ended and the semiclassical approximation is valid) at 
some time $\tc$ during inflation, and as a consequence,  
$\Psi_{\nk} \neq 0$ for all $\nk$, i.e. the primordial curvature 
perturbation has been generated. Then, one can focus on some particular 
Fourier's mode $\Psi_{\nk}$ and ask if its associated proper wavelength 
$\lambda_P = a/k$ (with $k$ the comoving wavenumber), at the time of collapse, 
is smaller or bigger than the Hubble radius $H^{-1}$, which we know is 
approximately constant during inflation. The answer is simply to compare which 
of the inequalities gets satisfied $k \gg a(\tc) H$ or $k \ll a(\tc) H$, if it 
is the former then the mode is still ``inside the horizon,'' and we know that 
it will eventually cross the horizon, during inflation,   and then ``freezes;'' 
however, if it is the latter then the mode is ``bigger than the horizon'' and 
is 
already ``frozen,'' consequently for this last type of modes it would make no 
sense to evaluate the power spectrum at the ``horizon crossing.''

The second important aspect of the collapse power spectrum is the function 
$Q(|z_k|)$, which is in principle a function of $k$. If one assumes that the 
time of collapse is of the form $\tc \propto k^{-1}$, then $z_k \equiv k\tc = 
z$ 
is independent of $k$; consequently, $Q(|z|)$ would be  also $k$ independent; 
thus, the collapse power spectrum would depend on $k$ as $P(k) \propto 
k^{3-2\nu}$, which for all phenomenological purposes would be indistinguishable 
from the prediction given by the traditional inflationary approach. 
Consequently, in our model, if the dependence on $k$ of the time of collapse 
$\tc$ is slightly different from $\tc \propto 1/k$, then our proposal will 
yield different predictions from the standard inflationary paradigm.

Summarizing the above discussion, if $\tc \propto k^{-1}$, the collapse power 
spectrum can be expressed as $P(k) = \mathcal{A} k^{n_s-1}$, with the amplitude 
given by

\beq\label{ampcolapso}
\mathcal{A} = \frac{2^{2\nu-11} |\Gamma(\nu-1)|^2 H^2 |\eta|^{3-2\nu} }{\pi 
M_P^2 \epsilon_H} Q(|z|);
\eeq
we emphasize, that $\mathcal{A}$ is constant in time, up to second order in the 
slow-roll parameters, and independent of $k$; additionally, the scalar spectral 
index obtained from the collapse model is 

\beq\label{nscolapso}
n_s - 1 = -2 \epsilon_H + 2 \delta_V, 
\eeq
which is a little different from the standard prediction [see Eq. \eqref{Ayns}].

One can relax the condition on the form of the time of collapse by allowing a 
small dependence on $k$, namely, assuming that the time of collapse is  $\tc = 
A/k +B$ and use the observational data to constraint the parameters $A$ and $B$ 
(a similar analysis has been done  for the perfect de Sitter universe, see Ref. 
\cite{LSS12}).  In such case, the collapse power spectrum is of the form $P(k) 
= \mathcal{C}/\pi^2 Q(|z_k|)k^{3-2\nu}$, with $z_k = A + Bk$ [Eq. 
\eqref{pscolapso}]. This is, the collapse power spectrum depends on $k$ in two 
ways, one as $k^{3-2\nu}$ and another through the time of collapse $z_k \equiv 
k\tc = A +Bk$; however, the former dependence is due to the inflationary 
dynamics of the curvature perturbation, and, thus, interpreted as the scalar 
spectral index, while the latter reflects the consideration of the collapse 
hypothesis. Additionally, one could also reinterpret the dependence on $k$, 
introduced by the collapse proposal through the $B$ parameter, as something 
resembling to a running of the scalar spectral index; this would be an 
interesting idea to pursuit since it has been pointed out before that the 
lowest multipoles of the temperature anisotropies prefer models with $n_{\rm 
run} \neq 0$ \cite{wmap9cosmo,Planckcosmo}.
However,  the physical interpretation of the departure of a power law in $k$ 
corresponding to the primordial power spectrum is different. In the collapse 
proposal, it arises from the inclusion of the self-induced collapse of the wave 
function while in the standard approach it arises from considering the second 
order approximation in the slow roll parameters (see 
Ref.\cite{Planckinflation15})

As mentioned before, the time of collapse can occur at any time during the 
inflationary regime; in particular, it can happen when the proper wavelength of 
the mode is bigger or smaller than the Hubble radius, which is approximately 
constant. 

If the proper wavelength of the mode, at the time of collapse, is bigger than 
the Hubble radius, then $k \ll a(\tc) H = \mH (\tc) \simeq -1/\tc$ which is 
equivalent to $-k\tc \\ 
\ll 1$. Then, for modes with a proper wavelength bigger 
than the Hubble radius at the time of collapse,  the collapse power spectrum 
can 
be approximated by

\beq\label{pscolapsoz0}
P(k) \simeq \frac{\mathcal{C}}{\pi^2} \Xi (|z_k|) k^{3-2\nu},
\eeq
where $\Xi(|z_k|)$ is obtained by expanding the function $Q(|z_k|)$, when 
$|z_k| 
\to 0$  [i.e., expanding  $M(|z_k|), N(|z_k|),$ and $
W(|z_k|)$], to the lowest order 
in $|z_k|$. Thus, for each scheme the function $\Xi(|z_k|)$ is

\begin{subequations}\label{pfuera3esquemas}
\beq\label{pindfuera}
\Xi(|z_k|)^{\textrm{ind}} \equiv \frac{4}{\pi^2} \left[  1 + \frac{|z_k|^2}{2} 
\left(  \frac{1}{\nu-1} - \frac{1}{\nu}    \right)      \right],
\eeq
\beq\label{pnewtfuera}
\Xi(|z_k|)^{\textrm{newt}} \equiv \frac{|z_k|^4}{4 \pi^2 \nu^2 (\nu-1)^2},
\eeq
\barr\label{pwigfuera}
& & \Xi(|z_k|)^{\textrm{wig}} \equiv \frac{16}{\pi} \nn
&\times& \left[ \frac{5}{4} 
+\frac{1}{4\zk^2} \left( 1-\sqrt{1+10\zk^2 + 9\zk^4}     \right)   \right]^{-2} 
\nonumber \\ 
&\times& \left[   \frac{\zk^{\nu-1/2}}{\Gamma(\nu) 2^{\nu-1}} \cos \Theta_k  + 
\frac{\zk^{\nu+1/2}}{\Gamma(\nu-1) 2^\nu} \sin \Theta_k     \right]^2,
\earr
\end{subequations}
where $\tan 2\Theta_k \simeq 4\zk/(1-3\zk^2)$.

 On the other hand, if the proper wavelength associated to the mode is smaller 
than the Hubble radius, at the time of collapse, then $k \gg a(\tc) H$, which 
is 
equivalent to $-k \tc \gg 1$. Then the approximated collapse power spectrum, 
when $-k \tc = |z_k| \to \infty$, is
 
 \beq\label{pscolapsozinfty}
P(k) \simeq \frac{\mathcal{C}}{\pi^2} \Upsilon (|z_k|) k^{3-2\nu},
\eeq
where $\Upsilon(|z_k|)$ is obtained by considering the asymptotic behavior of  
the function $Q(|z_k|)$ [i.e., the asymptotic behavior of  $M(|z_k|), N(|z_k|), 
W(|z_k|)$] when $|z_k| \to \infty$. Thus, for each scheme the 
function $\Upsilon(|z_k|)$ is

\begin{subequations}\label{pdentro3esquemas}
\barr
& & \Upsilon(|z_k|)^{\textrm{ind}} \equiv   \frac{4}{\pi^2} \bigg\{ \left[  1 + 
\frac{1}{4 |z_k|^2} \left( \frac{\Gamma(\nu + 3/2)}{ \Gamma(\nu - 1/2)}  
\right)^2       \right] \nn
&\times& \bigg[ \sin \beta(\nu,|z_k|) + \frac{\cos \beta(\nu, |z_k|) }{|z_k|} 
\nn
&\times& \left( -2\nu + \frac{\Gamma(\nu + 
5/2)}{2 \Gamma(\nu + 1/2)}  \right)     \bigg]^2  + \bigg[  1 +
\frac{1}{|z_k|^2} \nn
&\times&  \left( -2\nu + \frac{\Gamma(\nu + 5/2)}{2 \Gamma(\nu + 
1/2)}  
\right)^2       \bigg] \nonumber \\
&\times& \left[ \cos \beta(\nu,|z_k|) - \frac{\sin \beta(\nu, |z_k|) }{2|z_k|} 
\frac{\Gamma(\nu+3/2)}{\Gamma(\nu - 1/2)}     \right]^2 \bigg\},
\label{pinddentro}
\earr
\barr
& & \Upsilon(|z_k|)^{\textrm{newt}} \equiv \frac{4}{\pi^2} \nn
&\times& \left[  1 + 
\frac{1}{|z_k|^2} \left( -2\nu + \frac{\Gamma(\nu + 5/2)}{2 \Gamma(\nu + 1/2)}  
\right)^2       \right] \nonumber \\
&\times& \left[ \cos \beta(\nu,|z_k|) - \frac{\sin \beta(\nu, |z_k|) }{2|z_k|} 
\frac{\Gamma(\nu+3/2)}{\Gamma(\nu - 1/2)}     \right]^2,
\label{pnewtdentro}
\earr
\barr
& & \Upsilon(|z_k|)^{\textrm{wig}} \equiv \frac{16}{\pi^2} \bigg\{  \bigg[ 
\frac{2\nu}{\zk^{3/2}} \nn
&\times& \left( \cos \beta(\nu,|z_k|) - \frac{\sin \beta(\nu, 
|z_k|) }{2|z_k|} \frac{\Gamma(\nu+3/2)}{\Gamma(\nu - 1/2)}     \right)   
\nonumber \\
&-& \left( \sin \beta(\nu,|z_k|) + \frac{\cos \beta(\nu, |z_k|) }{2|z_k|} 
\frac{\Gamma(\nu+5/2)}{\Gamma(\nu + 1/2)}     \right)            \bigg] \cos 
\Theta_k   \nonumber \\
&+&  \left[ \cos \beta(\nu,|z_k|) - \frac{\sin \beta(\nu, |z_k|) }{2|z_k|} 
\frac{\Gamma(\nu+3/2)}{\Gamma(\nu - 1/2)}     \right] \sin \Theta_k  \bigg\}^2, 
\nn
\label{pwignerdentro}
\earr
\end{subequations}
where $\beta(\nu,|z_k|) \equiv |z_k| - (\pi/2) (\nu+1/2)$ and $\tan 2\Theta_k 
\simeq -4/3\zk$.

Expressions \eqref{pfuera3esquemas} and \eqref{pdentro3esquemas} are useful for 
performing the comparisons between  the theoretical prediction of our model and 
the observational data.

\section{Primordial power spectrum in quasi-de Sitter}\label{plots}

In this section, we will show the primordial power spectrum  in the 
quasi-de Sitter case for the different collapse schemes analyzed in this paper. 
 
In particular, we will focus on the cases in which the proper wavelength 
associated 
to the mode is bigger and smaller than the Hubble radius at the time of 
collapse, i.e we analyze the cases such that the comoving $k$, associated to 
the 
curvature perturbation, is bigger or smaller than  $a(\tc)H=\mH(\tc)$; these 
cases correspond to  Eqs. \eqref{pscolapsoz0} and \eqref{pscolapsozinfty} 
respectively. 

This preliminary qualitative analysis indicates that the aforementioned 
collapse 
schemes are good candidates to account for the observational data of the CMB 
collected by the \emph{Planck} \cite{Planckcls13} and WMAP \cite{wmap9cls} 
collaborations.  However, we will not perform here the statistical analyses to 
compare the theoretical predictions with the observational data in order to 
constrain the value of the free parameters of the collapse model ($A$ and $B$). 
We will leave this task for a forthcoming paper \cite{Picci15}. 

First, let us define a fiducial model with a primordial power spectrum $P(k) = 
A_s k^{n_s-1}$ with $n_s=0.96$ that will be taken just as a reference to 
discuss the plots we obtain for the collapse models.  The value of $n_s$, for 
our fiducial model, is the mean value obtained by the \emph{Planck} and WMAP 
collaborations.  Let us remind the reader that the free parameters of all 
collapse schemes ($A$ and $B$) are related to the time of collapse of each mode 
of the inflaton field $\tc=A/k+B$. 

\begin{figure}
\begin{center}
\includegraphics[scale=0.68]{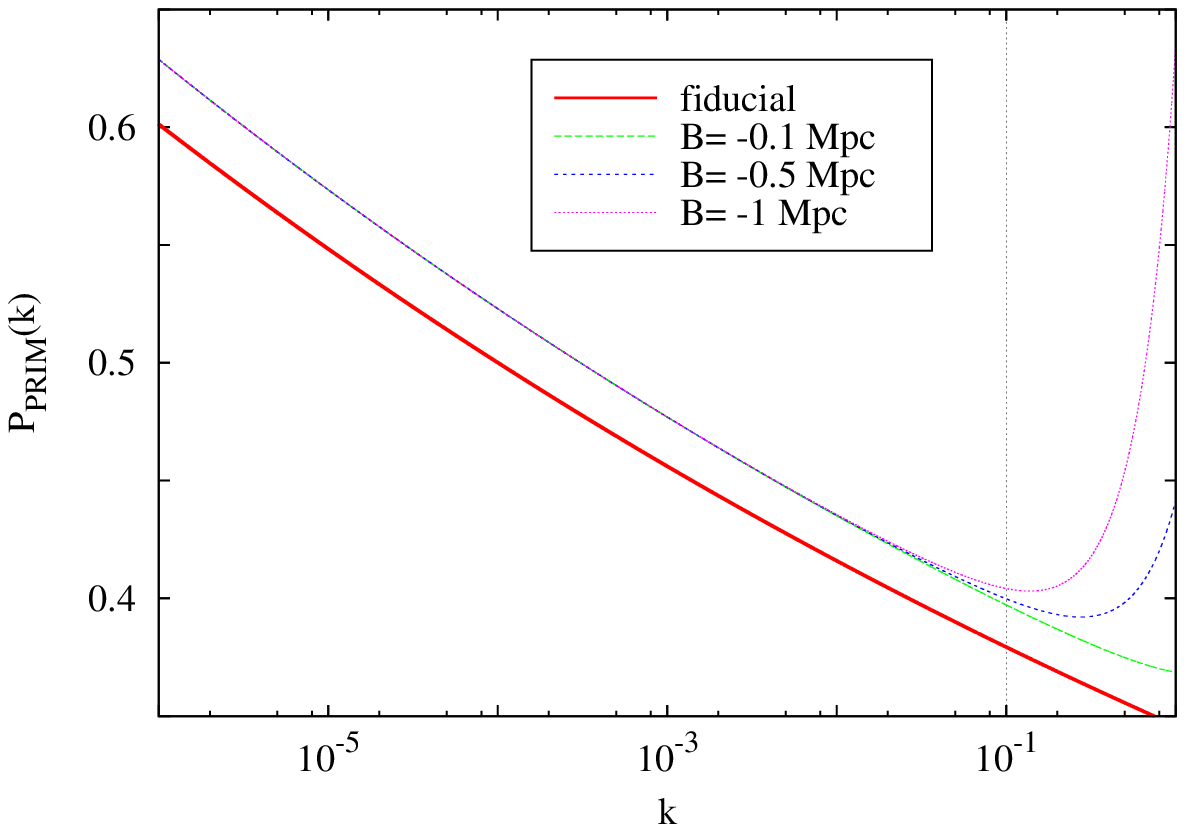}
\includegraphics[scale=0.68]{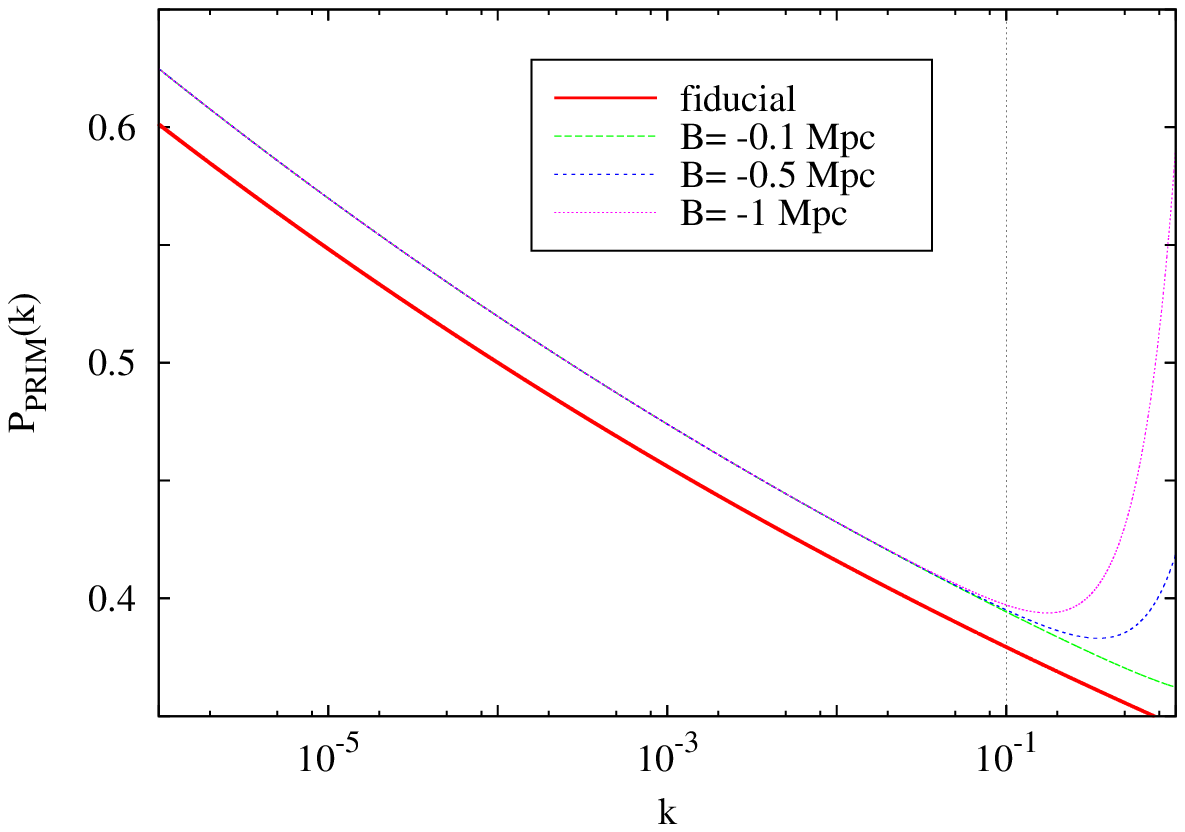}
\end{center}
\caption{Primordial power spectra for the \textit{independent} scheme in the 
case where $k \ll \mH(\tc)$.  The power 
spectrum of the fiducial model is also shown in red. Different values of the 
collapse time $\tc=A/k+B $ are considered, the scalar spectral index  
$n_s=0.96$; Top: $A=-10^{-1}$, Bottom: $A=-10^{-2}$; for both figures 
$A_s=0.39$. The difference in the amplitude between the collapse models and the
fiducial model is artificially set to show the functional form of both
models. }
\label{powerindfuera}
\end{figure}

\begin{figure}
\begin{center}
\includegraphics[scale=0.68]{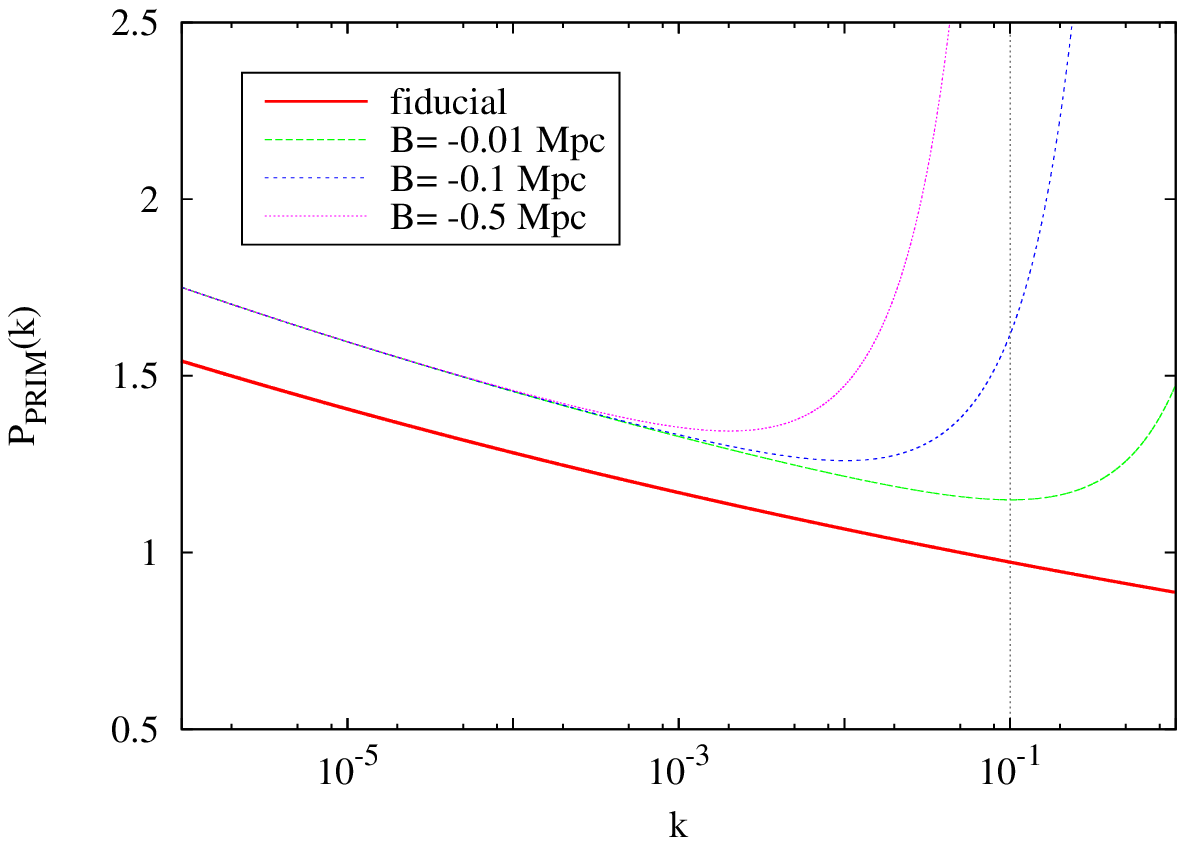}
\includegraphics[scale=0.68]{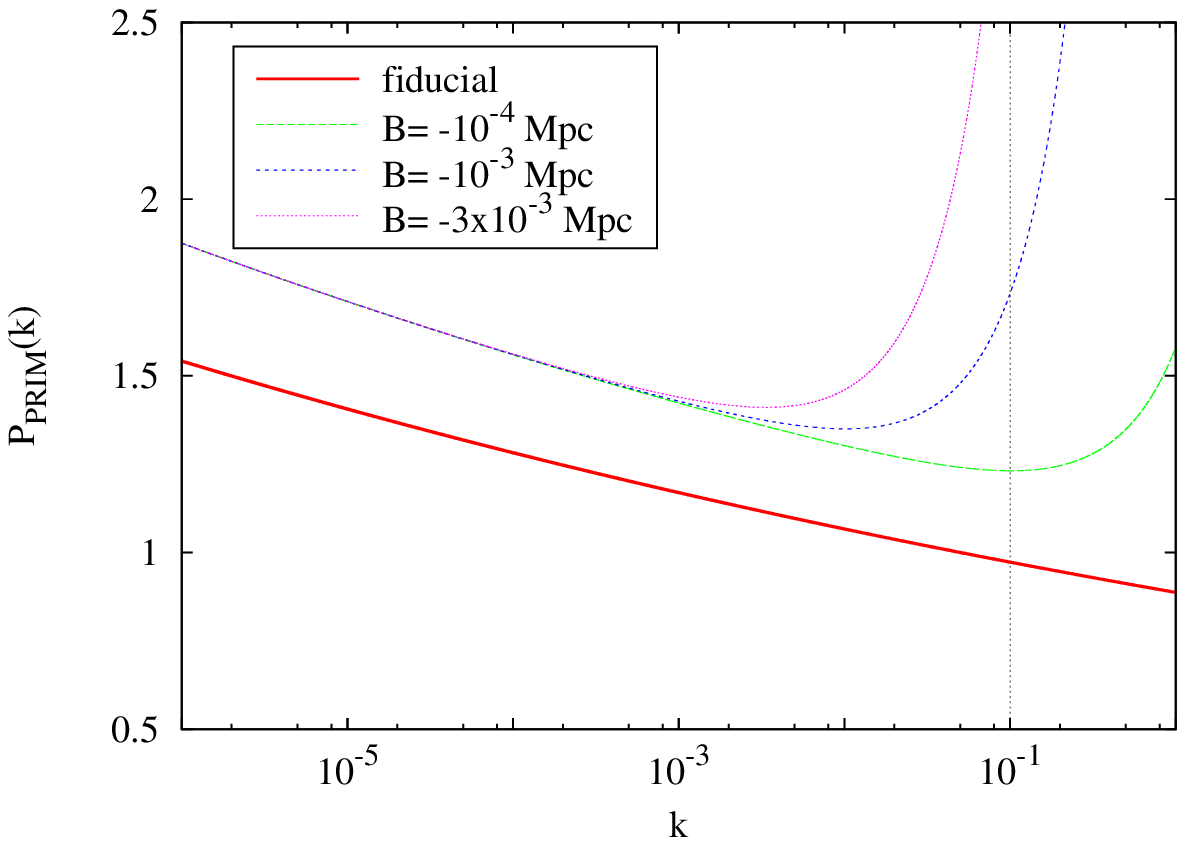}
\end{center}
\caption{Primordial power spectra for the \textit{Newtonian} collapse scheme in 
the case where $k\ll\mH(\tc)$.  The 
power spectrum of the fiducial model is also shown in red. Different values of 
the collapse time $\tc=A/k+B $ are considered, the scalar spectral index  
$n_s=0.96$; Top: $A=-10^{-1}$ and $A_s = 2.8 \times 10^{5}$, Bottom: $A=-10^{-3}$ and 
$A_s=3 \times 10^{13}$. 
The difference in the amplitude between the  collapse models  and the fiducial 
model is artificially set to show the shape of both models. }
\label{powernewtfuera}
\end{figure}

\begin{figure}
\begin{center}
\includegraphics[scale=0.68]{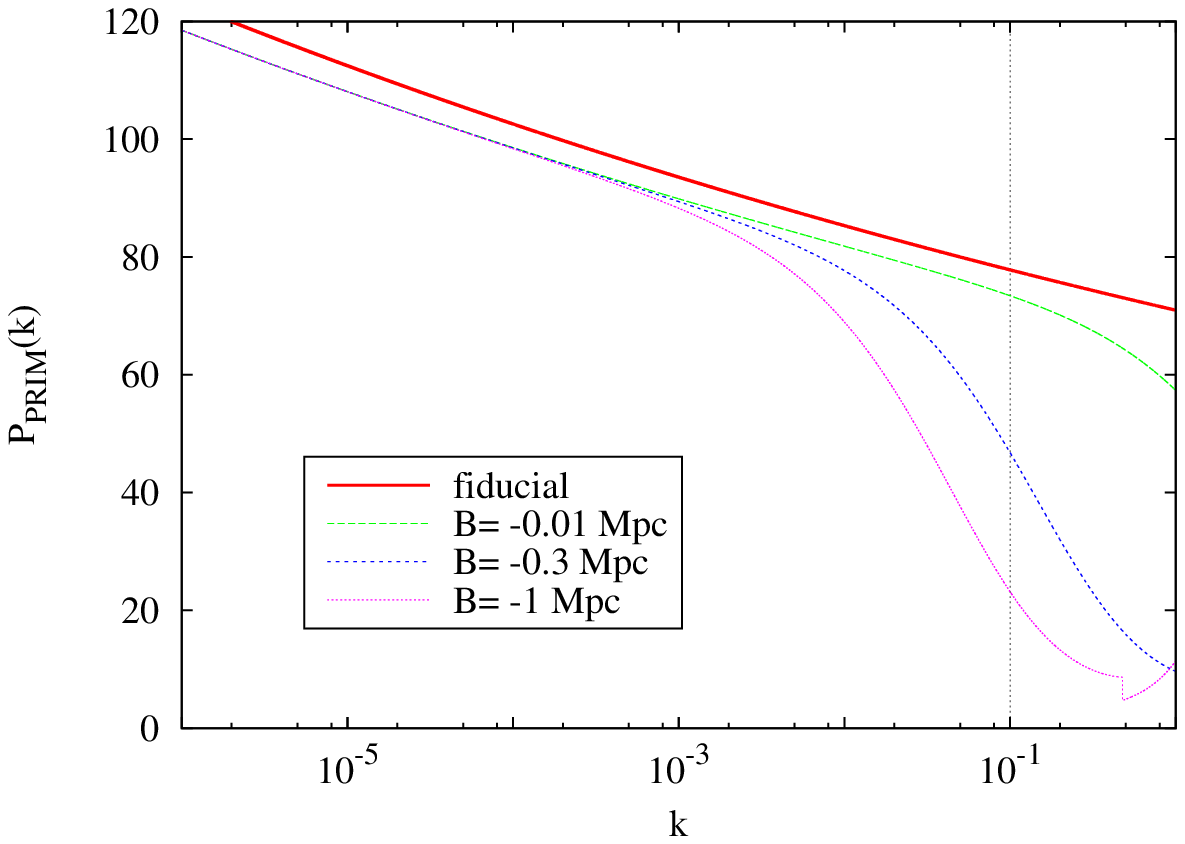}
\includegraphics[scale=0.68]{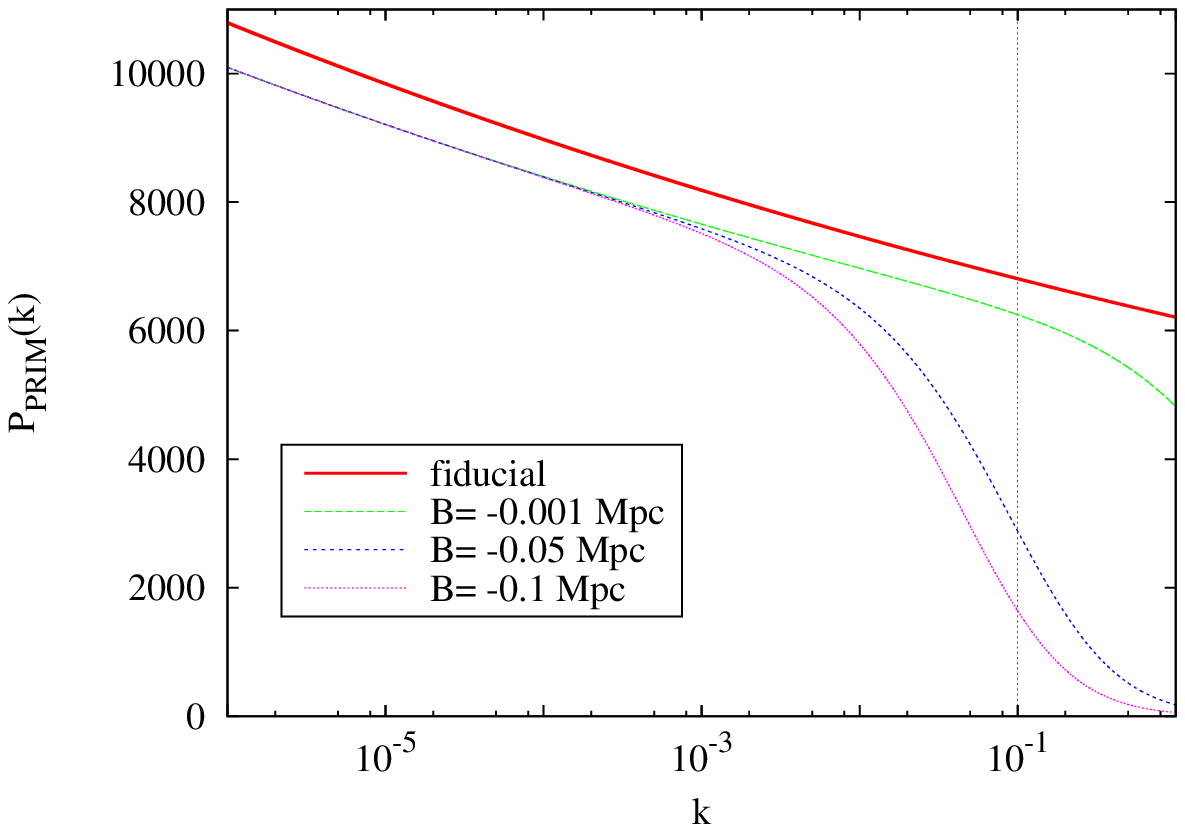}
\end{center}
\caption{Primordial power spectra for the \textit{Wigner} collapse scheme in 
the case where $k \ll \mH(\tc)$.  The 
power spectrum of the fiducial model is also shown in red. Different values of 
the collapse time $\tc=A/k+B $ are considered, the scalar spectral index  
$n_s=0.96$; Top: $A=-10^{-1}$ with $A_s=80$, Bottom: $A=-10^{-2}$ with $A_s = 
7000$. The difference in the amplitude between the collapse models and the
fiducial model is artificially set to show the functional form of both
models.}
\label{powerwigfuera}
\end{figure}

Figures \ref{powerindfuera}, \ref{powernewtfuera} and \ref{powerwigfuera} show 
the primordial spectra for the different collapse schemes (\textit{independent, 
Newtonian, Wigner}), in the case where the proper wavelength associated to the 
mode is bigger than the Hubble radius at the time of collapse, i.e. $k \ll 
\mH(\tc)$ [Eqs. \eqref{pfuera3esquemas}], for different values of the 
collapse parameters 
$A$ and $B$.  The primordial power spectrum of the fiducial model is also shown 
in each Figure; the value of $A_s$ (detailed in the caption of each Fig.)  is 
settled in each case in order to provide the reader a clear idea regarding the 
differences in the form of the different spectra shown in the plot. For 
example, 
in Fig. \ref{powerindfuera} the fiducial power spectrum could be normalized in 
such a way that it almost overlaps the primordial spectrum for $B=-0.1$ {\rm 
Mpc}. However, for $B=-0.5$ {\rm Mpc} and $B=-1$ {\rm Mpc} there is no value of 
$A_s$ that makes both spectra (collapse and fiducial model) overlap.
Hereinafter, we discuss the effects of introducing the collapse hypothesis for 
the different collapse schemes in the primordial power spectrum. The relevant 
values 
for 
$k$ that will affect the prediction of the CMB temperature and polarization 
anisotropy are $10^{-6}{\rm Mpc}^{-1} < k < 10^{-1}{\rm Mpc}^{-1}$.     
Therefore, we have drawn a vertical line in each Figure, at $k=0.1$   
Mpc$^{-1}$. 

Figure \ref{powerindfuera} shows no change in the slope of the primordial power 
spectrum (with respect to the fiducial model one) for the \textit{independent} 
scheme in the range $k < 0.1$ Mpc$^{-1}$ and the considered values of 
$B$ and the two values of $A$; a notorious deviation in shape from the fiducial 
one occurs for $k > 0.1$ Mpc$^{-1}$ . However, we have already mentioned 
that the relevant values of $l$ that affect the CMB spectrum correspond to  
$10^{-6}{\rm Mpc}^{-1} < k < 10^{-1}{\rm Mpc}^{-1}$, thus we expect no 
deviation 
of the CMB temperature spectrum respect the fiducial one for this scheme.

{Again, Figure \ref{powernewtfuera} shows no change in the slope of the 
power spectrum, but unlike the previous case, there is a deviation 
 from the fiducial power spectrum  for increasing values of $B$, and this 
deviation happens within the relevant range of $k$. Furthermore, it should be 
noted the huge value of $A_s$ considered  in order to make comparisons 
between the collapse and fiducial power spectrum. A 
different value of this parameter with respect to the the value of the standard 
model could result in a different value for the scale energy of the inflationary 
period. However, the complete determination of this scale energy depends not 
only on the temperature fluctuations of the CMB but also on the $B$-mode 
polarization. As mentioned before, a recent joint paper by the BICEP2/Keck and  
Planck collaborations concludes that there is no evidence for primordial 
$B$-modes at low angular multipoles.}

 { Figure \ref{powerwigfuera} shows the differences in the primordial spectra among 
different values of $B$ 
for the two values of $A$ considered and also with respect to the fiducial 
model for the \textit{Wigner's} scheme. It is interesting to note that the 
deviation with respect to the fiducial model goes to the opposite side that in 
the \textit{Newtonian} scheme discussed previously.  Furthermore, as in the 
\textit{Newtonian} scheme, it was necessary to consider large values of $A_s$ 
in order to make the collapse and fiducial model spectra comparable. Thus, the discussion 
made for the \textit{Newtonian} scheme is also valid for this case.

In summary, in this preliminary analysis for the case where $k \ll \mH(\tc)$, 
we could find values of the collapse 
parameters ($A$ and $B$) that make the primordial power spectrum of the 
collapse 
models to be almost equal to the fiducial primordial power spectrum for  the 
all the schemes considered in this paper.} In consequence, we expect
that the corresponding CMB spectrum will not deviate
much from the fiducial one.

\begin{figure}
\begin{center}
\includegraphics[scale=0.68]{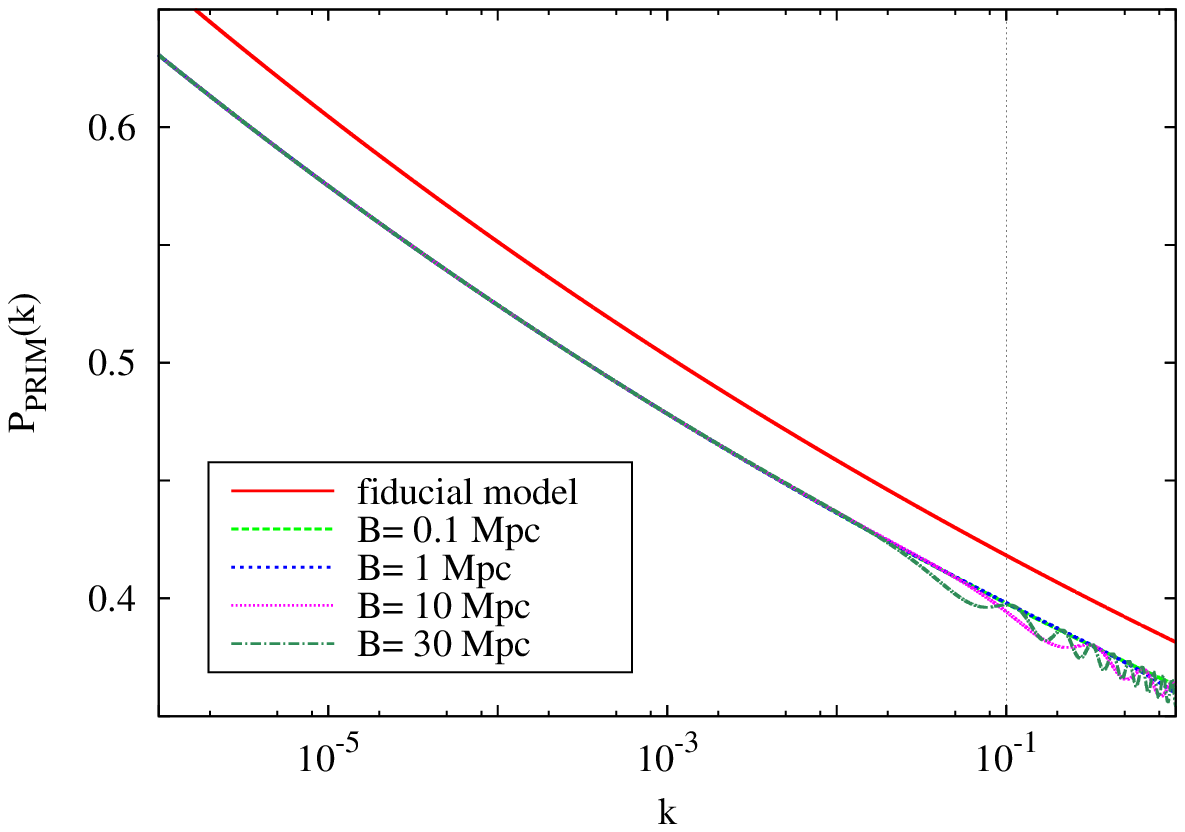}
\includegraphics[scale=0.68]{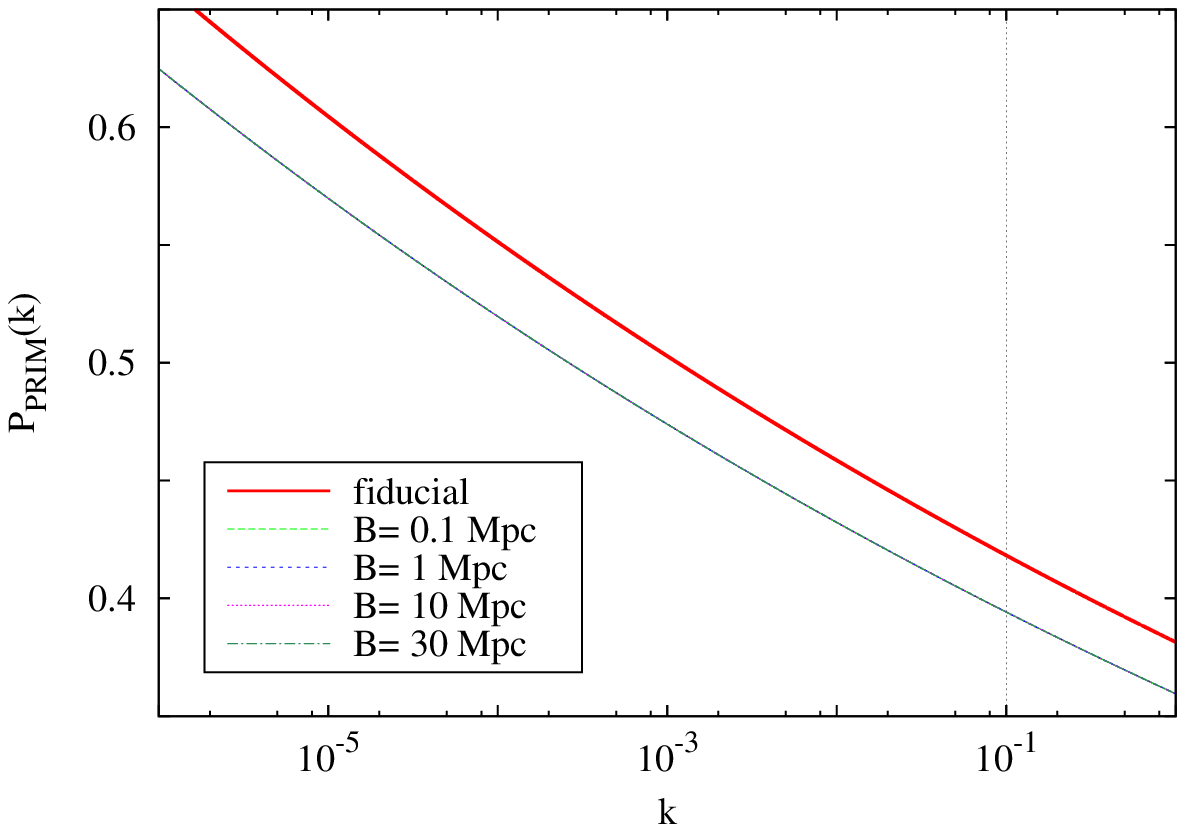}
\end{center}
\caption{Primordial power spectra for the \textit{independent} scheme in the 
case where $k \gg \mH(\tc)$. The power 
spectrum of the fiducial model is also shown in red. Different values of the 
collapse time $\tc=A/k+B $ are considered, the scalar spectral index  
$n_s=0.96$; Top: $A=-10^{2}$, Bottom: $A=-10^{6}$, for both figures $A_s=0.43$. 
The difference in the amplitude between the  collapse models  and the fiducial 
model is artificially set to show the shape of both models.}
\label{powerind}
\end{figure}

\begin{figure}
\begin{center}
\includegraphics[scale=0.68]{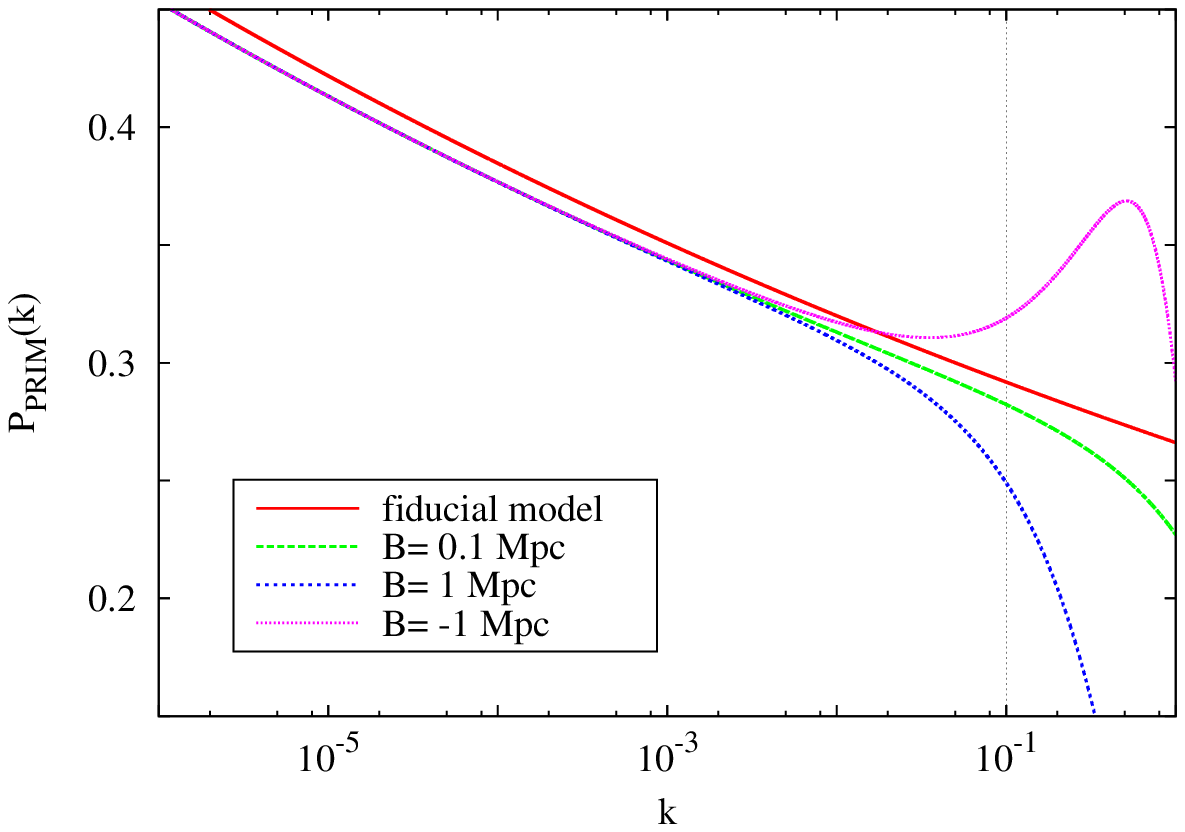}
\includegraphics[scale=0.68]{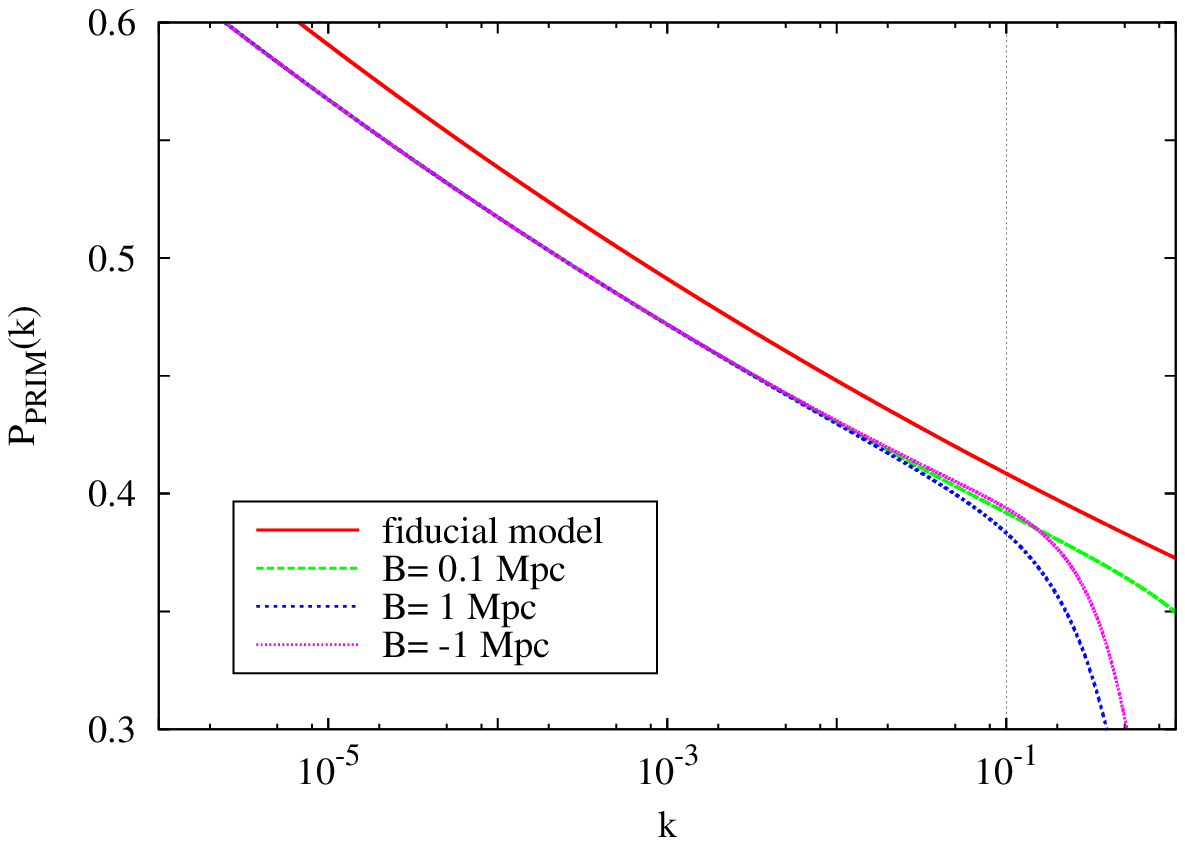}
\end{center}
\caption{Primordial power spectra for the \textit{Newtonian} collapse scheme in 
the case where $k \gg \mH(\tc)$.  The 
power spectrum of the fiducial model is also shown in red. Different values of 
the collapse time $\tc=A/k+B $ are considered, the scalar spectral index  
$n_s=0.96$; Top: $A=-10^{2}$ with $A_s=0.3$, Bottom: $A=-10^{5}$ with 
$A_s=0.42$. 
The difference in the amplitude between the  collapse models  and the fiducial 
model is artificially set to show the shape of both models. }
\label{powernewt}
 \end{figure}

\begin{figure}
\begin{center}
\includegraphics[scale=0.68]{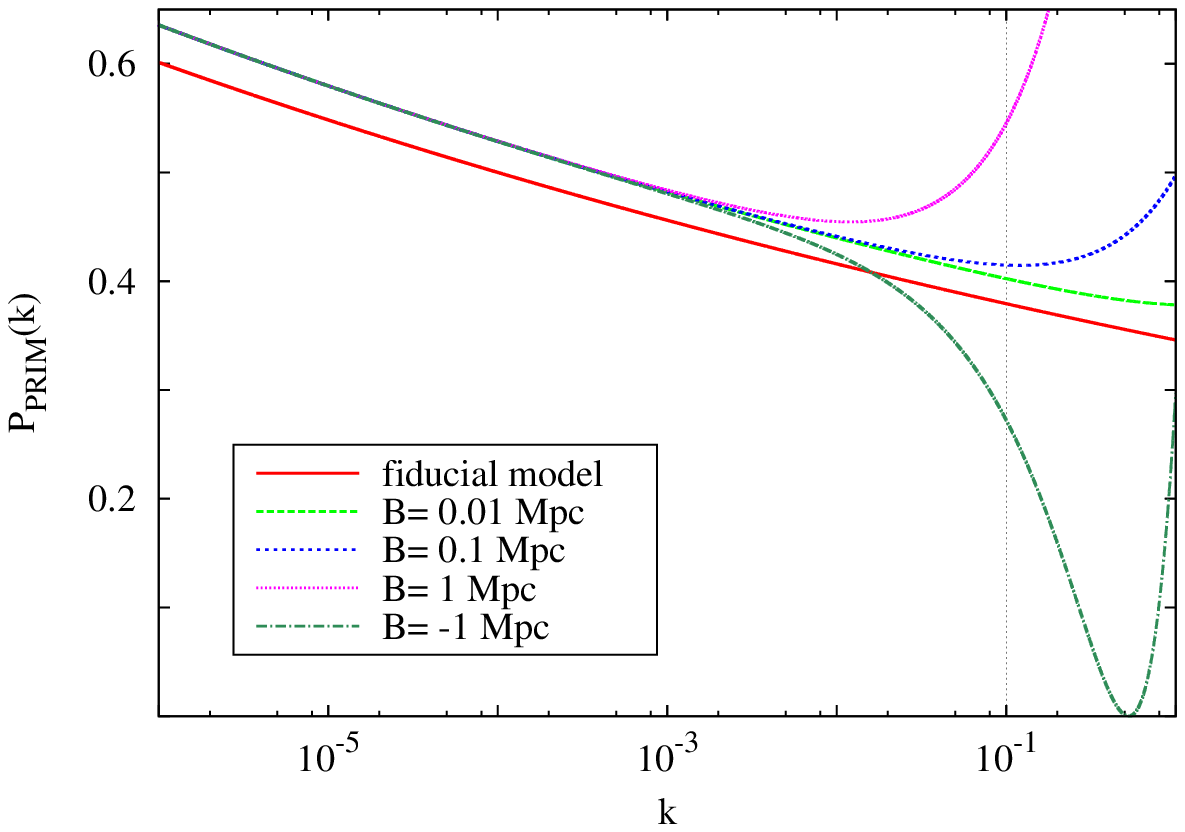}
\includegraphics[scale=0.68]{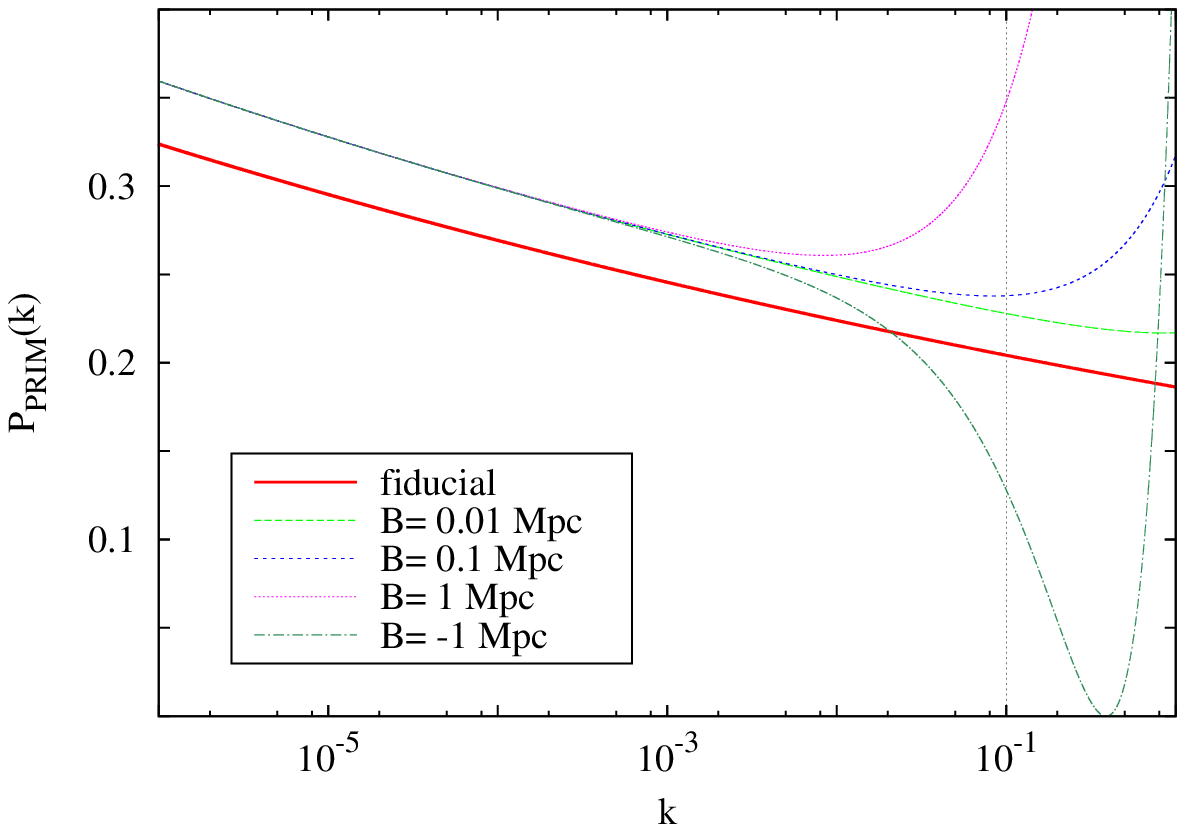}
\end{center}
\caption{Primordial power spectra for the \textit{Wigner} collapse scheme in 
the case where $k \gg \mH(\tc)$.  The 
power spectrum of the fiducial model is also shown in red. Different values of 
the collapse time $\tc=A/k+B $ are considered, the scalar spectral index  
$n_s=0.96$; Top: $A=-10^{2}$ with $A_s=0.39$, Bottom: $A=-10^{6}$ with 
$A_s=0.21$. 
The difference in the amplitude between the  collapse models  and the fiducial 
model is artificially set to show the shape of both models.  }
\label{powerwig}
\end{figure}

Figures \ref{powerind}, \ref{powernewt} and \ref{powerwig} show the primordial 
spectra for the different collapse schemes (\textit{independent, Newtonian, 
Wigner}) in the case where the proper wavelength associated to the mode is 
smaller than the Hubble radius at the time of collapse, i.e. $k \gg \mH(\tc)$  
[Eqs. \eqref{pdentro3esquemas}] 
for different values of the collapse 
parameters $A$ and $B$.  The 
primordial power spectrum of the fiducial model is also shown in each Figure  
and the value of $A_s$ (detailed in the caption of each Figure)  is settled in 
each case in order to provide the reader a clear idea regarding the differences 
in the form of the different spectra shown in the plot.   The relevant values 
for 
$k$ that will affect the prediction of the CMB temperature and polarization 
anisotropy are $10^{-6}{\rm Mpc}^{-1} < k < 10^{-1}{\rm Mpc}^{-1}$.     
Therefore, we have drawn a vertical in each Figure, at $k=0.1$ Mpc$^{-1}$. 
  For the 
\textit{independent} scheme (Fig. \ref{powerind}), it should be noted that for 
$A=-10^{2}$ the power spectrum deviates  from the fiducial model for 
increasing values of $B$, however, it is a small change compared to the 
deviations of other schemes (see Figs. \ref{powernewt} and \ref{powerwig}),  
while for $A=-10^{6}$ there is no difference in the spectrum among different 
values of $B$. This is due to the fact that for large values of $A$, the value 
of $z_k$ becomes also very large, and therefore the value of $B$ does not 
affect the final form of the power spectrum. This behavior is similar to the 
case where the scale factor is exactly de Sitter (see Ref. \cite{LSS12}). 

For the \textit{Newtonian} scheme (Figure \ref{powernewt}) it should be 
mentioned that in both cases ($A=-10^{2}$ and $A=-10^{5}$) considered  for 
increasing values of $B$  the spectrum deviates  from the fiducial one for 
$k>0.01$ Mpc$^{-1}$.
For the \textit{Wigner} scheme (Figure \ref{powerwig}), it should be mentioned 
that  for increasing values of $B$ the power spectrum deviates more from the 
fiducial model in both cases ($A=-10^{2}$ and $A=-10^{6}$)  for $k>0.005$
Mpc$^{-1}$.
In summary,  increasing the value of $B$ results in a deviation from the 
fiducial model spectrum; the deviation is more drastically for  the 
\textit{Newtonian} and \textit{Wigner} scheme. 

 \section{Predictions of the collapse schemes on the CMB temperature 
spectrum}\label{Cls}

The aim of this section is to show, that introducing the collapse of the 
inflaton wave function during inflation has observable consequences on the CMB 
fluctuation spectrum. 
In this paper, we will limit ourselves to the analysis of the temperature 
auto-correlation spectrum; however, from a previous analysis  of similar models 
\cite{LSS12} we might expect that  the $E$-mode polarization and 
temperature-$E$-mode cross correlation will be also modified as a consequence 
of the collapse hypothesis.  
As we will see, the effect is different for the three
collapse schemes proposed in this paper and it also depends on the value of the 
time of collapse. This is not a surprise since we have shown in the previous 
section that there are differences between the primordial power spectrum 
in each collapse scheme and the standard inflationary model one. We want
to stress that, in this paper, we will only perform a preliminary analysis of 
the CMB spectrum predicted by the collapse models for some particular cases; a 
complete  analysis, including 
statistical analysis, in which recent CMB data are confronted with the 
predictions from all collapse schemes, is in progress \cite{Picci15}. In order 
to perform our analysis, let us 
define the cosmological parameters of our fiducial model: baryon density in 
units of the critical density  $\Omega_B h^2= 0.02212$, dark matter density in 
units of the critical density
$\Omega_{CDM} h^2=0.1187 $, Hubble constant in 
units of ${\rm Mpc^{-1} km\: s^{-1} }$ $H_0 = 67.75$, reionization optical 
depth, 
$\tau = 0.092$, and the scalar spectral index, $n_s = 0.96$. These are the 
best-fit
values presented by the Planck collaboration \cite{Planckcosmo2013}.  The value 
of $A_s$ is settled in each case in order to match the maximum of the first 
Doppler peak with the fiducial model one. 

\begin{figure}
\begin{center}
\includegraphics[scale=0.68]{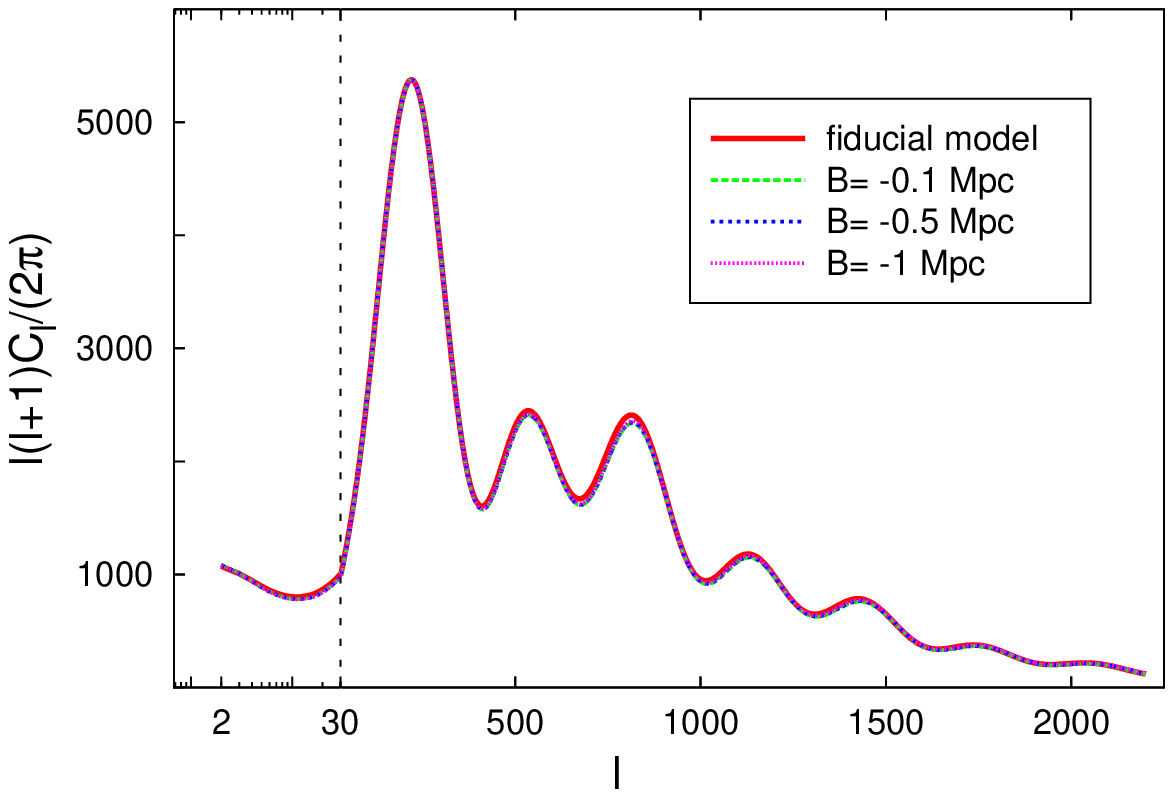}
\includegraphics[scale=0.68]{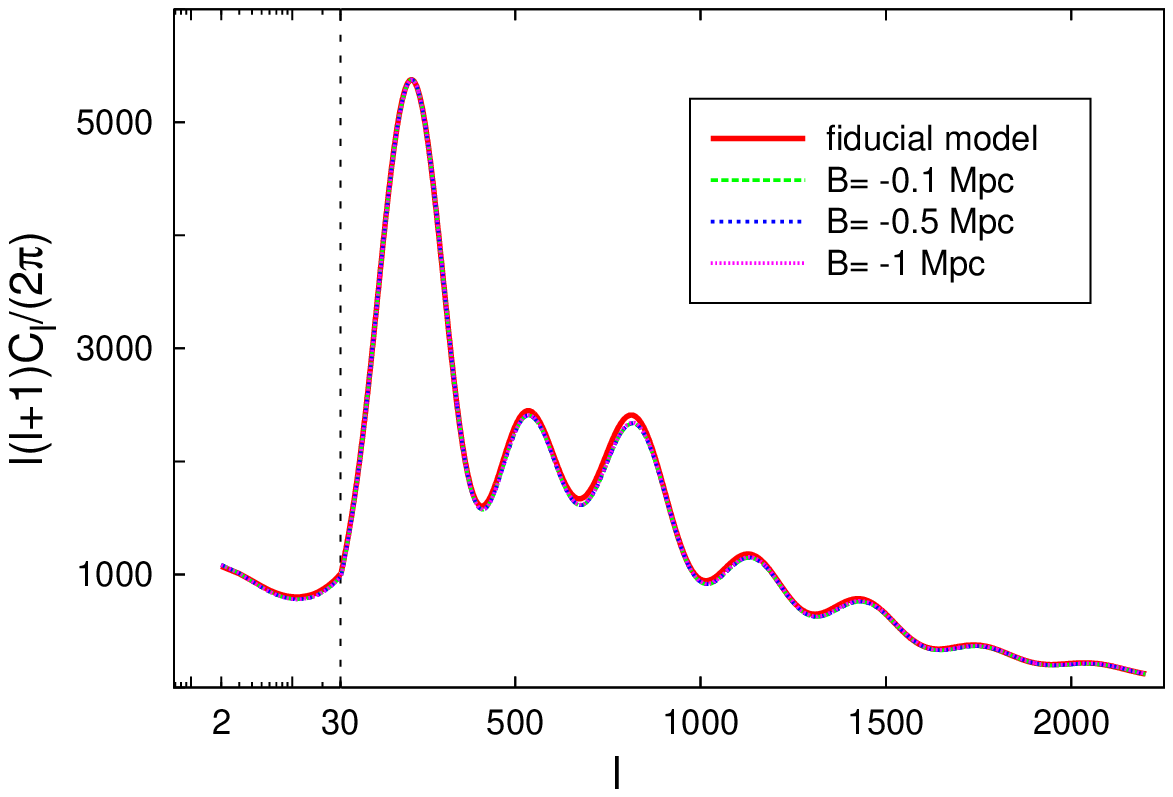}
\end{center}
\caption{The temperature auto-correlation (TT) power spectrum for the 
\textit{independent} scheme in 
the case where $k \ll \mH(\tc)$.   All models
are normalized to the maximum of the first peak of the fiducial model.  The 
power spectrum of the fiducial model is also shown in red. Different values of 
the collapse time $\tc=A/k+B $ are considered, the scalar spectral index  
$n_s=0.96$; Top: $A=-10^{-1}$, Bottom: $A=-10^{-2}$. }
\label{cls-ind-fuera}
\end{figure}

\begin{figure}
\begin{center}
\includegraphics[scale=0.68]{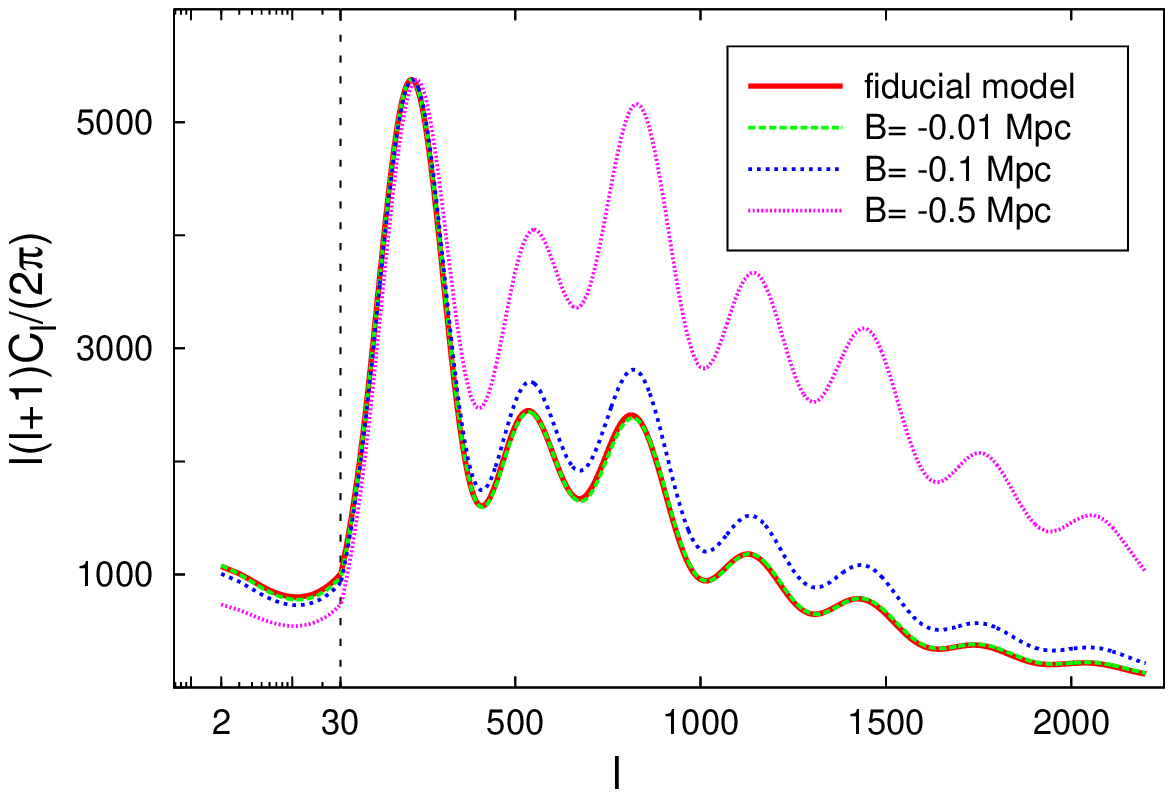}
\includegraphics[scale=0.68]{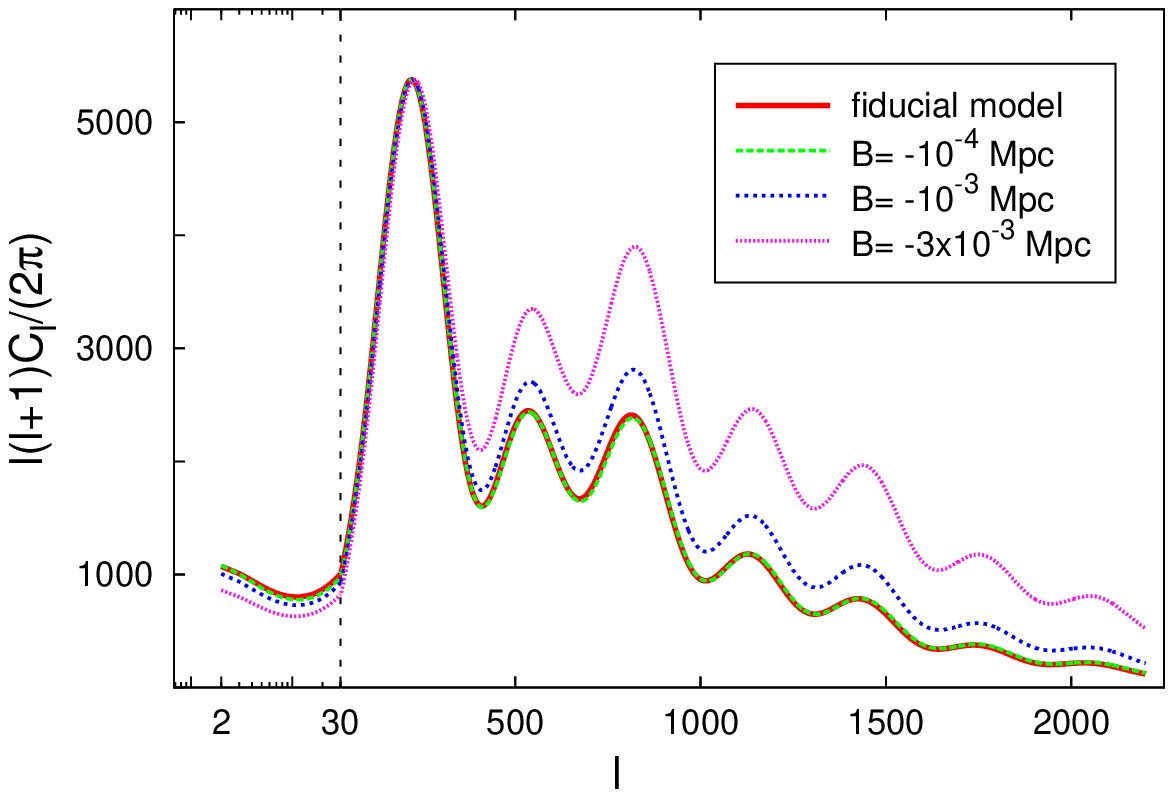}
\end{center}
\caption{The temperature auto-correlation (TT) power spectrum for the 
\textit{Newtonian} scheme in 
the case where $k \ll \mH(\tc)$.   All 
models
are normalized to the maximum of the first peak of the fiducial model.  The 
power spectrum of the fiducial model is also shown in red. Different values of 
the collapse time $\tc=A/k+B $ are considered, the scalar spectral index  
$n_s=0.96$; Top: $A=-10^{-1}$, Bottom: $A=-10^{-3}$.  }
\label{cls-newt-fuera}
\end{figure}

\begin{figure}
\begin{center}
\includegraphics[scale=0.68]{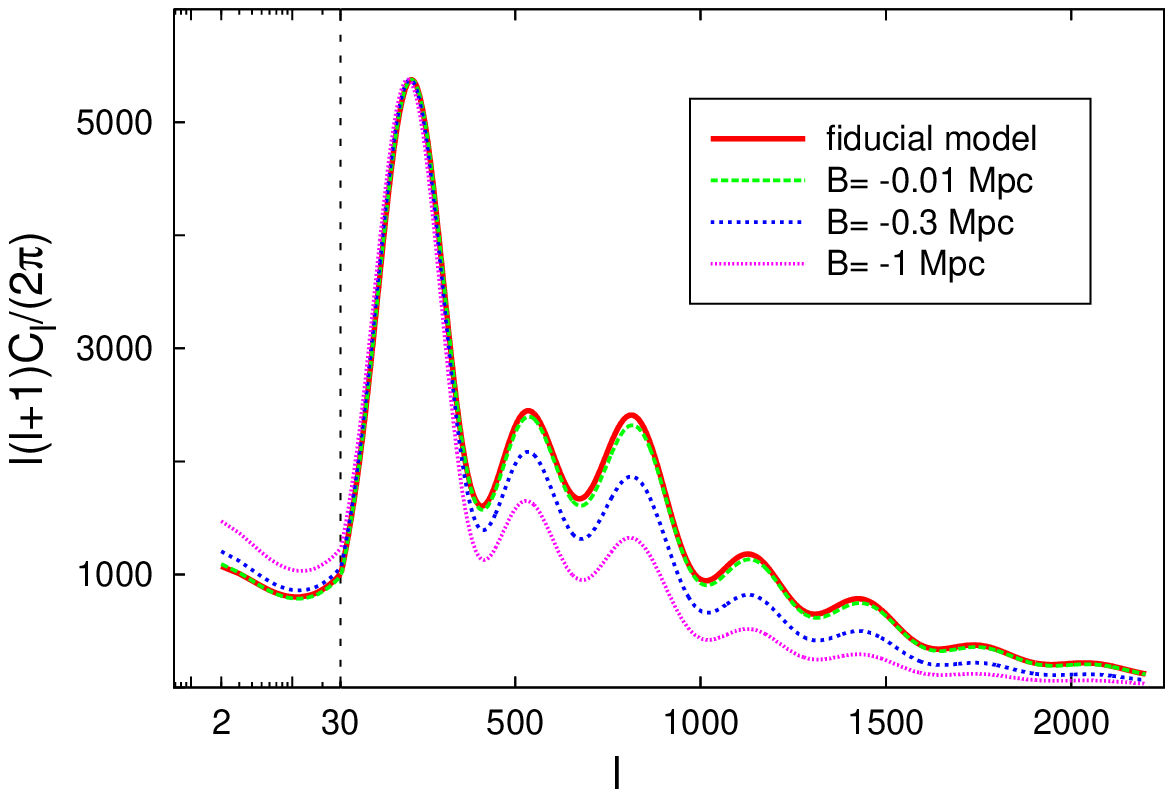}
\includegraphics[scale=0.68]{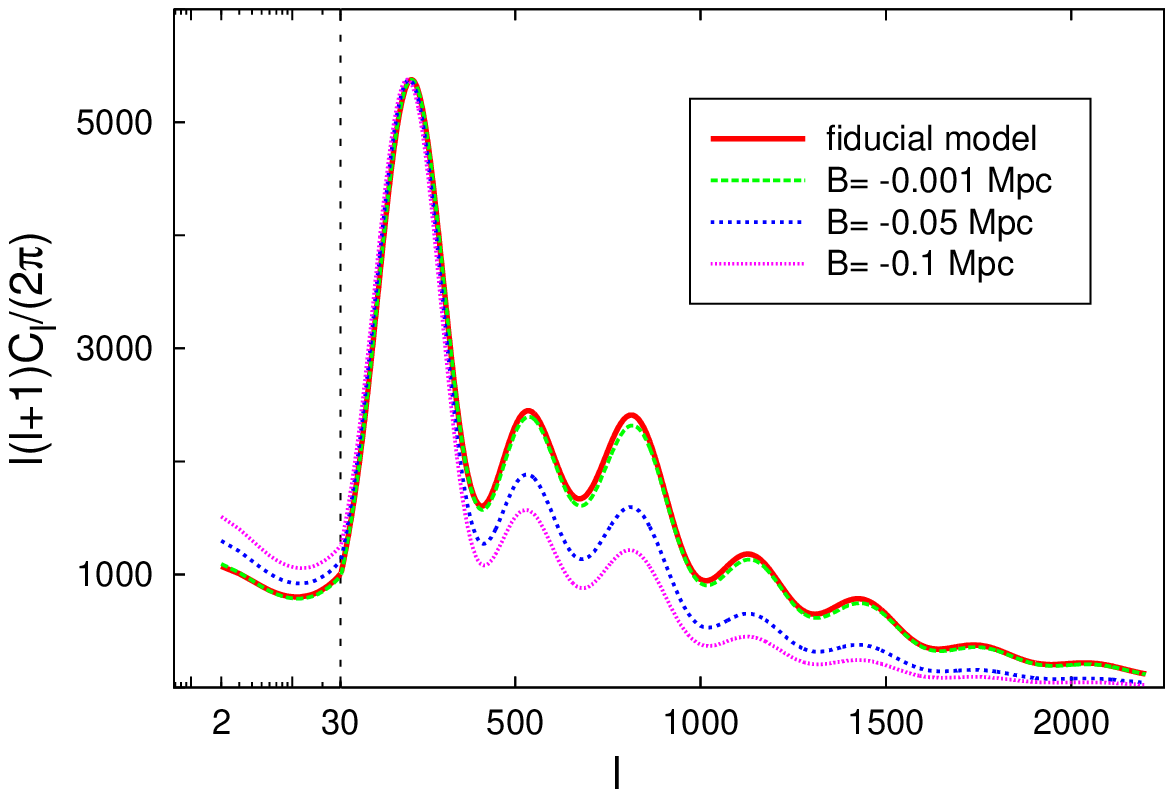}
\end{center}
\caption{The temperature auto-correlation (TT) power spectrum for the 
\textit{Wigner} scheme in 
the case where $k \ll \mH(\tc)$.   All 
models
are normalized to the maximum of the first peak of the fiducial model.  The 
power spectrum of the fiducial model is also shown in red.  Different values of 
the collapse time $\tc=A/k+B $ are considered, the scalar spectral index  
$n_s=0.96$; Top: $A=-10^{-1}$, Bottom: $A=-10^{-2}$.  }
\label{cls-wig-fuera}
\end{figure}

Figs.  \ref{cls-ind-fuera}, \ref{cls-newt-fuera} and \ref{cls-wig-fuera} show 
the prediction for the CMB temperature fluctuation spectrum for the 
\textit{independent}, \textit{Newtonian} and \textit{Wigner} schemes in the 
case 
where   $k \ll \mH(\tc)$.  Let us first analyze the case of the 
\textit{independent} scheme; for the two cases shown in this paper 
($A=-10^{-1}$ 
and $A=-10^{-2}$) there is only a very tiny difference between the fiducial 
spectrum and the one  including the self-induce collapse of the inflaton's 
wave function. We have calculated the $\chi^2$ using  the latest release of 
Planck data, and the difference between the value of the fiducial model and the 
collapse model is within the expected values for this quantity. In consequence, 
we expect that any value of $B$ can explain recent observational data.

{ Conversely, Figs. \ref{cls-newt-fuera} and  \ref{cls-wig-fuera} show large 
deviations for increasing values of $B$ in the temperature power spectrum 
respect to the fiducial one.  On the other hand, we would like to stress that 
the values of $A_s$ required to match the collapse and the fiducial spectrum are 
large and, thus, these cases could be severely constrained with future 
measurements of the $B$-polarization mode.

In such way, we expect that a future statistical analysis using latest Planck 
data, will allow us to constrain the values of $B$ for the Newtonian and 
Wigner's scheme in the case where the collapse occurs when the proper 
wavelength of the mode is bigger than the Hubble radius.}

\begin{figure}
\begin{center}
\includegraphics[scale=0.68]{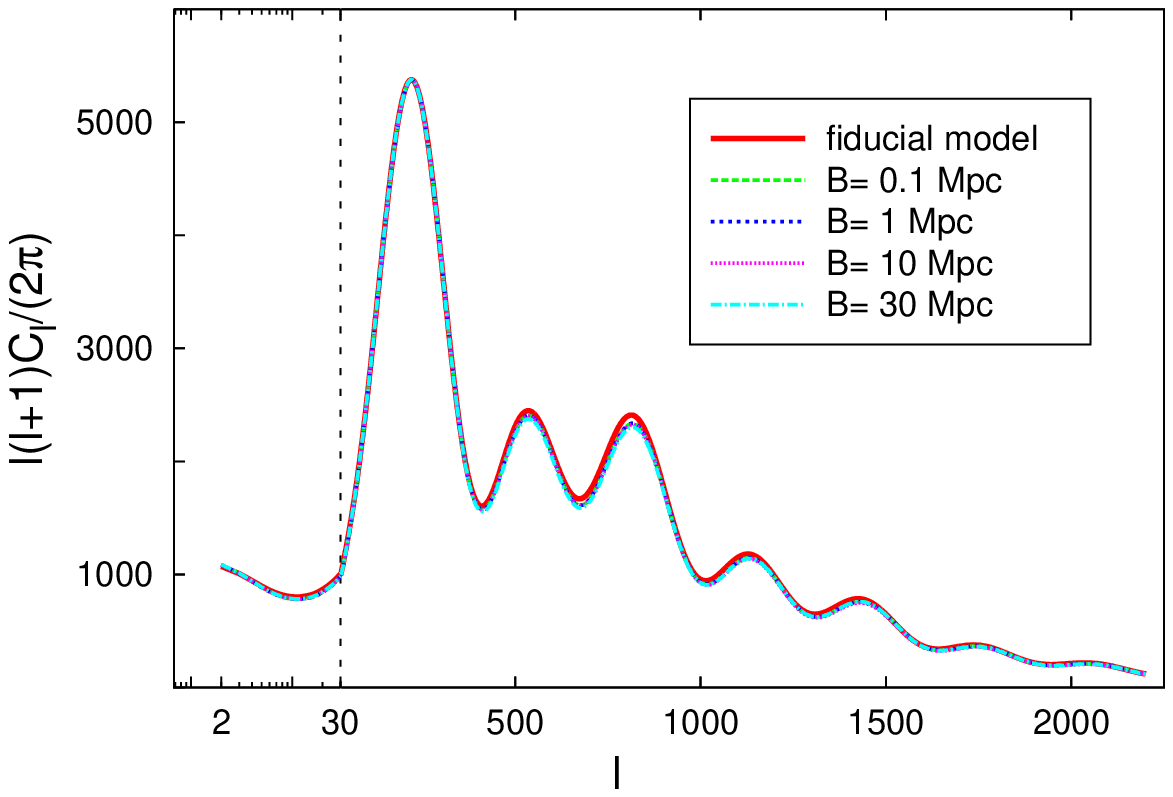}
\includegraphics[scale=0.68]{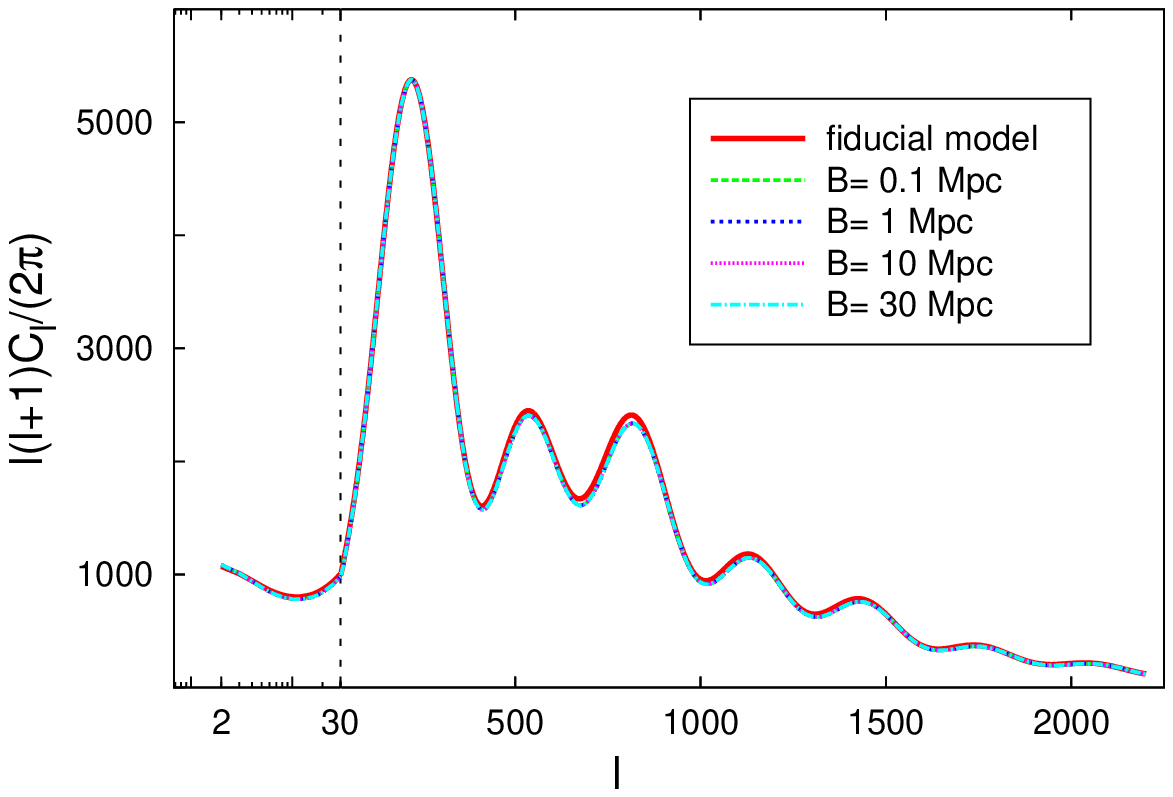}
\end{center}
\caption{The temperature auto-correlation (TT) power spectrum for the 
\textit{independent} scheme in 
the case where $k \gg \mH(\tc)$.   All 
models
are normalized to the maximum of the first peak of the fiducial model.  The 
power spectrum of the fiducial model is also shown in red. Different values of 
the collapse time $\tc=A/k+B $ are considered, the scalar spectral index  
$n_s=0.96$; Top: $A=-10^{2}$, Bottom: $A=-10^{6}$.  }
\label{cls-ind-dentro}
\end{figure}

\begin{figure}
\begin{center}
\includegraphics[scale=0.68]{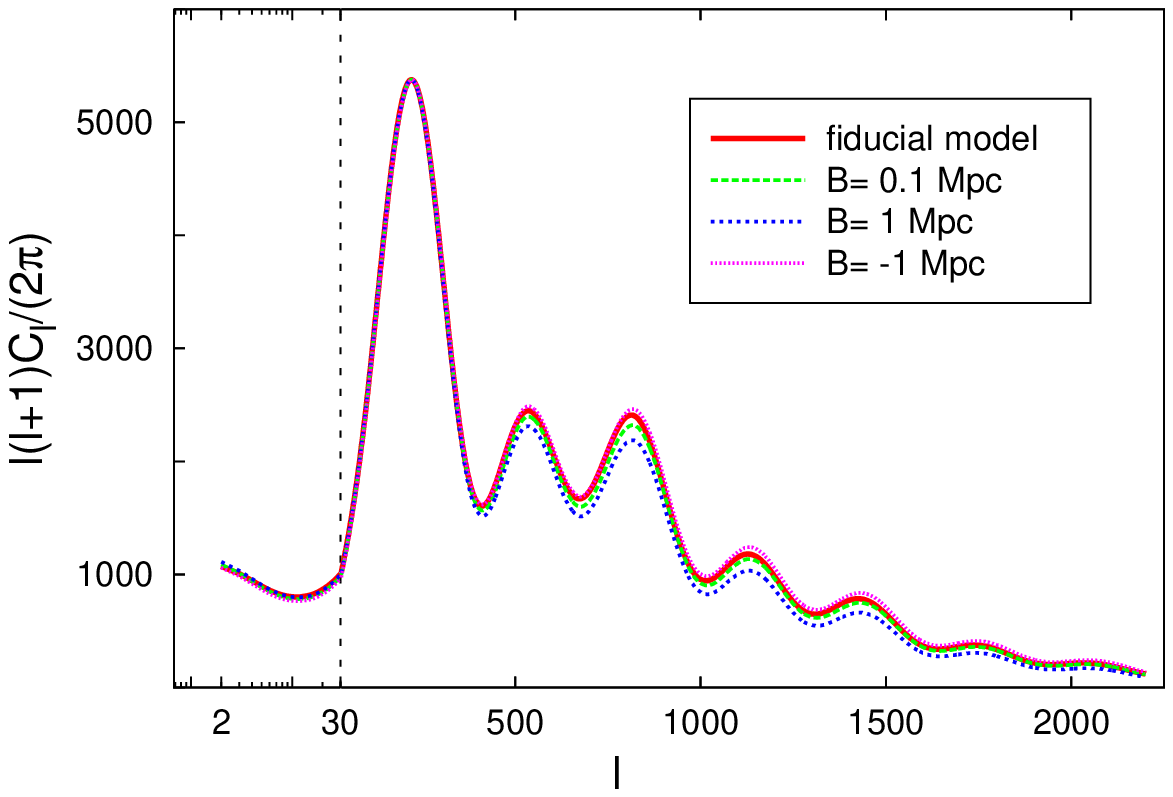}
\includegraphics[scale=0.68]{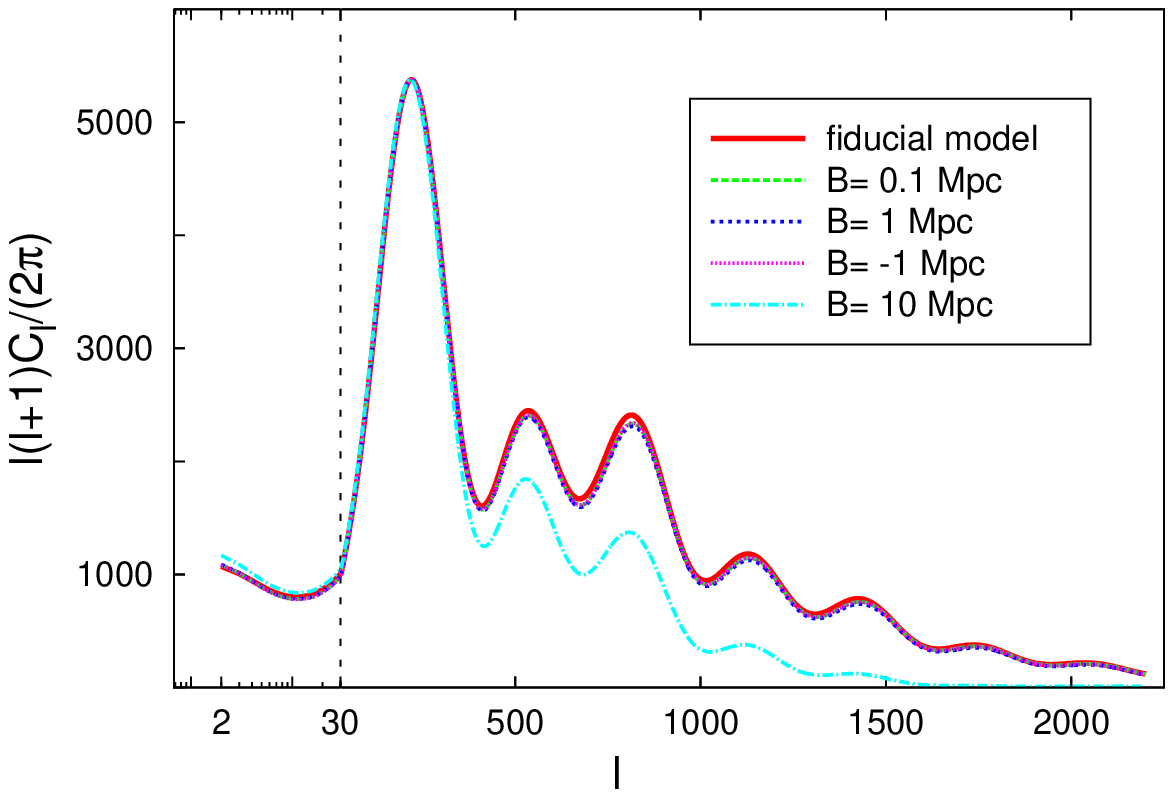}
\end{center}
\caption{The temperature auto-correlation (TT) power spectrum for the  
\textit{Newtonian} scheme in 
the case where $k \gg \mH(\tc)$.   All 
models
are normalized to the maximum of the first peak of the fiducial model.  The 
power spectrum of the fiducial model is also shown in red. Different values of 
the collapse time $\tc=A/k+B $ are considered, the scalar spectral index  
$n_s=0.96$; Top: $A=-10^{2}$, Bottom: $A=-10^{5}$.  }
\label{cls-newt-dentro}
\end{figure}

\begin{figure}
\begin{center}
\includegraphics[scale=0.68]{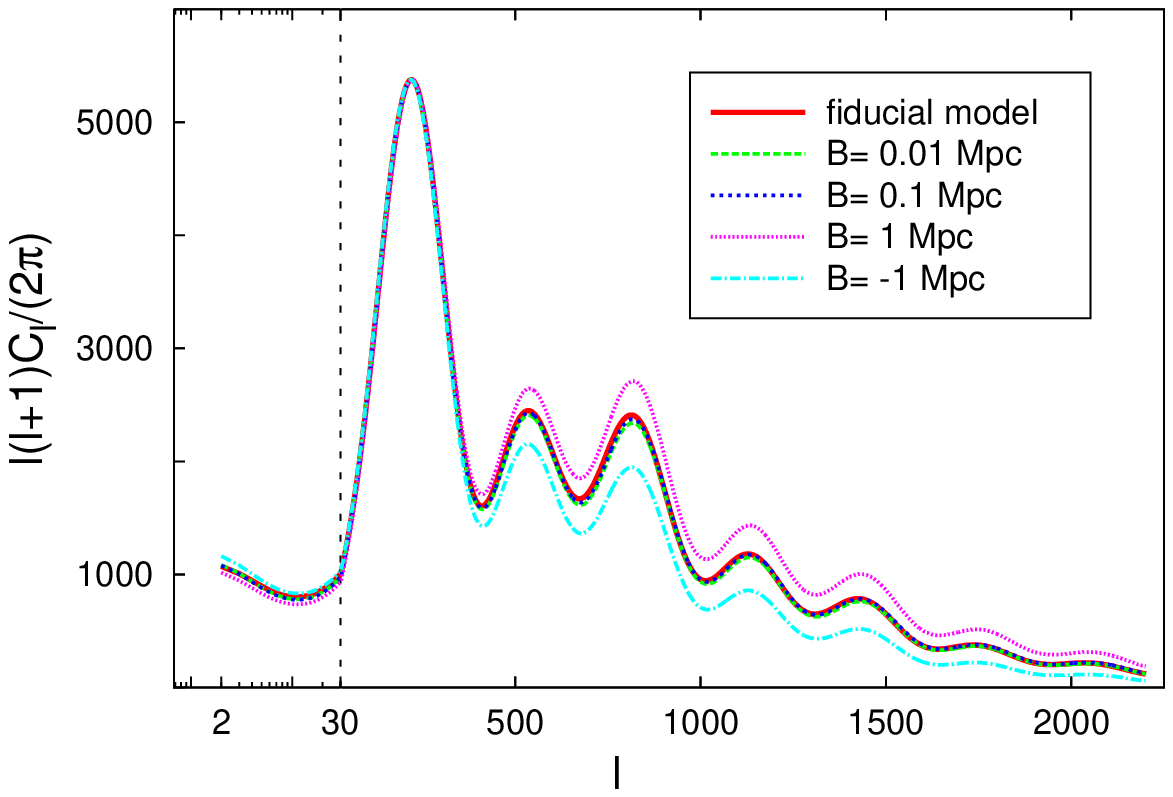}
\includegraphics[scale=0.68]{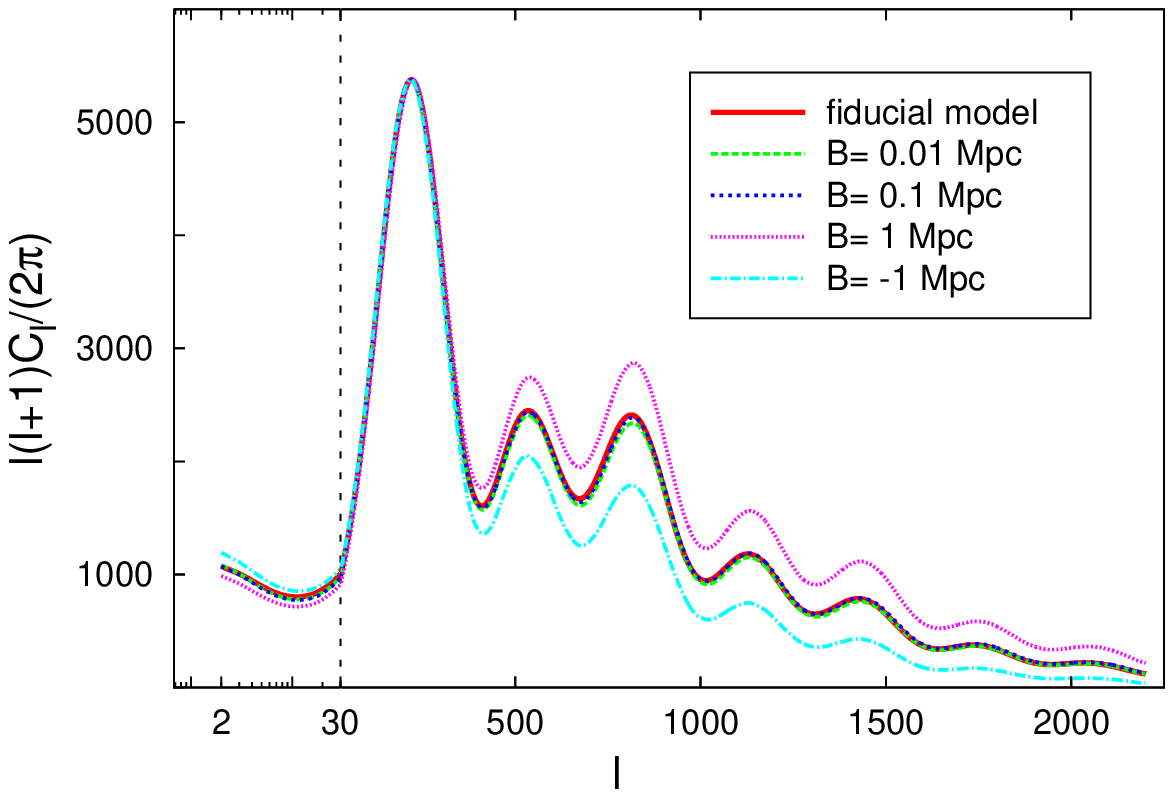}
\end{center}
\caption{The temperature auto-correlation (TT) power spectrum for the 
\textit{Wigner} scheme in 
the case where $k \gg \mH(\tc)$.   All 
models
are normalized to the maximum of the first peak of the fiducial model.  The 
power spectrum of the fiducial model is also shown in red. Different values of 
the collapse time $\tc=A/k+B $ are considered, the scalar spectral index  
$n_s=0.96$; Top: $A=-10^{2}$, Bottom: $A=-10^{6}$.  }
\label{cls-wig-dentro}
\end{figure}

Figures  \ref{cls-ind-dentro}, \ref{cls-newt-dentro} and \ref{cls-wig-dentro} 
show the prediction for the temperature fluctuation spectrum for the 
\textit{independent}, \textit{Newtonian} and \textit{Wigner} schemes in the 
case where $k \gg \mH(\tc)$. In the \textit{independent} scheme (Fig. 
\ref{cls-ind-dentro}), we obtain no changes in the CMB temperature power 
spectrum for the different values of $B$ considered and only a small  
difference with respect to the 
fiducial model (in agreement with Fig \ref{powerind}). Therefore, we expect 
that 
any value of $B$ will explain recent observational data for the values of $A$ 
analyzed in this paper. In the 
\textit{Newtonian} and \textit{Wigner} schemes (Figs. \ref{cls-newt-dentro} and 
\ref{cls-wig-dentro} respectively), for increasing value of $B$ we can observe 
an increase in the value of the secondary peaks and a decrease of the value at 
the valleys; the magnitude of the changes depending on the value of $B$, 
the change is greater for the \textit{Wigner} scheme. Therefore, we expect that 
a statistical analysis comparing the model predictions with observational data 
can give good constraints on the value of $B$ and in consequence on the time of 
collapse. In particular, from this preliminary analysis, we can expect that the 
value of $B$ should be smaller than $1$ Mpc in order to fit recent 
observational data. To get a more stringent bound we need to perform a 
statistical analysis using Planck and other CMB data and this is left for a 
forthcoming work \cite{Picci15}.
{
In summary, from the predictions for the CMB temperature power spectrum shown 
in this section, we can expect that the comparison with recent observational 
data will constrain the values of $B$  for the 
\textit{Newtonian} and \textit{Wigner} schemes in both cases analyzed in this 
paper (i.e. when the proper wavelength of the mode is smaller/bigger than the 
Hubble radius at the time of collapse). In contrast, we can expect that any 
value of $B$ will explain recent 
observational data  for the \textit{independent} scheme in the case where $k \gg 
\mH(\tc)$ and $k \ll \mH(\tc)$. }   However, we remind the reader that this analysis is 
valid only for the values of $A$ 
considered in this section; a complete analysis studying all allowed values of 
$A$ is 
in progress \cite{Picci15}.

 \section{Summary and Conclusions}\label{conclusions}
 
In this paper, we have calculated the primordial power spectra for the 
simplest inflationary model, i.e. a single-scalar field in the slow-roll 
approximation, but taking into account a collapse of the inflaton wave function 
(for each mode); the motivation for considering an objective collapse is 
to provide a precise explanation for the emergence of an inhomogeneous and 
anisotropic universe. Unlike previous works, we have considered a quasi-de 
Sitter background to perform our calculations. Even though, we have not 
assumed a particular mechanism for the collapse to happen, we have  
chosen three different collapse schemes, in which the difference between them 
is due to the variable that is affected by the collapse and their correlations. 

The three collapse schemes induced a modification to the standard 
scalar power spectrum of the form $P(k) = (\mathcal{C}/\pi^2) k^{n_s-1} 
Q(|z_k|)$, with $n_s- 1= -2\epsilon_H+2 \delta_V$ and $Q(|z_k|)$ a function of 
the time of collapse $\tc$ (recall the definition $z_k \equiv k\tc$).

Moreover, we have characterized the time of collapse as $\tc = A/k +B$.  If 
$B=0$, in all schemes, we can recover an equivalent power spectrum that is, for 
all practical purposes, indistinguishable from the standard one. The reason for 
this, is that in this case, $z_k=A$ and the dependence of primordial power 
spectrum with $k$ is the same than in  standard inflationary models.  The 
only difference in this  case is the relation between $n_s$ and the slow-roll 
parameters of the inflationary model [see Eqs. \eqref{Ayns} and 
\eqref{nscolapso}]. However, we would like to stress that present constraints 
on $n_s$ (obtained by comparison with observational data)  that affect the 
slow-roll parameters can be fulfilled by our model as well as the standard one, 
since in both cases the requirement for an efficient inflationary stage is 
$\epsilon_V \simeq \delta_V \ll 1 $.

On the other hand, by assuming $B\neq 0$, we have found small departures from 
the standard prediction that are uniquely  determined by the collapse proposal; 
this is in contrast with previous works, in which the introduction of the $B$ 
parameter was used primarily to depart from an exact scale invariant spectrum. 
Moreover, given that those works were based on assumptions that led to a 
power spectrum with an spectral index $n_s=1$, one could not tell the precise 
difference between the dependence on $k$ given purely by the dynamics in the 
aforementioned background or by the collapse proposal. 

The primordial power spectrum for the collapse models obtained in this paper 
considering  $B \neq 0$ has an additional dependence on $k$, which is similar 
to 
the one that is obtained in  standard inflationary models with a running 
spectral index.\footnote{The primordial power spectrum with a running spectral 
index has the form: $P(k) = A_s (\frac{k}{k_0})^{n_s-1+\frac{1}{2} n_{\rm run} 
\log(k/k_0)}$.} However, as it follows from Eqs. 
\eqref{pfuera3esquemas} and \eqref{pdentro3esquemas} the dependence 
on $k$ of the collapse primordial spectrum is different from the standard model 
one. Additionally, as we have mentioned before,  the lowest multipoles of the 
temperature anisotropy are best fitted by models with  $n_{\rm run} \neq 0$ 
\cite{wmap9cosmo,Planckcosmo}.

  We have also shown  some plots of the primordial spectrum resulting from 
our schemes for some specific values of the collapse parameters and compared 
them with 
the 
standard inflationary model spectrum. We have considered the case where the 
associated proper wavelength of the modes is  bigger and smaller than the 
Hubble 
radius  at the time of collapse. Finally, we have shown the effects of 
introducing a self-induced collapse of the inflaton wave function on the CMB 
temperature fluctuation spectrum. For this preliminary analysis, most of the 
collapse models proposed in this paper seem to be good candidates to explain 
present data of the CMB fluctuation spectrum. In particular, in some cases 
there 
are no differences between the prediction of the collapse models with respect 
to 
standard inflationary model. However, in other cases, for an increasing value 
of $B$ (affecting the time of collapse) there are important departures from the 
standard model prediction. Therefore, by performing a statistical analysis 
using 
all present observational data from the CMB (which is left for future work 
\cite{Picci15}) we will be able to constrain the value of $B$ and thus the 
acceptable values for the time of collapse. Finally, we have also shown that 
some particular cases could be discarded without performing any statistical 
analysis.

We conclude that a more detailed analysis involving recent observational data 
can be used to discriminate between the three collapse schemes presented. 
Nevertheless, from the theoretical point of view, we think that the 
\emph{Wigner} collapse scheme should be preferred over the other two schemes 
since it takes into account the natural correlations between the canonical 
variables that are present in the pre-collapse vacuum state. The choice of the 
Wigner distribution to describe these correlations in the present setting is 
justified by some of its standard properties regarding the ``classical limit,'' 
and, by the fact that there is a precise sense in which it is known to encode 
the correlations in question.

\begin{acknowledgements}
The authors thank D. Sudarsky for useful discussions. Support for this work was 
provided by PIP 0152/10 CONICET. GL's research funded by Consejo Nacional de 
Investigaciones Cient\'{i}ficas y T\'{e}cnicas (CONICET), Argentina and Consejo 
Nacional de Ciencia y Tecnolog\'{i}a 
\\ (CONACYT), Mexico.
\end{acknowledgements}

\appendix
\section{Deduction of Eq. \eqref{25x}}\label{appA}

Einstein equations at first order in the perturbations,  $\delta G_0^0 = 8 
\pi G \delta T_0^0$, $\delta G_i^0 = 8 \pi G \delta T_i^0$ and $\delta G^i_j = 
8 \pi G \delta T^i_j$, are given respectively by

\beq\label{00inf1}
\nabla^2 \Psi -3\mH(\mH\Phi + \Psi') = 4 \pi G [-\phi_0'^2 \Phi + \phi_0' 
\dphi' 
+ \partial_\phi V a^2 \dphi],
\eeq

\beq\label{0iinf1}
\partial_i (\mH \Phi + \Psi') = 4 \pi G \partial_i ( \phi_0' \dphi),
\eeq

\barr\label{ijinf1}
 & & [\Psi'' + \mH(2\Psi+\Phi)' + (2\mH' + \mH^2)\Phi + \textstyle{\half} 
\nabla^2 (\Phi - \Psi)] \delta^i_j \nonumber \\
&-&\textstyle{\half} \partial^i \partial_j (\Phi - \Psi) =   4 \pi G 
[\phi_0' \dphi' -\phi_0'^2 \Phi  - \partial_\phi V a^2 
\dphi]\delta^i_j. \nonumber \\
\earr
The perturbations considered $\Psi,\Phi$ correspond to the gauge-invariant 
quantities known as the Bardeen potentials; the perturbation $\dphi$ and 
$\dphi'$ are the gauge invariant perturbations associated to the inflaton 
field. 
For the case $i\not=j$ in Eq. \eqref{ijinf1}, together with appropriate  
boundary conditions (more easily seen in the Fourier transformed  version), 
leads to  $\Psi = \Phi$; from now on we will use this result.

Equations \eqref{00inf1}, \eqref{0iinf1}, \eqref{ijinf1} above, together with 
Friedmann equations, can be manipulated to yield the following expression: 

\beq\label{nabla2psi}
\nabla^2 \Psi + \mu \Psi = 4\pi G ( \omega \dphi + \phi_0' \dphi'),
\eeq
where $\mu \equiv \mH^2 - \mH'$ and $\omega \equiv 3 \mH \phi_0' + a^2 
\partial_\phi V$. Upon use of the motion equation for $\phi_0$ in the slow-roll 
approximation, i.e. $3\mH \phi_0' + a^2 \partial_\phi V \approx 0$, implies 
that 
$\omega \approx 0$. Thus \eqref{nabla2psi} becomes:

\beq\label{25b}
\nabla^2 \Psi +\mu \Psi =  4 \pi G \phi_0' \dphi'.
\eeq

\section{Explicit equations of Sec. \ref{quantum}}\label{appB}

\subsection{Equations of Sec. \ref{esquemas-colapso}}\label{appB1}

The mode function $y_k(\eta)$ of Eq. \eqref{nucolapso}, can be expressed as  
$y_k(\eta) = \textrm{Re} [y_k (\eta)] + i \textrm{Im}[y_k 
(\eta)]$; similarly, the mode function $g_k(\eta)$ can be expressed as
$g_k(\eta) = \textrm{Re} [g_k (\eta)] + i \textrm{Im}[g_k (\eta)]$ (recall that 
$g_k = y_k'-\mH y_k$). Thus, the real and imaginary parts of $y_k$ and $g_k$ 
are:

\beq\label{yreal}
 \textrm{Re} [y_k (\eta)] = \left( \frac{\pi}{4} \right)^{1/2} 
\frac{\sqrt{-k\eta}}{k^{1/2}} J_\nu (-k\eta), 
\eeq

\beq\label{yim}
\textrm{Im} [y_k (\eta)] = \left( \frac{\pi}{4} \right)^{1/2} 
\frac{\sqrt{-k\eta}}{k^{1/2}} Y_\nu (-k\eta),
\eeq

\barr\label{greal}
\textrm{Re} [g_k (\eta)] &=& k^{1/2} \left( \frac{\pi}{4} \right)^{1/2} \bigg( 
\frac{-\alpha J_{\nu} (-k\eta)}{\sqrt{-k\eta}} \nonumber \\
&+& \sqrt{-k\eta} J_{\nu+1} (-k\eta) \bigg), 
\earr

\barr\label{gim}
\textrm{Im} [g_k (\eta)] &=& k^{1/2} \left( \frac{\pi}{4} \right)^{1/2} \bigg( 
\frac{-\alpha Y_{\nu} (-k\eta)}{\sqrt{-k\eta}} \nonumber \\
&+& \sqrt{-k\eta} Y_{\nu+1} (-k\eta) \bigg),
\earr
where $\alpha \equiv 1/2 + \nu + 1/(1-\epsilon_H)$. 

With the previous expressions at hand, we can now compute $\left(\Delta 
\hat{y}^{R,I}_{\nk} (\tc) \right)^2_0$ and $\left(\Delta 
\hat{\pi}^{R,I}_{\nk} (\tc) \right)^2_0$ within the \emph{independent} and 
\emph{Newtonian} collapse schemes. The exact form of the quantum uncertainties 
can be obtained from Eqs. \eqref{operadoresRI}, this is

\barr
\left(\Delta \hat{y}^{R,I}_{\nk} (\tc) \right)^2_0 &=& \frac{L^3}{4} 
|y_k(\tc)|^2 \nonumber \\
&=& \frac{L^3 \pi |z_k|}{16 k} \left[ J_\nu^2 (|z_k|) + Y_\nu^2 (|z_k|) 
\right], 
\earr

\barr
\left(\Delta \hat{\pi}^{R,I}_{\nk} (\tc) \right)^2_0 &=& \frac{L^3}{4} 
|g_k(\tc)|^2 = \frac{L^3 \pi k }{16} \nonumber \\
&\times& \bigg[ \left( \frac{-\alpha 
J_\nu 
(|z_k|)}{\sqrt{|z_k|}} + \sqrt{z_k|} J_{\nu+1} (|z_k|) \right)^2 \nonumber \\
&+& \left( \frac{-\alpha Y_\nu (|z_k|)}{\sqrt{|z_k|}} + \sqrt{|z_k|} Y_{\nu+1} 
(|z_k|) \right)^2 \bigg], \nonumber \\
\earr
with $z_k \equiv k \tc$ and $\tc$ the time of collapse for each mode. Note 
that this quantities have the information that the background space-time is 
quasi-de Sitter

In the case of the \emph{Wigner} scheme, the explicit formulas characterizing 
the parameters $\Lambda_k (z_k)$ and $\Theta_k (z_k)$ are:

\barr\label{lambdak}
&\Lambda_k& = (2L)^{3/2} \sqrt{\frac{\pi |z_k|}{4k}} \left[ J_\nu^2 (|z_k|) + 
Y_\nu^2 (|z_k|) \right]^{1/2} \bigg[ S(|z_k|)  \nonumber \\
&-& \sqrt{S^2 (|z_k|) -  \left(\frac{\pi 
|z_k|}{2}\right)^2    (J_\nu^2 (|z_k|) + Y_\nu^2 (|z_k|) )^2} \bigg]^{-1/2}, 
\nonumber \\
\earr

\barr\label{2thetak}
&\tan 2\Theta_k& =  -\frac{\pi^2 |z_k|}{4}  \left[ J_\nu^2 (|z_k|) + Y_\nu^2 
(|z_k|) \right] \nonumber \\
&\times& \left[  S(|z_k|) -  \frac{\pi |z_k|}{8} \left( 
J_\nu^2 (|z_k|) 
+ 
Y_\nu^2 (|z_k|) \right)^2 \right]^{-1} \nonumber \\
&\times& [ -2\nu \left( J_\nu^2 (|z_k|) + Y_\nu^2 (|z_k)  \right) + |z_k| 
\nonumber \\
&\times& \left( J_\nu (|z_k|) J_{\nu+1} (|z_k|) +  Y_\nu (|z_k|) Y_{\nu+1} 
(|z_k|)       
\right)  ], \nonumber \\  
\earr
where

\barr\label{defS}
&S(|z_k|)& \equiv 1 + \frac{\pi^2}{16} \bigg\{  |z_k|^2 (J_\nu^2 (|z_k|) + 
Y_\nu^2 (|z_k|))^2 
\nonumber \\ 
&+& 4 \bigg[ J_\nu^2 (|z_k|) + Y_\nu^2 (|z_k|) - |z_k| ( J_\nu (|z_k|) 
J_{\nu+1} 
(|z_k|) \nonumber \\ 
&+& Y_\nu (|z_k|) Y_{\nu+1} (|z_k|) )   \bigg]^2  \bigg\}. 
\nonumber \\
\earr

\subsection{Equations of Sec. \ref{curvatura3esquemas}}\label{appB2}

In order to deduce Eq. \eqref{expecpitexto}, we introduce the quantity  
$d^{R,I}_{\nk} \equiv 
\langle\Theta|\ann_{\nk}^{R,I}|\Theta\rangle $ that   determines   the  
expectation value of the field and momentum  operator for the mode  $\nk$ at 
all 
times after the collapse. That is, from Eq. \eqref{operadoresRI}, we have

\beq\label{expeceta}
\bra \hat{\pi}_{\nk}^{R,I} (\eta) \ket_{\Theta} = \sqrt{2} 
\mathcal{R}[g_k(\eta) 
d_{\nk}^{R,I}],
\eeq
which corresponds to expectation values at any time after the collapse in the 
post-collapse state $| \Theta \ket$. One can then relate the value of 
$d_{\nk}^{R,I}$ with the value of the expectation value of the fields operators 
at the time of collapse $\bra \hat{y}_{\nk}^{R,I} (\tc) \ket_{\Theta} = 
\sqrt{2} \mathcal{R}[y_k(\tc) d_{\nk}^{R,I}]$,  $\bra 
\hat{\pi}_{\nk}^{R,I} 
(\tc) \ket_{\Theta} = \sqrt{2} \mathcal{R}[g_k(\tc) d_{\nk}^{R,I}]$. 
Using the latter relations to express $d_{\nk}^{R,I}$ in terms of the 
expectation values at the time of collapse and substituting it in 
\eqref{expeceta}, we obtain an expression for the expectation value of the 
momentum field operator in terms of the expectation value at the time of 
collapse

\barr\label{expecpi}
\bra \hat{\pi}_{\nk}^{R,I} (\eta) \ket_\Theta &=& F(k\eta,z_k)  \bra 
\hat{y}_{\nk}^{R,I} (\tc) \ket_\Theta \nn 
&+& G(k\eta,z_k) \bra \hat{\pi}_{\nk}^{R,I} (\tc) \ket_\Theta,
\earr
with

\barr\label{F}
F(k\eta,z_k) &\equiv& \frac{k\pi}{4} \bigg\{   \left(  \frac{-\alpha Y_\nu 
(|z_k|)}{\sqrt{|z_k|}} + \sqrt{|z_k|} Y_{\nu+1} (|z_k|)  \right) \nn
&\times& \left( 
\frac{-\alpha J_\nu (|k\eta|)}{\sqrt{|k\eta|}} + \sqrt{|k\eta|} J_{\nu+1} 
(|k\eta|)  \right) \nonumber \\
&-& \left( \frac{-\alpha J_\nu (|z_k|)}{\sqrt{|z_k|}} + \sqrt{|z_k|} J_{\nu+1} 
(|z_k|)  \right) \nn
&\times& \left( \frac{-\alpha Y_\nu (|k\eta|)}{\sqrt{|k\eta|}} + 
\sqrt{|k\eta|} Y_{\nu+1} (|k\eta|) \right) \bigg\}, \nn
\earr

\barr\label{G}
& & G(k\eta,z_k) \equiv \frac{\pi \sqrt{|z_k|}}{4} \bigg\{ J_{\nu} (|z_k|) 
\nn
&\times& \left[ 
 \frac{-\alpha Y_\nu (|k\eta|)}{\sqrt{|k\eta|}} + \sqrt{|k\eta|} Y_{\nu+1} 
(|k\eta|)  \right] \nonumber \\
&-&  Y_{\nu} (|z_k|) \left[   \frac{-\alpha J_\nu (|k\eta|)}{\sqrt{|k\eta|}} + 
\sqrt{|k\eta|} J_{\nu+1} (|k\eta|)  \right]  \bigg\}.
\earr

Furthermore, given Eqs. \eqref{expecpi} and \eqref{masterpi}, we can find the 
curvature perturbation in the longitudinal gauge within the three collapse 
schemes. 

For the \textit{independent scheme} the curvature perturbation is:

\barr\label{psiindep}
& & \psiind (\eta) =  \frac{ \sqrt{L^3 \pi \epsilon_V} H }{2^{5/2} M_P k^2} \nn
&\times& \bigg\{ F(k\eta,z_k) X_{\nk,1} \sqrt{\frac{|z_k|}{k}}  \left[ J_\nu^2 
(|z_k|) + 
Y_\nu^2 (|z_k|) \right]^{1/2} \nonumber \\
&+& G(k\eta,z_k) X_{\nk,2} \sqrt{k}  \nn
&\times& \bigg[ \left( \frac{-\alpha J_\nu 
(|z_k|)}{\sqrt{|z_k|}} + \sqrt{|z_k|} J_{\nu+1} (|z_k|) \right)^2   \nonumber \\
&+& \left( \frac{-\alpha Y_\nu (|z_k|)}{\sqrt{|z_k|}} + \sqrt{|z_k|} Y_{\nu+1} 
(|z_k|) \right)^2    \bigg]^{1/2}   \bigg\}. 
\earr
Meanwhile, for the \textit{Newtonian scheme} we have

\barr\label{psinewt}
\psinew (\eta) &=&   \frac{ \sqrt{L^3 \pi \epsilon_V} H }{2^{5/2} M_P k^2}  
G(k\eta,z_k) X_{\nk,2} \sqrt{k}    \nn
&\times& \bigg[ \left( \frac{-\alpha J_\nu 
(|z_k|)}{\sqrt{|z_k|}} + \sqrt{|z_k|} J_{\nu+1} (|z_k|) \right)^2   \nonumber \\
&+&  \left( \frac{-\alpha Y_\nu (|z_k|)}{\sqrt{|z_k|}} + \sqrt{|z_k|} Y_{\nu+1} 
(|z_k|) \right)^2    \bigg]^{1/2} 
\earr
and finally for the \textit{Wigner scheme}

\barr\label{psiwigner}
\psiwig (\eta) &=&   \sqrt{\frac{\epsilon_V}{2}} \frac{H}{M_P k^2}  \Lambda_k 
[ F(k\eta,z_k) \cos \Theta_k \nn 
&+& k G(k\eta,z_k) \sin \Theta_k ] 
X_{\nk},
\earr
with $X_{\nk} \equiv x_{\nk}^R + i x_{\nk}^I$.

\subsection{Definitions of the functions $M(|z_k|)$, $N(|z_k|)$ and $W(|z_k|)$ }
\label{appB3}

\barr\label{mzk}
M(|z_k|) &\equiv& -\sqrt{|z_k|} \bigg[ \frac{-\alpha J_\nu 
(|z_k|)}{\sqrt{|z_k|}} 
+ \sqrt{|z_k|} J_{\nu+1} (|z_k|) \bigg]  \nn
&\times& \left[ J_\nu^2 (|z_k|) + Y_\nu^2 
(|z_k|) \right]^{1/2},
\earr

\barr\label{nzk}
& & N(|z_k|) \equiv \sqrt{|z_k|} J_{\nu} (|z_k|) \nn
&\times& \bigg[ \left( \frac{-\alpha 
J_\nu 
(|z_k|)}{\sqrt{|z_k|}} + \sqrt{|z_k|} J_{\nu+1} (|z_k|) \right)^2   \nonumber \\
&+& \left( \frac{-\alpha Y_\nu (|z_k|)}{\sqrt{|z_k|}} + \sqrt{|z_k|} Y_{\nu+1} 
(|z_k|) \right)^2    \bigg]^{1/2},
\earr

\barr\label{wzk}
& & W (|z_k|) \equiv  \frac{2k^{1/2}}{\pi^{1/2}L^{3/2}} \nn
&\times& \bigg[- \bigg( 
\frac{-\alpha J_\nu (|z_k|)}{\sqrt{|z_k|}} + \sqrt{|z_k|} J_{\nu+1} (|z_k|) 
\bigg) \Lambda_k  \cos \Theta_k  \nonumber \\
&+& \sqrt{|z_k|} J_{\nu} (|z_k|)  \Lambda_k \sin \Theta_k \bigg].
\earr

\section{Equations of Sec. \ref{oquantities}}\label{appC}

The explicit expressions for $|a_{lm}|^2_{\text{ML}}$ can be found by 
substituting $\mR_k$, given in Eq. \eqref{R3esquemas} into Eq. \eqref{alm2}, 
and then making the identification $|a_{lm}|^2_{\text{ML}}= 
\overline{|a_{lm}|^2}$

\barr\label{alm2indep}
& & |a_{lm}|^{2 \text{$ $ ind} }_{\text{ML}} = 16 \pi^2  
\frac{\mathcal{C}}{L^3} 
\sum_{\nk,\nk'}  \frac{  j_l (kR_D) j_l(k'R_D)}{k^{3/2} k'^{3/2}} 
\nn
&\times& Y_{lm}^\star 
(\hat{k}) Y_{lm} (\hat{k}') T(k) T(k') \bigg( M(|z_{k}|)  M(|z_{k'}|) 
\overline{X_{\nk,1} X_{\nk',1}^\star} \nonumber \\ 
&+&  N(|z_{k}|)   N(|z_{k'}|) \overline{X_{\nk,2} X_{\nk',2}^\star} \bigg) 
(kk')^{3/2-\nu},
\earr

\barr\label{alm2newt}
& & |a_{lm}|^{2 \text{$ $ newt} }_{\text{ML}} = {16 \pi^2} 
\frac{\mathcal{C}}{L^3} 
\sum_{\nk,\nk'}  \frac{ j_l (kR_D) j_l(k'R_D)}{k^{3/2} k'^{3/2}} \nn
&\times& Y_{lm}^\star 
(\hat{k}) Y_{lm} (\hat{k}') T(k) T(k') N(|z_{k}|)   N(|z_{k'}|) \nn
&\times&  \overline{X_{\nk,2} X_{\nk',2}^\star} 
(kk')^{3/2-\nu},
\earr

\barr\label{alm2wig}
& & |a_{lm}|^{2 \text{$ $ wig} }_{\text{ML}} = {16 \pi^2} 
\frac{\mathcal{C}}{L^3} 
\sum_{\nk,\nk'}  \frac{ j_l (kR_D) j_l(k'R_D)}{k^{3/2} k'^{3/2}} 
\nn
&\times& Y_{lm}^\star (\hat{k}) Y_{lm} (\hat{k}') T(k) T(k')  \nonumber \\ 
&\times& W  (|z_{k}|)  W (|z_{k'}|)  \overline{X_{\nk} X_{\nk'}^\star} 
(kk')^{3/2-\nu},
\earr
with

\beq\label{Capp}
\mathcal{C}\equiv \frac{\pi}{ M_P^2 \epsilon_H} \left( 2^{\nu-11/2} 
\Gamma(\nu-1) H |\eta|^{3/2-\nu}    \right)^2,
\eeq
where we used the fact that in slow-roll inflation $\epsilon_V \simeq 
\epsilon_H$.

\section{On the meaning of the power spectrum within the collapse 
  proposal}\label{appD}

In order to obtain the power spectrum in the traditional inflationary scenario 
one needs to compute the quantum two-point correlation function. That is, if $\hat{\Psi}$ 
represents the quantum operator associated to the metric perturbation, then the (scalar) 
power spectrum is obtained from

\beq
\bra 0 | \hat \Psi_{\nk} \hat \Psi_{\nk'} | 0 \ket = \frac{2 \pi^2}{k^3} P(k) 
\delta(\nk-\nk')
\eeq

On the other hand, let us recall that in general, the definition of the power 
spectrum is given in terms of $\Psi_{\nk}$, i.e. a classical stochastic field 
not a quantum operator. Therefore, the standard approach is based on the 
identification

\beq\label{relacion}
\bra 0 | \hat \Psi_{\nk} \hat \Psi_{\nk'} | 0 \ket = \overline{\Psi_{\nk} 
  \Psi_{\nk'}}
\eeq
with $\overline{\Psi_{\nk}\Psi_{\nk'}}$ denoting an average over an ensemble of 
classical stochastic fields. The justification for the relation above relies on 
arguments based on decoherence and the squeezing nature of the evolved vacuum 
state \cite{grishchuk,kiefer} (although we do not subscribe to such arguments 
for the reasons exposed in \cite{LLS13,Shortcomings}). Nevertheless, the 
explanation about the fact that a single outcome or realization $\Psi_{\nk}$ 
has 
emerged is incomplete.

Therefore, in order to show explicitly the manner in which we can obtain an 
equivalent power spectrum within the collapse proposal, we start by 
focusing on the temperature 
anisotropies of the CMB observed today on the celestial
two-sphere and its relation to the scalar metric perturbation $\Psi$, which can 
be approximated by
\begin{equation}\label{deltaT}
\frac{\delta T}{T_0} (\theta,\varphi) \simeq \frac{1}{3} \Psi.
\end{equation}
On the other hand, the observational data are  described  in terms of the 
coefficients  $a_{lm}$
of the multipolar series expansion
\begin{equation}\label{expansion.alpha}
\begin{split}
\frac{\delta T}{T_0}(\theta,\varphi)=\sum_{lm}a_{lm}Y_{lm}(\theta,\varphi),
\\
a_{lm}=\int 
\frac{\delta T}{T_0}(\theta,\varphi)Y^*_{lm}(\theta,\varphi)d\Omega,
\end{split}
\end{equation}
here $\theta$ and $\varphi$ are the coordinates on the celestial two-sphere, 
with $Y_{lm}(\theta,\varphi)$ as the spherical harmonics.

The  value  for the  quantities $a_{lm}$  are then given by

\begin{equation}\label{alm2p}
a_{lm} = \frac{4 \pi i^l}{3}   \int \frac{d^3{k}}{(2 \pi)^3} j_l (kR_D) 
Y_{lm}^* (\hat{k}) T (k) \Psi_{\vec{k}},
\end{equation}
with $j_l (kR_D)$ as the spherical Bessel function of order $l$ and $R_D$ is 
the comoving radius of the last scattering surface. We have
explicitly included the modifications associated with latetime physics encoded 
in the transfer functions $T(k)$. The metric perturbation $\Psi_{\nk}$ is 
the primordial curvature perturbation.

Now, the problem is that, if we compute the expectation value of the right-hand 
side (i.e., identifying $\langle \hat{\Psi} \rangle = \Psi$) in
the vacuum state $| 0  \rangle$, we obtain 0, while it is clear that for
any given $l, m$, the measured value of this quantity is not 0. That is, if we 
rely in this case on the one-point function and
the standard identification of quantum averages with classical ensemble 
average, we find a large conflict between
expectation and observation. Nevertheless, in the standard approach 
somehow (e.g. by invoking decoherence, squeezing of the vacuum, many-world 
interpretation of quantum mechanics, etc.) occurs the transition $\hat 
\Psi_{\nk} \to \Psi_{\nk} = A e^{i \alpha_{\nk}}$ with $\alpha_{\nk}$ a random 
phase and $A$ is identified with the quantum uncertainty of $\hat \Psi_{\nk}$, 
i.e. $A^2=\bra 0 | \hat \Psi_{\nk}^2 |0\ket$, but the random nature of 
$\Psi_{\nk}$ remains unclear. 

In our approach, the random nature comes directly 
from the stochastic aspects of the quantum dynamical-reduction, i.e. from the 
self-induced collapse. Thus, using Eq. \eqref{masterpi} in Eq. \eqref{alm2p} we 
obtain

\barr\label{almd1}
a_{lm} &=& \frac{4 \pi i^l}{3}   \int \frac{d^3{k}}{(2 \pi)^3} j_l (kR_D) 
Y_{lm}^* (\hat{k}) \Delta (k) \Psi_{\nk} \nn
&=& \frac{4 \pi i^l }{3} \sqrt{\frac{\epsilon_H}{2}}  \frac{H}{M_P} \int 
\frac{d^3{k}}{(2 \pi)^3} j_l (kR_D) 
Y_{lm}^* (\hat{k}) T (k) \frac{\bra \hat \pi_{\nk} \ket}{k^2}. \nn
\earr
The previous expression shows how the expectation value of the momentum field 
in the post-collapse state acts as a source for the coefficients $a_{lm}$. 

Furthermore, the angular power spectrum is defined by

\beq
C_{l} = \frac{1}{2l+1} \sum_m |a_{lm}|^2.
\eeq
For the reasons presented in Sec. \ref{oquantities}, we can identify the 
observed value $|a_{lm}|^2$ with the most likely value of $|a_{lm}|^2_{ML}$ and 
in turn, assume that the most likely value coincides approximately with the 
average $\overline{|a_{lm}|^2}$. Thus, in our approach, the observed $C_l$ 
coincides with 

\beq
C_{l} \simeq  \frac{1}{2l+1} \sum_m \overline{|a_{lm}|^2}.
\eeq
From Eq. \eqref{almd1} we obtain

\barr\label{almd2}
 \overline{|a_{lm}|^2} &=& \left (\frac{4 \pi}{3} \right)^2   \int 
 \frac{d^3{k} d^3{k'}}{(2 \pi)^6} j_l (kR_D) j_l (k'R_D) 
\nn
 &\times&  Y_{lm}^* (\hat{k}) Y_{lm} (\hat k')  T (k) T(k') 
\overline{\Psi_{\nk} \Psi_{\nk'}} \nn 
 &=& \left (\frac{4 \pi}{3} \right)^2   \int 
 \frac{d^3{k} d^3{k'}}{(2 \pi)^6} j_l (kR_D) j_l (k'R_D) 
 \nn
 &\times&  Y_{lm}^* (\hat{k}) Y_{lm} (\hat k') T(k) T(k') \bigg[ 
\frac{\epsilon}{2} \frac{H^2}{M_P^2} 
 \overline{\frac{\bra \hat \pi_{\nk} \ket \bra \hat \pi_{\nk'} \ket }{k^2 
k^{'2}}} \bigg]. \nn 
\earr
Consequently using the generic definition of the power spectrum [i.e. not 
relying on the identification \eqref{relacion}]

\beq
\overline{\Psi_{\nk} \Psi_{\nk'}} \equiv \frac{2 \pi^2}{k^3} P(k) 
\delta(\nk-\nk'),
\eeq
and also using Eq. \eqref{almd2}, the power spectrum, associated to 
$\Psi_{\nk}$, in our approach is given by

\beq
P(k) = \frac{\epsilon}{2} \frac{H^2}{M_P^2} 
 \overline{\bra \hat \pi_{\nk} (\eta) \ket \bra \hat \pi_{\nk'} (\eta) \ket }.
\eeq

The quantity $\overline{\bra \hat \pi_{\nk} (\eta) \ket \bra 
\hat \pi_{\nk'} (\eta) \ket }$ is obtained by using Eq. \eqref{expecpitexto} in 
the limit   $-k\eta \to 0$, i.e. when the proper wavelength of the modes of 
interest are bigger than the Hubble radius, but taking into account that Eq. 
\eqref{expecpitexto} is different for each collapse scheme.

\bibliography{bibliografia}
\bibliographystyle{apsrev}

\end{document}